\documentclass[preprint,linenumbers]{aastex7}
\usepackage{graphicx} 
\usepackage{natbib}
\usepackage{comment}
\usepackage{rotating}
\usepackage{lineno}
\usepackage{url}
\usepackage{soul}
\linenumbers

\newcommand{\Fermi}{{\sl Fermi}}
\newcommand{\be}{\begin{equation}}
\newcommand{\ee}{\end{equation}}
\newcommand{\bi}{\begin{itemize}}
\newcommand{\ei}{\end{itemize}}
\newcommand{\ben}{\begin{enumerate}}
\newcommand{\een}{\end{enumerate}}

\newcommand{\new}[1]{\textcolor{blue}{#1}}
\newcommand{\old}[1]{\textcolor{red}{\st{#1}}}

\date{ }
\begin{document}
\title{Exploring the nature of Galactic unassociated sources detected by the {\sl Fermi-LAT}}
\author[0000-0002-6606-2816]{F.~Acero}
\affiliation{Universit\'e Paris-Saclay, Universit\'e Paris Cit\'e, CEA, CNRS, AIM, F-91191 Gif-sur-Yvette Cedex, France}
\affiliation{FSLAC IRL 2009, CNRS/IAC, La Laguna, Tenerife, Spain}
\email{}
\author[0000-0002-2028-9230]{A.~Acharyya}
\affiliation{Center for Cosmology and Particle Physics Phenomenology, University of Southern Denmark, Campusvej 55, DK-5230 Odense M, Denmark}
\email{}
\author[0009-0004-3923-9884]{A.~Adelfio}
\affiliation{Istituto Nazionale di Fisica Nucleare, Sezione di Perugia, I-06123 Perugia, Italy}
\email{}
\author[0000-0002-6584-1703]{M.~Ajello}
\affiliation{Department of Physics and Astronomy, Clemson University, Kinard Lab of Physics, Clemson, SC 29634-0978, USA}
\email{}
\author[0009-0001-2927-8968]{E.~Aviano}
\affiliation{Dipartimento di Fisica, Universit\`a di Trieste, I-34127 Trieste, Italy}
\affiliation{Istituto Nazionale di Fisica Nucleare, Sezione di Trieste, I-34127 Trieste, Italy}
\email{}
\author[0000-0002-9785-7726]{L.~Baldini}
\affiliation{Universit\`a di Pisa, Dipartimento di Fisica E. Fermi, I-56127 Pisa, Italy}
\affiliation{Istituto Nazionale di Fisica Nucleare, Sezione di Pisa, I-56127 Pisa, Italy}
\email{}
\author[0000-0002-8784-2977]{J.~Ballet}
\email[show]{jean.ballet@normalesup.org}
\affiliation{Universit\'e Paris-Saclay, Universit\'e Paris Cit\'e, CEA, CNRS, AIM, F-91191 Gif-sur-Yvette Cedex, France}
\author[0000-0001-7233-9546]{C.~Bartolini}
\affiliation{Istituto Nazionale di Fisica Nucleare, Sezione di Bari, I-70126 Bari, Italy}
\affiliation{Universit\`a degli studi di Trento, via Calepina 14, 38122 Trento, Italy}
\email{}
\author[0000-0002-6954-8862]{D.~Bastieri}
\affiliation{Istituto Nazionale di Fisica Nucleare, Sezione di Padova, I-35131 Padova, Italy}
\affiliation{Dipartimento di Fisica e Astronomia ``G. Galilei'', Universit\`a di Padova, Via F. Marzolo, 8, I-35131 Padova, Italy}
\affiliation{Center for Space Studies and Activities ``G. Colombo", University of Padova, Via Venezia 15, I-35131 Padova, Italy}
\email{}
\author[0000-0002-6729-9022]{J.~Becerra~Gonzalez}
\affiliation{Instituto de Astrof\'isica de Canarias and Universidad de La Laguna, Dpto. Astrof\'isica, 38200 La Laguna, Tenerife, Spain}
\email{}
% ...
\old{\email{david.smith@u-bordeaux.fr}}
\new{\email{}}
\affiliation{Laboratoire d'Astrophysique de Bordeaux, Universit\'e de Bordeaux, CNRS, B18N, all\'ee Geoffroy Saint-Hilaire, F-33615 Pessac, France}
\author[0000-0003-0802-3453]{G.~Spandre}
\affiliation{Istituto Nazionale di Fisica Nucleare, Sezione di Pisa, I-56127 Pisa, Italy}
\email{}
\author[0000-0001-6688-8864]{P.~Spinelli}
\affiliation{Dipartimento di Fisica ``M. Merlin" dell'Universit\`a e del Politecnico di Bari, via Amendola 173, I-70126 Bari, Italy}
\affiliation{Istituto Nazionale di Fisica Nucleare, Sezione di Bari, I-70126 Bari, Italy}
\email{}
\author[0000-0003-3799-5489]{A.~W.~Strong}
\affiliation{Max-Planck Institut f\"ur extraterrestrische Physik, D-85748 Garching, Germany}
\email{}
\author[0000-0003-2911-2025]{D.~J.~Suson}
\affiliation{Purdue University Northwest, Hammond, IN 46323, USA}
\email{}
\author[0000-0002-1721-7252]{H.~Tajima}
\affiliation{Nagoya University, Institute for Space-Earth Environmental Research, Furo-cho, Chikusa-ku, Nagoya 464-8601, Japan}
\affiliation{Kobayashi-Maskawa Institute for the Origin of Particles and the Universe, Nagoya University, Furo-cho, Chikusa-ku, Nagoya, Japan}
\email{}
\author[0000-0002-9051-1677]{J.~B.~Thayer}
\affiliation{W. W. Hansen Experimental Physics Laboratory, Kavli Institute for Particle Astrophysics and Cosmology, Department of Physics and SLAC National Accelerator Laboratory, Stanford University, Stanford, CA 94305, USA}
\email{}
\author[0000-0001-5217-9135]{D.~J.~Thompson}
\affiliation{Astrophysics Science Division, NASA Goddard Space Flight Center, Greenbelt, MD 20771, USA}
\affiliation{Department of Astronomy, University of Maryland, College Park, MD 20742, USA}
\email{}
\author[0000-0001-7523-570X]{L.~Tibaldo}
\affiliation{IRAP, Universit\'e de Toulouse, CNRS, UPS, CNES, F-31028 Toulouse, France}
\email{}
\author[0000-0002-1522-9065]{D.~F.~Torres}
\affiliation{Institute of Space Sciences (ICE, CSIC), Campus UAB, Carrer de Magrans s/n, E-08193 Barcelona, Spain and Institut d'Estudis Espacials de Catalunya (IEEC), E-08034 Barcelona, Spain and Instituci\'o Catalana de Recerca i Estudis Avan\c{c}ats (ICREA), E-08010 Barcelona, Spain}
\email{}
\author[0000-0002-8090-6528]{J.~Valverde}
\affiliation{Department of Physics, Marquette University, Milwaukee, WI 53201, USA}
\email{}
\author[0000-0002-7376-3151]{K.~Wood}
\email[show]{kentswood@gmail.com}
\affiliation{Praxis Inc., Alexandria, VA 22303, resident at Naval Research Laboratory, Washington, DC 20375, USA}
\author[]{Q.~Yu}
\affiliation{Institut f\"ur Astro- und Teilchenphysik, Leopold-Franzens-Universit\"at Innsbruck, A-6020 Innsbruck, Austria}
\email{}
\author[0000-0001-8484-7791]{G.~Zaharijas}
\affiliation{Center for Astrophysics and Cosmology, University of Nova Gorica, Nova Gorica, Slovenia}
\email{}
\author[0000-0003-2839-1325]{W.~Zhang}
\affiliation{Key Laboratory for Particle Astrophysics, Institute of High Energy Physics, Beijing 100049, China}
\affiliation{Institute of Space Sciences (ICE, CSIC), Campus UAB, Carrer de Magrans s/n, E-08193 Barcelona, Spain; and Institut d'Estudis Espacials de Catalunya (IEEC), E-08034 Barcelona, Spain}
\email{}

\date{February 2026}
\begin{abstract}
We investigate the nature of the unassociated sources detected by the Fermi-LAT close ($|b|<10^\circ$) to the Galactic plane, representing 16\% of all sources in the 4FGL-DR4  catalog. The bulk of these sources (referred to as soft Galactic unassociated sources, SGUs) exhibit properties not found in known classes of gamma-ray emitters, as confirmed by a machine-learning classification approach.  In particular, these properties include  a steep, curved spectrum peaking below 1 GeV and a specific Galactic-latitude distribution with both a narrow  and a broad component (dubbed the spike and the shoulder, respectively). Some source clusters are highlighted. New plausible source classes are explored, but only star-forming regions are found to account for a significant fraction (at most 10\%) of the unassociated population. A thorough search for counterparts to the 175 brightest sources brings out a number of plausible counterparts but does not reveal clues about the nature of the whole population. We investigate the possibility that SGUs originate from mismodeled clumps of diffuse emission. Using Monte Carlo simulations, the SGU spectra can be reproduced in this scenario under an ad hoc condition concerning the clump spatial extension. The possible connection between the SGUs and gas not accounted for by the $^{12}$CO tracer is explored using the $^{13}$CO MOPRA data but leads to inconclusive results. The origin of SGUs being related to diffuse emission remains plausible. 
However, a scenario whereby SGUs represent a new class of gamma-ray emitters cannot be fully excluded.

\end{abstract}

{\tableofcontents}

\section{Introduction}
\label{sec:intro}

The fourth and most recent release of the fourth {\sl Fermi}-LAT source catalog \citep[][hereafter 4FGL-DR4]{LAT23_4FGL_DR4} includes 7114 gamma-ray point sources detected above 50 MeV over the first 14 years of operation.
While AGNs (active galactic nuclei, mostly blazars) dominate the extragalactic sky, pulsars and their nebulae, supernova remnants, and X-ray binaries represent the main classes of Galactic sources 
established so far. Firm identifications are based on periodic variability for LAT-detected pulsars or X-ray binaries, correlated variability at other wavelengths for AGNs or spatial morphology related to that found in another band for extended sources. Associations, on the other hand, are simply based on spatial coincidences.  Overall, about one third of the detected sources remain unassociated, i.e., no high-confidence counterparts could be found in catalogs of known classes of gamma-ray emitters. This fraction has remained more or less constant for the various {\Fermi}-LAT catalogs released since 2FGL, when the current association scheme was introduced. The large unassociated population has triggered numerous works aiming to identify possible counterparts using existing or new dedicated multiwavelength data \citep{Ace13,Pet13,Str13,Lan15,Sch17, Ker21,Bru23,May24}.   
A very active effort has been devoted to classifying the sources into established classes on the basis of their gamma-ray properties using machine-learning techniques  \citep{Ack12,Mir16,Saz16,Lef17,Kov19,Luo:2020bbk,Fin20,Bha21,Ger21,Ker21,Jof22,Mal23,Zhu23}. A particular motivation was to identify the most promising pulsar candidates to guide radio observations and detect pulsations \citep[e.g., ][]{Saz16}.    

The observed unassociated fraction depends strongly on Galactic latitude. While it averages around 25\% in the high-latitude sky ($|b|>10^\circ$), it reaches 54\% closer to the Galactic plane: we call these 1129 point sources GUs, for Galactic Unassociated sources. The latter fraction 
is comparable to that in the third EGRET catalog \citep{3EGCatalog} despite the much improved position uncertainties \citep{Atw09} the LAT affords (over ten-fold when comparing 4FGL and 3EG for sources near the detection threshold). 
The vast majority of unassociated sources with $|b|>10^\circ$ exhibit properties akin to blazars', while a smaller population resembles millisecond pulsars \citep{LAT22_4FGL_DR3}. Works exploring the nature of the high-latitude sources confirm that blazars are the prime contenders \citep[e.g.,][]{Ulg24}. 

 In contrast, the bulk of the population of point-like unassociated sources located closer to the Galactic plane show peculiar features that set them apart from identified classes \citep{LAT22_4FGL_DR3}. In particular, the very soft energy spectra exhibited by many sources (837 with power-law photon indices, $\Gamma$, greater than 2.4 in 4FGL-DR4) earned them the designation ``soft Galactic unassociated sources'' (SGUs). In this paper, the GUs refer to the whole population of unassociated sources in the Galactic plane ($|b|<10^\circ$), while the SGUs designate the soft-spectra ($\Gamma>2.4$) subpopulation.  Along the line of 4FGL-DR4, we define the spike and the shoulder as the $|b|<1\arcdeg$ and $1<|b|<10\arcdeg$ GU spatial components, respectively. These definitions are purely driven by the data and do not correspond to particular Galactic structures.  
 Table \ref{tab:census} provides tallies of different populations considered in the following. Numbers for the two main  classes of  associated Galactic sources, namely millisecond pulsars (MSPs) and young pulsars (PSRs), are also given for comparison. 
\begin{table*}
\centering
\footnotesize
\begin{tabular}{|c|c|c|c|c|}
\hline
                  &  unassociated (GUs) & SGUs &  MSPs & PSRs\\
\hline
total             &  1129               & 837  &  139  & 137 \\
spike             &   259               &  208  & 2 & 72 \\
shoulder         &   870               &  629  &  36 & 50 \\
\hline
\end{tabular}

\caption{Tallies of different unassociated populations. The spike and shoulder correspond to the $|b|<1\arcdeg$ and $1\arcdeg <|b|<10\arcdeg$ sky regions, respectively.  Numbers for MSPs and PSRs are given for comparison.}
\label{tab:census}
\end{table*}

 The purpose of this paper is to present the avenues pursued to shed light on the GU nature, with emphasis on the SGUs. 
 Potentially, new classes of gamma-ray emitters could lurk in this population, in addition to the classes already discovered by the {\Fermi}-LAT in the Galaxy. 
 Although it seems very unlikely that one or a few up-to-now unknown classes could make up a sizeable fraction of the SGUs, given their large number, it is nevertheless useful to explore different physically-motivated  possibilities for such new classes. 
 The recent release of the {\sl eRosita} X-ray catalog offers an interesting opportunity to look for potential SGU counterparts in this nearby band, complementing the Swift catalog \citep{Str13}.    

The detection of LAT sources in the Galactic plane is carried out against the background of diffuse emission resulting from the interaction of cosmic rays with interstellar matter and low-energy radiation fields. 
We also consider the standpoint, opposite to that presented above,  that SGUs are mainly spurious and result from mismodeled interstellar emission.  We recall  that the presence of large-scale structures like Loop I or the {\sl Fermi} Bubbles (and other non-template bodies) entails  the need of incorporating \textit{ad hoc} components (``patchs") to the model to adequately represent the data. A visible spatial correlation between SGUs and the patches was already reported in \cite{LAT22_4FGL_DR3}, suggesting some connection. 
In this context, one expects that the SGU spectral properties relate to that of the interstellar gas  emission (keeping in mind that the analysis is performed under the assumption of a point source). 
In this paper, the possible spectral correlation between the SGUs and the diffuse emission is addressed via simulations. 
One possible limitation of the current Galactic interstellar emission model (IEM) may be related to unaccounted-for interstellar gas due to the saturation of $^{12}$CO lines. This scenario, reminiscent of the discovery of ``dark gas" using the EGRET data \citep{Gre05}  is tested by means of the $^{13}$CO MOPRA data \citep{2018PASA...35...29B}. 

Finally, we focus on the 175 most significant GUs ($>9.4\sigma$, corresponding to a Test Statistic~$>$~100). These sources, representing the ``tip of the iceberg",  are more likely to be real point sources and may help shed light on the nature of the whole SGU population. We report on the results of a thorough search for counterparts performed with multiwavelength (MW) data available for this subset.        

The paper is organized as follows.
Section \ref{sec:GU_prop} reviews the properties of the Galactic unassociated population. 
In Section \ref{sec:ML}, we further characterize the Galactic unassociated sources using multiclass classification with machine learning (ML). 
Section \ref{sec:new_classes} presents the possible new classes that have been proposed/explored in this context. Section \ref{sec:erosita} is devoted to the correlation of SGUs with the recently available {\sl eRosita}  DR1 catalog.  Section \ref{sec:diffuse} investigates the possibility that mismodeled diffuse emission drives the spurious detection of the SGUs. In Section \ref{sec:missing_gas}, the ``missing gas" hypothesis is examined. Section \ref{sec:bright_GUs} explores  the nature of the 175 brightest GUs.  A discussion and summary conclude the paper in Section \ref{sec:conclusions}.

\section{GU properties}
\label{sec:GU_prop}

\begin{figure}
\centering
\includegraphics[width=13cm]{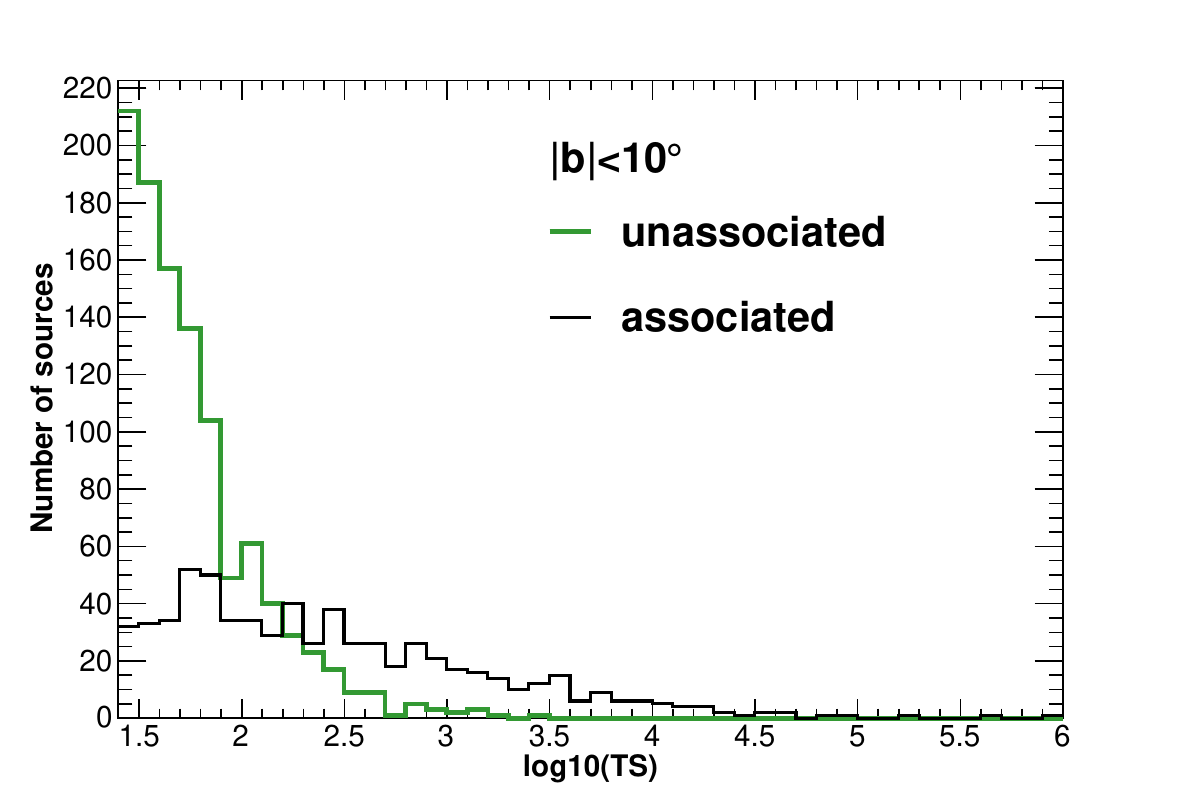}
\caption{Test-Statistic distributions for unassociated sources  and associated sources (omitting SPPs and UNKs) located at low latitudes.}
\label{fig:unassoc_TS}
\end{figure}

Some of these properties were already discussed in \cite{LAT22_4FGL_DR3}. We review them here while providing greater details. 
Figure \ref{fig:unassoc_TS} compares the Test Statistic distributions for associated and unassociated sources. We omitted SPP%
\setcounter{footnote}{0}
\footnote{These are sources of unknown nature but spatially overlapping with known SNRs or PWNe and thus candidate members of these classes.} 
and UNK%
\footnote{These are $|b|<10\arcdeg$ sources solely associated with the likelihood-ratio (LR) method from large radio and X-ray surveys. The nature of these sources, probably a mix of Galactic and extragalactic ones,  is unknown. } sources  in the associated sample due to their uncertain nature. Figure \ref{fig:unassoc_TS} illustrates the  steep rise in the number of unassociated sources as the detection significance decreases below TS~=~100.

\subsection{Spatial locations} \label{sec:GU_sploc}
\begin{figure}
\centering
\includegraphics[width=13cm]{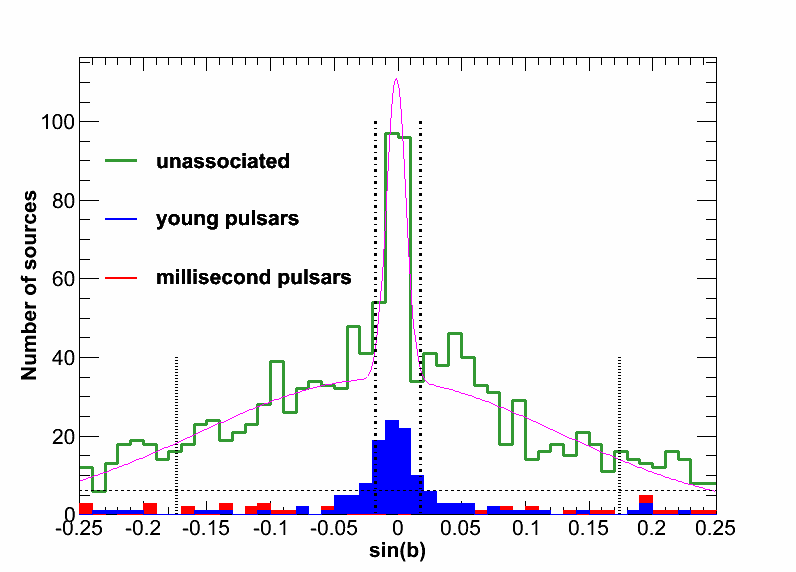}
\caption{Comparison between the Galactic-latitude distributions for unassociated sources, PSRs, and MSPs. The dashed horizontalline represents the average source number at latitudes~$>30^\circ$.  The dashdotted and dotted vertical lines represent the $|b|=1^\circ$ and $|b|=10^\circ$ limits, respectively. The magenta curve corresponds to a two-gaussian decomposition of the unassociated-source distribution. The mean sin(b) values of the broad and narrow components are $(-1.5\pm0.5)\times10^{-2}$ and $(-1.4\pm1.1)\times10^{-3}$, respectively.  }
\label{fig:latitude_1}
\end{figure}

The Galactic latitude distribution of the 4FGL-DR4 GUs is  displayed in Figure \ref{fig:latitude_1} along with those of young pulsars (PSRs) and millisecond pulsars (MSPs). 
As reported in \cite{LAT22_4FGL_DR3}, the Galactic-latitude distribution exhibits a narrow component (the ``spike", including 23\% of the sources), with a width of about $2^\circ$ and a much broader one (the ``shoulder") extending out to over $10^\circ$. 
The GU latitude distribution differs drastically from  those found for the two pulsar classes, the spike component being even narrower than the distribution of young pulsars\footnote{ A two-gaussian decomposition of the latitude distribution (displayed in Figure \ref{fig:latitude_1}) gives $\sigma_{sin(b)}$=0.008 for the narrow component,  while a gaussian fit to that of young pulsars for $|sin(b)|<0.03$ yields  $\sigma_{sin(b)}$=0.015.}.  
Young pulsars detected by the LAT typically have a characteristic age $<$ 3 Myr \citep{3PC}.  Older radio pulsars display a broader Galactic-latitude distribution because they had time to drift away from their native places in the plane. Those with a distribution compatible with the shoulder component would have an age $>$ 10 Myr and have thus spun down below gamma-ray detectability,  disfavoring this scenario.  

\begin{figure}
\centering
\includegraphics[width=13cm]{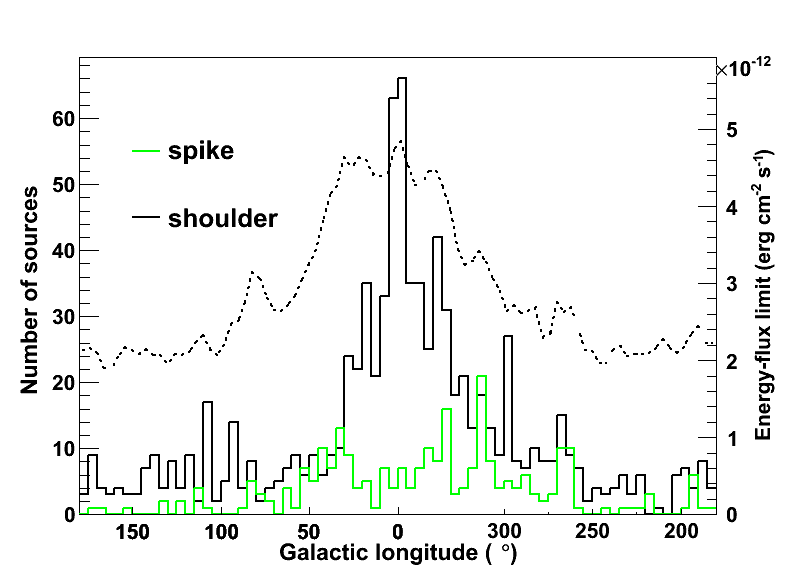}
\caption{GU Galactic-longitude distributions for the spike ($|b|<1^\circ$) and shoulder ($1^\circ<|b|<10^\circ$) components. The mean energy-flux limit  is depicted as the dashed curve (right-hand axis).}
\label{fig:longitude_1}
\end{figure}

The GU Galactic-longitude distributions for the spike and shoulder components are displayed in Figure \ref{fig:longitude_1}. While the shoulder shows a strong enhancement in the direction of the Galactic Center, the spike is more uniformly distributed across the inner Galaxy, similarly to the 4FGL-DR4 young pulsars, which trace the local star-forming activity. Both distributions for shoulder and spike are asymmetric, with more sources in the West hemisphere than in the East one (642 versus 487 in 4FGL-DR4, respectively,  when summing up the two components, a 27\% asymmetry). 
Taking into account the non-uniform flux limit over the Galactic plane (Figure \ref{fig:longitude_1}) and considering a flux-limited sample with an $E>100$~MeV energy flux greater than 4$\times 10^{-12}$ erg cm$^{-2}$ s$^{-1}$, the asymmetry West/East is 33\% 
(443 versus 317). 
For orientation, the asymmetry observed for 4FGL-DR4 young LAT pulsars is 25\% 

The clumpiness of the longitude distribution, clearly visible in Figure \ref{fig:longitude_1}, reveals the existence of several GU clusters.  
This clustering effect was already noted in \cite{LAT22_4FGL_DR3}, where a dedicated analysis flag was provided to indicate that an unassociated source is part of a cluster.  Some clusters will be studied in more details in Section \ref{sec:clusters}.  

It is interesting to look for possible spatial correlation between the GUs and the interstellar diffuse emission. We recall that the Pass8 IEM\footnote{{https://fermi.gsfc.nasa.gov/ssc/data/analysis/software/aux/4fgl/Galactic\_Diffuse\_Emission\_Model\_for\_the\_4FGL\\\_Catalog\_Analysis.pdf}} includes nine components: molecular hydrogen (traced by CO), atomic hydrogen (HI), Inverse Compton (IC),  Dark Neutral Medium (a correction based on infrared tracers of dust column density in directions where the combination of HI and CO is either under or over-estimating the data), positive (DNMp) or negative (DNMn), unresolved sources, the ``patch",  and the isotropic component. 
The CO, HI and IC components were fitted to the data over 10 Galactocentric rings of increasing radii.
 
The patch is an \textit{ad hoc} component derived from the all-sky residuals and  encapsulates the contributions of non-template components like the Galactic Center excess, Loop I, or the {\Fermi} Bubbles. The patch was smoothed out in order not to absorb features with spatial scales smaller than 4$^\circ$. 

\begin{figure}
\centering
\includegraphics[width=13cm]{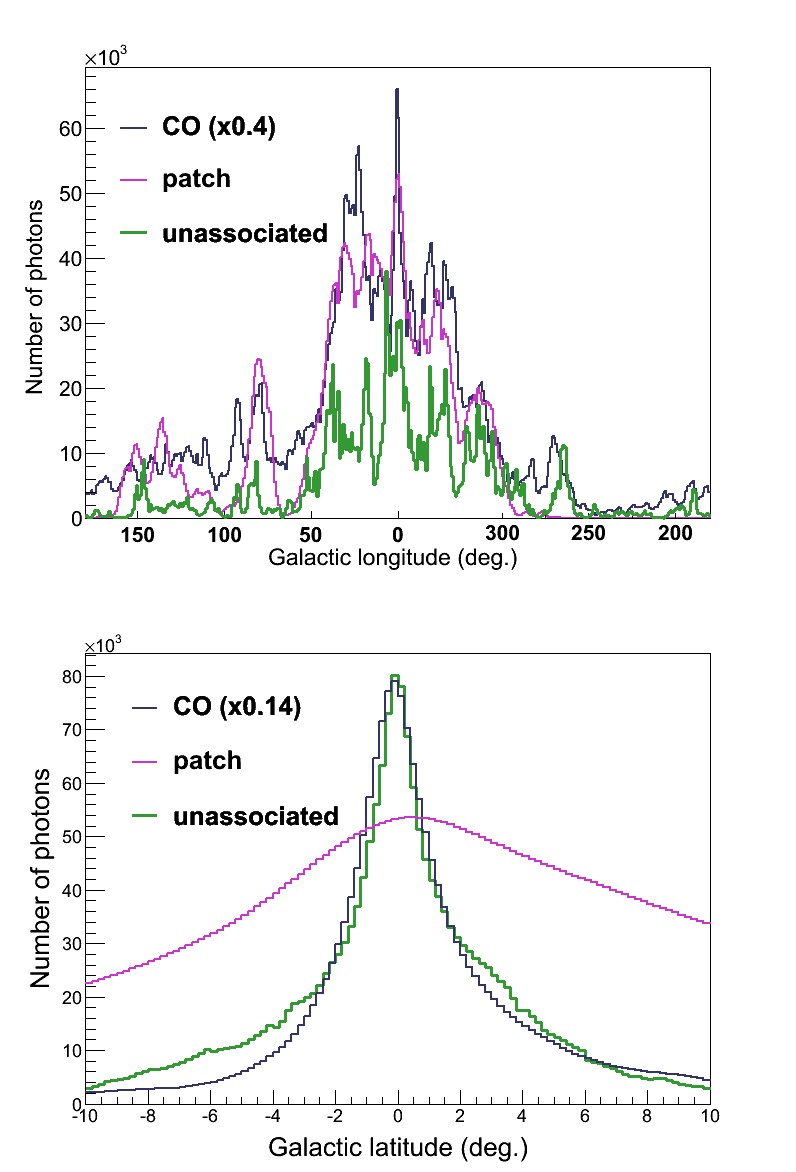}
\caption{Top: Galactic-longitude distributions of the photons from unassociated sources
estimated by the 4FGL analysis model in the $|b|<10^\circ$ area (black histogram). The blue (red) histogram corresponds to the photon counts predicted for the IEM patch (CO) component in the same area. The CO component has been rescaled to facilitate the comparison. Bottom: same as top, for the Galactic latitude.
}
\label{fig:longitude_patch_CO_npred}
\end{figure}

The Galactic-longitude distributions of $E>100$~MeV photons from the unassociated sources and the IEM patch component as estimated by the analysis model are compared in Figure  \ref{fig:longitude_patch_CO_npred}. 
In the inner Galactic region, the photon counts are commensurable, making a common origin plausible. The same exercise is repeated in the same Figure  for the CO component.  This component has been chosen over the other gas component, HI, because of i) its larger clumpiness, more prone to generate point-like spatial features, ii) a stronger intensity asymmetry between the inner and outer Galaxy,  more in line with the asymmetry observed for the GUs.   In the central Galactic region ($b<60^\circ$, $b>300^\circ$), the GU photons represent about 20\% of those predicted to originate from the CO component. 
The Galactic-latitude distributions of the photons are quite similar for the unassociated sources and the CO component. In contrast, the distribution for the patch component is much broader, so the unassociated sources do not map out the whole structures encapsulated in the patch.

\begin{figure}
\centering
\includegraphics[width=13cm]{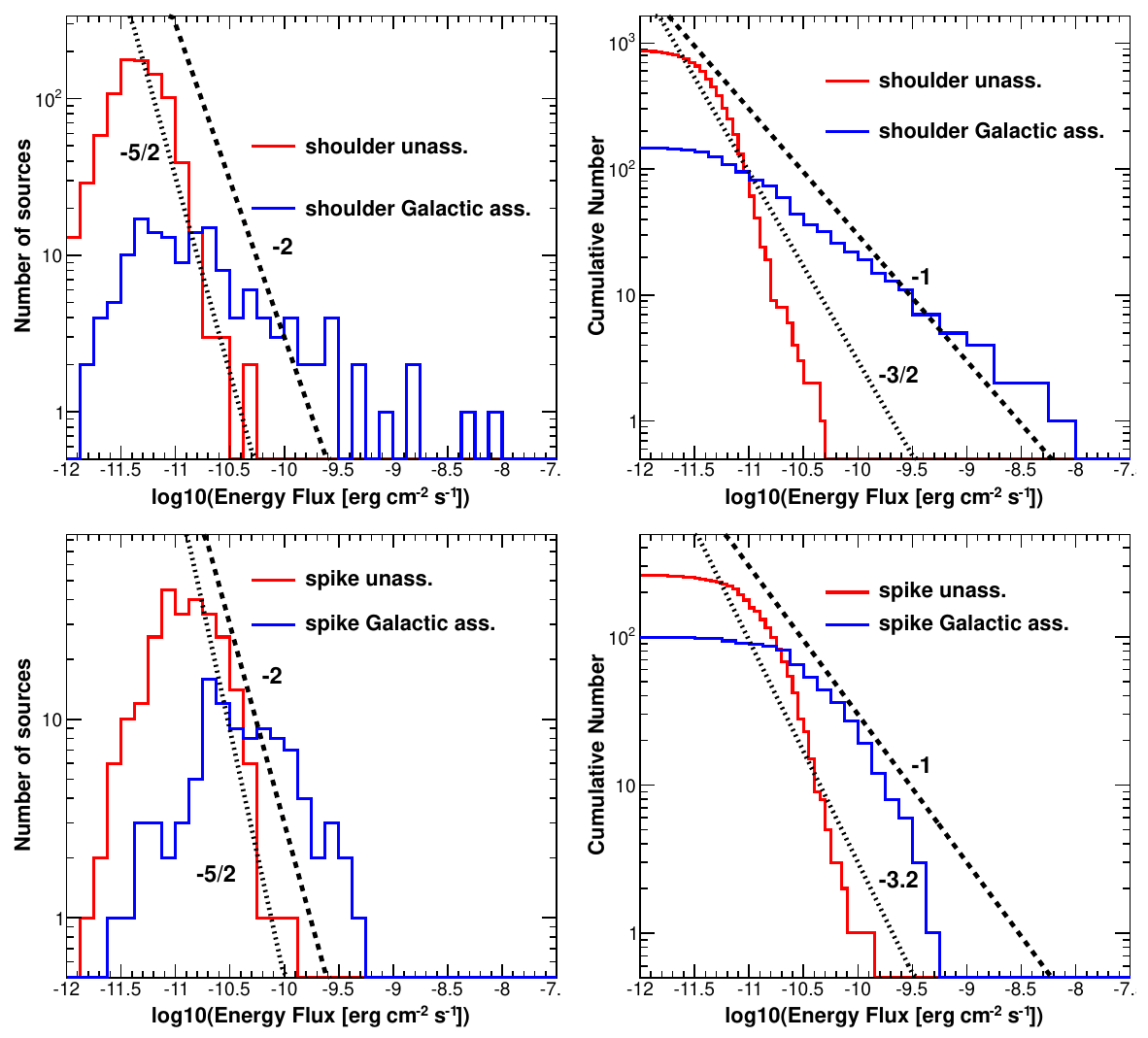}
\caption{Left: Flux distributions of the associated and non-associated sources in the shoulder (top) and spike (bottom) populations. The dotted and dashed lines depict powerlaw functions with slopes of $-$2 and $-$5/2, corresponding to the expected dependence for disk-like and isotropic populations, respectively. Right: corresponding log N - log S for the different selections displayed at left. The dotted and dashed lines depict powerlaw functions with slopes of $-$1 and $-$3/2, corresponding to the expected dependence for disk-like and isotropic populations, respectively.}
\label{fig:logN_logS}
\end{figure}

\subsection{Fluxes  and Log N - log S }  \label{sec:flux}

The energy-flux distributions and the related log N - log S \footnote{These distributions are uncorrected for coverage, the flux limits being estimated at $10^{-11}$ erg cm$^{-2}$ s$^{-1}$ and 6$\times 10^{-12}$ erg cm$^{-2}$ s$^{-1}$ for the spike and the shoulder, respectively.} are given separately for the spike and the shoulder in Figure \ref{fig:logN_logS}, with a comparison to the corresponding ones for the associated sources (excluding AGNs).    The association rate is about 100\% for energy fluxes greater than $10^{-10}$ erg cm$^{-2}$ s$^{-1}$ and 3$\times 10^{-11}$ erg cm$^{-2}$ s$^{-1}$ for the spike and the shoulder, respectively. Below these fluxes, the association rate drops dramatically for both the spike and the shoulder. The unassociated sources outnumber the associated ones by a factor of 2.6 and 5.9 for the spike and shoulder, respectively. The slopes of the log N - log S diagrams are quite different for the associated and unassociated populations. This is especially the case of the shoulder where the steep rise of the unassociated sources for lower S seems difficult to reconcile with the scenario of an emerging new population of Galactic gamma-ray emitters.  The expected slopes of the log N - log S for isotropic (slope=$-$3/2) and disk-like (slope=$-$1) populations are displayed in Figure \ref{fig:logN_logS} for orientation.      

\subsection{Spectral properties \label{sec:spectra}}

The SGUs  show soft and curved energy spectra \citep{LAT22_4FGL_DR3}. 
In 4FGL, this population was defined from the condition that the power-law photon index is greater than 2.4. 
Since the spectra are curved, the measured power-law photon index depends on the detection significance. This is because the pivot energy moves to a higher energy for a fainter source, leading to a larger photon index (indicative of a softer spectrum). Here we consider the peak-energy parameter derived from the LogParabola  fit (LP\_EPeak), which is insensitive to the above effect.  
Of the 1129 GUs, 384 have a low ($<2 \sigma$)  spectral-curvature significance. Only 72 of these 384 have no measured LP\_EPeak and 26 others have LP\_EPeak $<$ 30 MeV  or  $>$ 100 GeV. The remaining 286 of 384 sources  show an LP\_EPeak distribution very similar to the rest of the GUs, so  LP\_EPeak appears to be a meaningful parameter for this subset as well.

\begin{figure}[h]
\centering
\includegraphics[width=13cm]{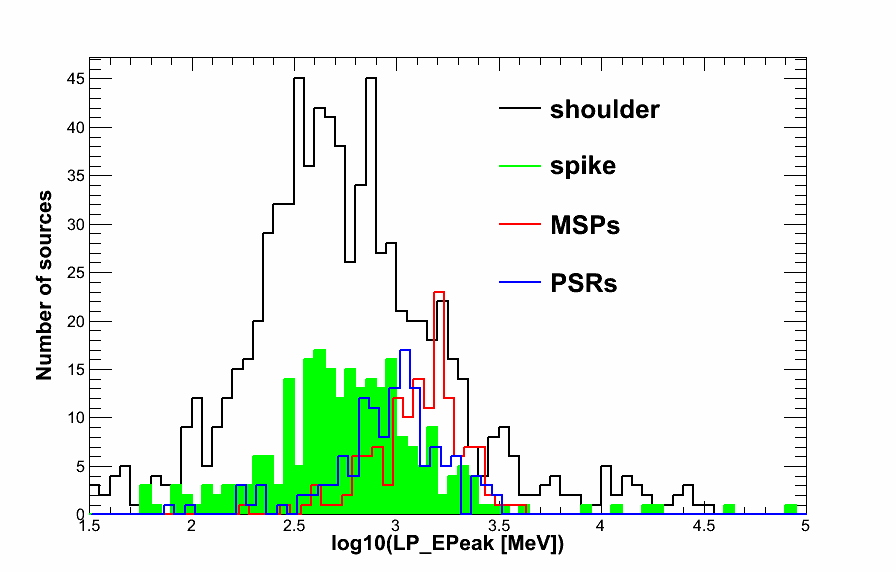}
\caption{ LP\_EPeak distributions. Black: GU shoulder component, green: GU spike component, red: millisecond pulsars, blue: young pulsars.  }
\label{fig:Epeak}
\end{figure}

\begin{figure}[h]
\centering
\includegraphics[width=13cm]{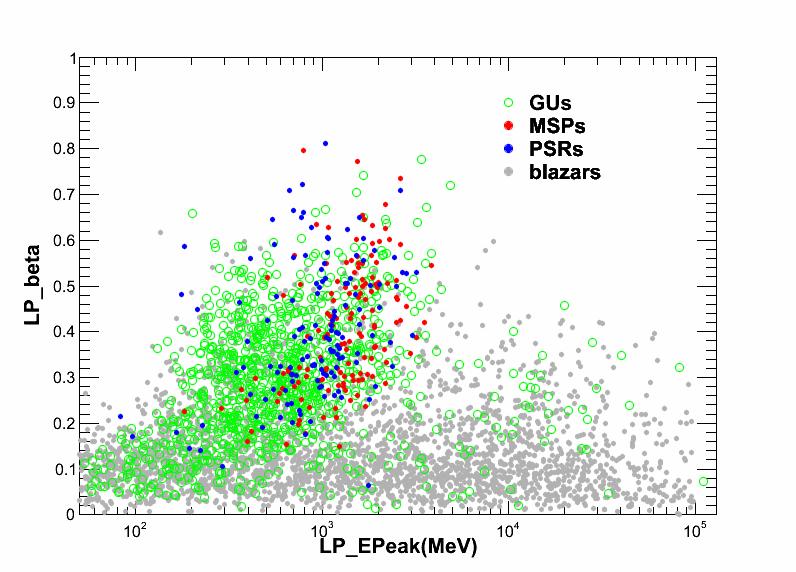} 
\caption{ Spectral-curvature parameter beta (half the curvature) plotted versus LP\_EPeak for different source populations.  No condition on the significance of the spectral curvature 
has been imposed. 
}
\label{fig:epeak_beta}
\end{figure}
\begin{figure}[h]
\centering
\includegraphics[width=13cm]{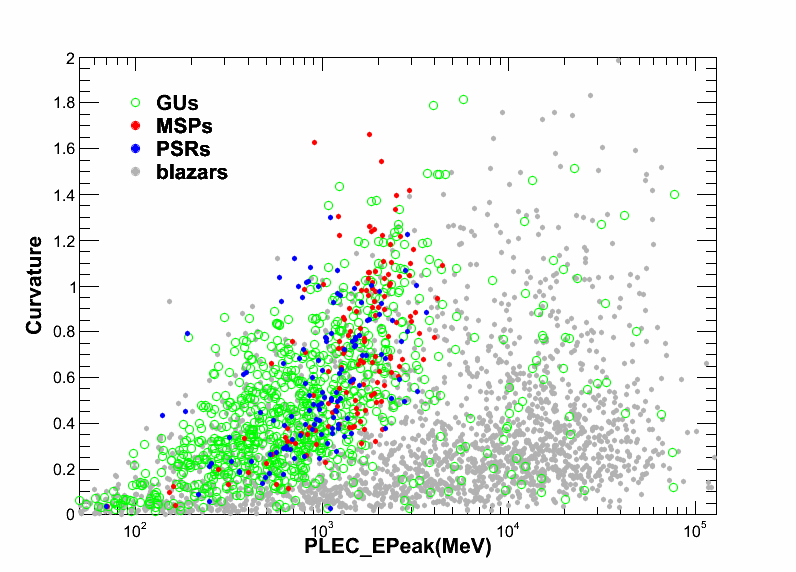} 
\caption{ Similar to Figure \ref{fig:epeak_beta}, the peak position and curvature being computed with a PLEC function. The curvature is that at EPeak.
}
\label{fig:PLECepeak_curv}
\end{figure}

The GU LP\_EPeak distributions for the spike and shoulder are compared in the top panel of Figure \ref{fig:Epeak}. They look very similar, peaking around 500 MeV. A comparison between the whole GU LP\_EPeak distribution and  those found for the MSPs and PSRs is given in the bottom panel. The latter two distributions are shifted to higher LP\_EPeak values w.r.t. the GU one. The trend that, for gamma-ray pulsars, LP\_EPeak decreases with larger spin down power $\dot{E}$ was found in the 3PC catalog \citep[see Figure 19 in][]{3PC}, hence the larger LP\_EPeak on average for MSPs than for PSRs. 
A Kolmogorov-Smirnov test yields a Test Statistic of 0.51 (0.36) when comparing the GU and MSP (PSR) LP\_EPeak distributions (p-values$<10^{-10}$, confirming that they are unlikely to derive from the same underlying distribution.

Another common property of the GUs is a pronounced spectral curvature. 
About 66\% of the GUs present significant curvature (LP\_SigCurv~$>$~2). GUs occupy a different region in the LP\_EPeak-LP\_beta plane than pulsars and blazars, where LP\_beta is the LogParabola curvature parameter (Figure \ref{fig:epeak_beta}).
LP\_beta is almost always greater than 0.1 for LP\_EPeak$>$ 300 MeV, in contrast to what is observed for blazars, and has values similar to those found for young and millisecond pulsars. Note that in 4FGL-DR4, a Gaussian prior (mean equal to 0.1 and width equal to 0.3) was introduced in the maximum-likelihood calculation of the beta parameter.  We checked using the PLEC\_ExpfactorS variable \footnote{The  powerlaw with exponential cutoff function (PLEC) is the most appropriate for the fitting of pulsars. } (which has a three times higher prior mean) instead that our conclusions are insensitive to this factor. Similar features to those seen in Figure \ref{fig:epeak_beta} are noted when the peak properties are  computed by means of  the PLEC function (Figure \ref{fig:PLECepeak_curv}). 

\begin{figure}
\centering
\includegraphics[width=18cm]{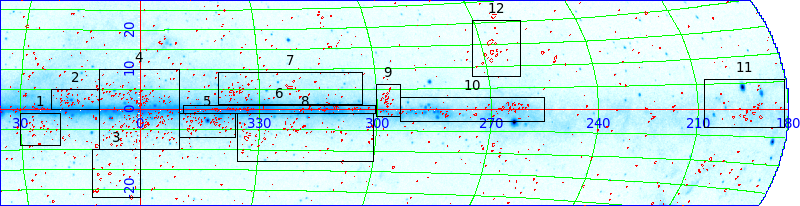} 
\caption{Sky map (Hammer-Aitoff projection, Galactic coordinates) highlighting the different regions with clusters of GUs discussed in the text. 
The 95\% error ellipses of the 4FGL-DR4 unassociated (plus SPP and UNK) sources are depicted in red. The background depicts the $E>1$ GeV intensity in log scale.}
\label{fig:map_ROIs}
\end{figure}


\subsection{Notable clusters}
\label{sec:clusters}

GUs are often found in clusters. Several regions with large densities of GUs are highlighted in Figure \ref{fig:map_ROIs}, outlined with rectangular contours, which were found by eye and are thus largely arbitratry . One notes a variety of situations, with a strong cluster centered around the Galactic Center, and fairly well defined clusters in the spike or shoulder components. 

Several regions in Figure \ref{fig:map_ROIs} are noteworthy.  
Region 4 encompasses the Galactic Center.
Region 9 is very compact and extends perpendicular to the Galactic plane. It lies close to the boundary between  Galactocentric rings 5 and 6 of the diffuse-emission model \footnote{This boundary, corresponding to a Galactic radius of 7 kpc, lies at a longitude of 299$^\circ$.}, which is a puzzling coincidence.
Region 10 includes the Vela Molecular Ridge, where the density of GUs is particularly high. 
Region 11 is coincident with the  Gemini OB1 Molecular Cloud Complex.   
Region 12 does not relate directly to the GUs as it lies further from the plane. 
It represents a quite unique case of a high-latitude cluster, to which no evident counterparts have been identified. 
The LP\_EPeak distributions of the different clusters are all quite similar.  The hardest spectra are found in  region 4,  located around the Galactic Center and region 6, in the spike.  
Although the statistics is low, sources in region 12 appear to be softer on average than those in other regions.  


In summary, the GUs show specific properties regarding their spatial locations and energy spectra compared to the established classes. While the LP\_EPeak  distributions of all GUs are similar and peaking at an unusually low value ($\sim 500$ MeV),  the longitude distributions of the spike and the shoulder are clearly different from each other. That of the shoulder strongly peaks towards the inner Galaxy and bears resemblance to the molecular gas distribution (so does the latitude distribution), while that of the spike is much flatter. Clusters of sources have been found in both components. The slopes of the LP\_EPeak log N - Log S distributions are steeper than those expected for  isotropic or disk-like populations. 

\section{Characterization of GU sources with machine learning} 
\label{sec:ML}


In this section, we continue the description of GUs presented in the previous section and investigate their properties using ML.
In particular, we estimate using multi-class classification the fraction of GU sources that can be attributed to the known classes of gamma-ray emitters (based on the spectral properties of associated sources in the 4FGL-DR4 catalog) and the fraction of GU sources that has, as a population, different spectral characteristics compared to the associated sources.
For training, we use sources over the whole sky, which significantly improves statistics for extragalactic sources and helps with the separation of the extragalactic sources from the Galactic ones in the Galactic plane.
We exclude bcu (blazars of uncertain type) and spp classes from the list of associated sources, as these classes are a combination of other classes of sources.
 In particular, bcu sources can be either FSRQs or BL Lacs, 
while spp sources can be a combination of SNRs, PWNe, and pulsars.
Provided that the physical nature of a counterpart for the unk sources is unknown, 
these sources are equivalent to unassociated sources from the point of view of classification, i.e., we add the unk sources to the unassociated sources.
The remaining physical classes are separated into four groups following the hierarchical definition of classes~\citep{Mal23, MalyshevGCEcov},
which puts classes of sources with similar distributions in the feature space (i.e., with similar spectral parameters in this work) in one group.
The groups are denoted by the largest physical class in the group\footnote{In this section we do not make a distinction between identified sources (upper-case class names) and associated sources (lower-case class names).
The class names are defined in \cite{LAT22_4FGL_DR3}.}:
``fsrq+'': fsrq, nlsy1, css;
``bll+'': bll, sey, sbg, agn, ssrq, rdg;
``psr+'': psr, snr, hmb, nov, pwn, gc;
``msp+'': msp, lmb, glc, gal, sfr, bin.
For the classification we use the following three spectral parameters of sources as input features:
log10(Energy\_Flux100), LP\_beta, log10(LP\_EPeak) (see \cite{LAT22_4FGL_DR3} for the definition of the source parameters).
On the one hand, as we show below, these parameters are sufficient for the separation of the four classes of associated sources and a possible new class of soft GU sources. On the other hand, including more features makes the classification less stable due to increased complexity of the model.
In this section, we subselect the sources, for which an estimate of LP\_EPeak exists and lies between 10 MeV and 1 TeV.

\begin{figure}
\centering
\includegraphics[width=0.47\textwidth]{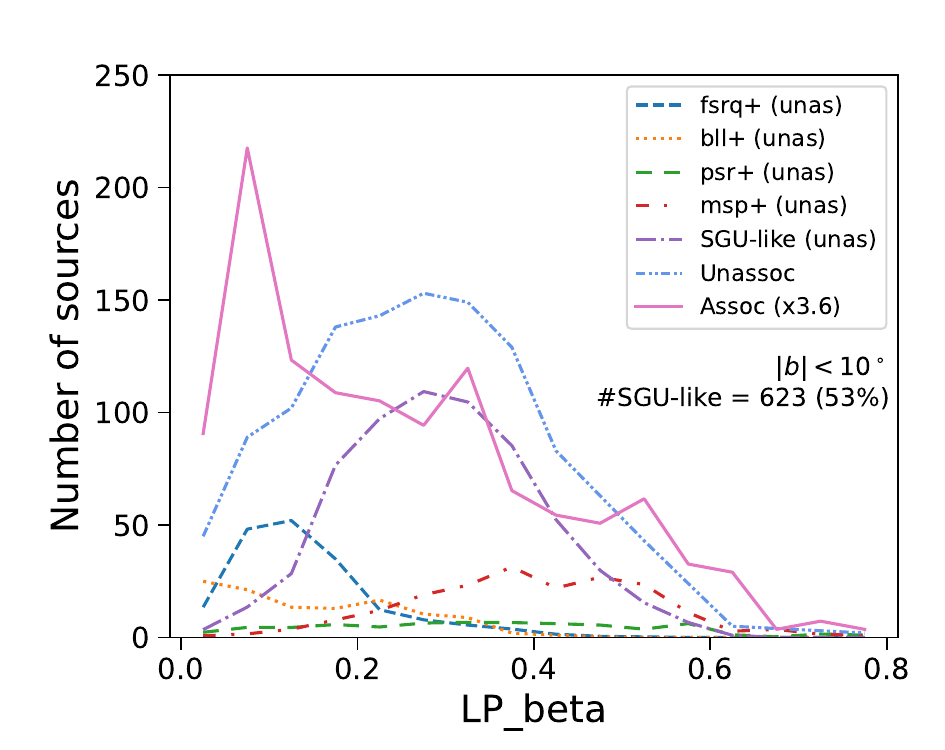}
\includegraphics[width=0.47\textwidth]{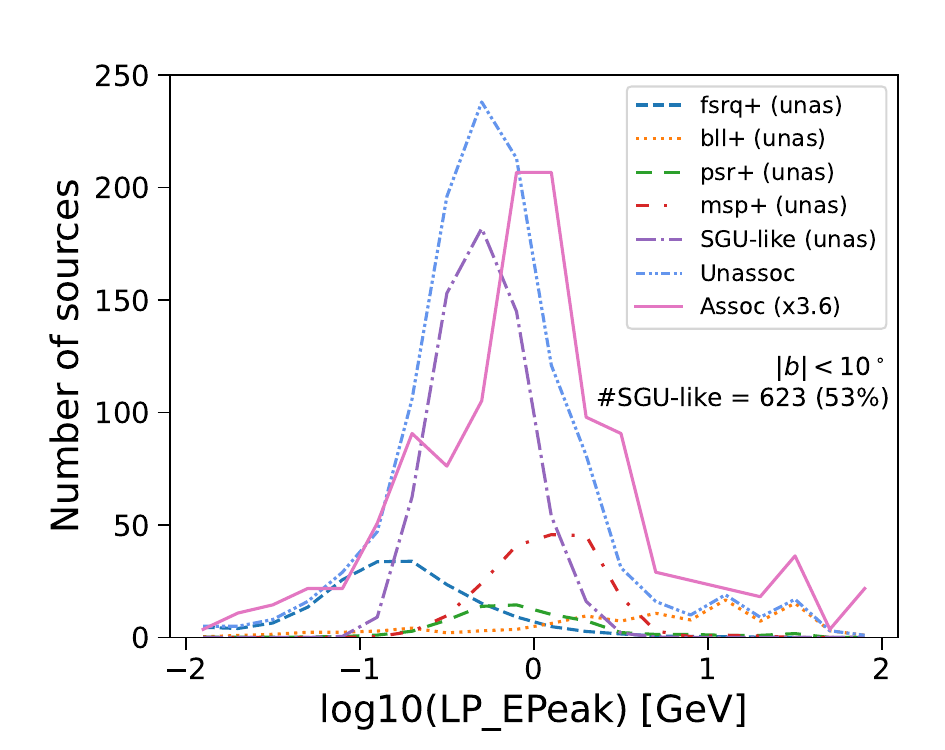}
\caption{Expected contributions of different source classes to the unassociated sources for $|b| < 10^\circ$, for the LP\_beta distribution (left) and the log10(LP\_EPeak) distribution (right).
The contribution of a new source component modeled as a Gaussian distribution in the LP\_beta and log10(LP\_EPeak) variables is shown by the purple dash-dotted line (labeled as SGU-like sources).
The overall distribution of unassociated plus unk sources is shown by the light blue dash-dot-dotted line.
The distribution of associated sources within $|b| < 10^\circ$ (without bcu and spp sources) scaled by 3.6 to match the total number of unassociated and unk sources in this range of latitudes is shown by the solid pink line.}
\label{fig:prior_shift}
\end{figure}

In this paper, we use the flux-dependent prior shift model of~\cite{Mal24CovPrior} with the additional Gaussian component to describe the GU sources.
In this model we allow the relative fractions of the four classes to change as a function of the energy flux above 100 MeV. We also introduce a new class modeled with a Gaussian distribution in the input features.
The parameters of the model including the Gaussian component are determined from the fit of the model to the distribution of unassociated sources.
The details of the model are presented in Appendix~\ref{app:cov_prior_models}. 
The probabilistic classification of the GU sources including the probability to belong to the new component is available online.
\footnote{\url{https://doi.org/10.5281/zenodo.20342245}}

The distributions of the components corresponding to the different classes in the best-fit model 
are shown in Figure~\ref{fig:prior_shift}
for LP\_beta and log10(LP\_EPeak) variables.
We notice that the center of the Gaussian component is around log10(LP\_EPeak [GeV]) = $-$0.3 or LP\_EPeak $\approx$ 500 MeV, which is consistent with the overall center of the GU sources in Figure~\ref{fig:Epeak}.
The center of the distribution for LP\_beta is around 0.3.
The number of SGUs (i.e., sources with $|b| < 10^\circ$ and $\rm PL\_Index < 2.4$, see Section \ref{sec:intro}) among unassociated and unk sources is 881, the number of sources with $|b| < 10^\circ$ classified to be in the Gaussian component is 710 (with Gaussian probability larger than 0.5), while the number of SGUs that are also classified to be in the Gaussian component is 649.
Since the fraction of sources in the Gaussian component which are SGUs is 0.91 = 649 / 710,
we denote the sources in this Gaussian component as SGU-like.
We also note that the fraction of SGUs classified to be in the Gaussian component is 0.74 = 649 / 881, i.e., about three quarters of SGUs are classified to be in the Gaussian component.
In Figure~\ref{fig:prior_shift}, the expected numbers of sources in the different classes among the unassociated sources are obtained by summing the corresponding class probabilities for sources within $|b| < 10^\circ$.
In particular, the expected number of SGU-like sources is 623, which is about 53\% of all unassociated and unk sources for $|b| < 10^\circ$.
We have performed several checks of systematics uncertainty in the model, including using two classes of associated sources (Galactic and extragalactic) instead of four classes and a model without the energy-flux dependence of the prior shift. The expected number of sources in the Gaussian component is within $\pm 100$ from the estimate in the model presented in this section.

The SGU-like sources modeled by the Gaussian distribution in Figure \ref{fig:prior_shift} have a similar distribution as the GU sources in Figure \ref{fig:epeak_beta}.
In particular, they 
have generally smaller spectral curvature (smaller LP\_beta parameter) than the Galactic sources in msp+ and psr+ classes, but larger curvature than extragalactic sources in fsrq+ and bll+ classes.
The LP\_EPeak values for SGU-like sources are smaller than for bll+, psr+, and msp+ sources but larger than for fsrq+ sources.
Thus, the distributions of spectral parameters of SGU-like sources are different from the distributions of parameters for the known classes of gamma-ray sources.


\begin{figure}
\centering
\includegraphics[width=\textwidth]{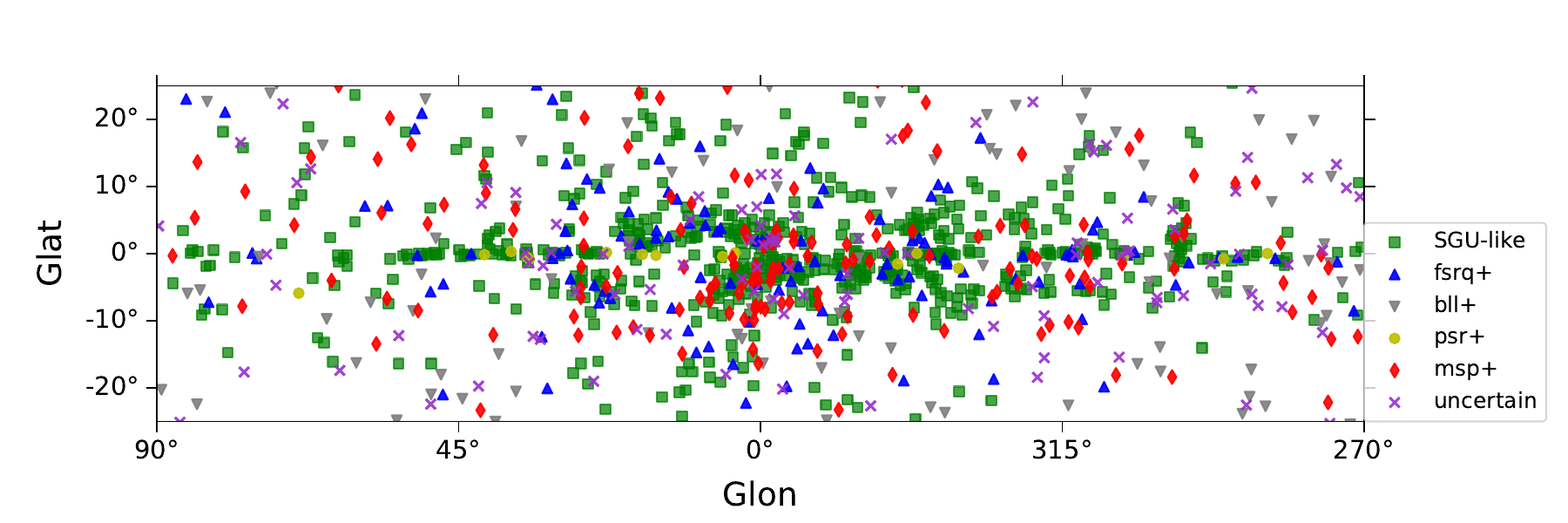}\\
\includegraphics[width=\textwidth]{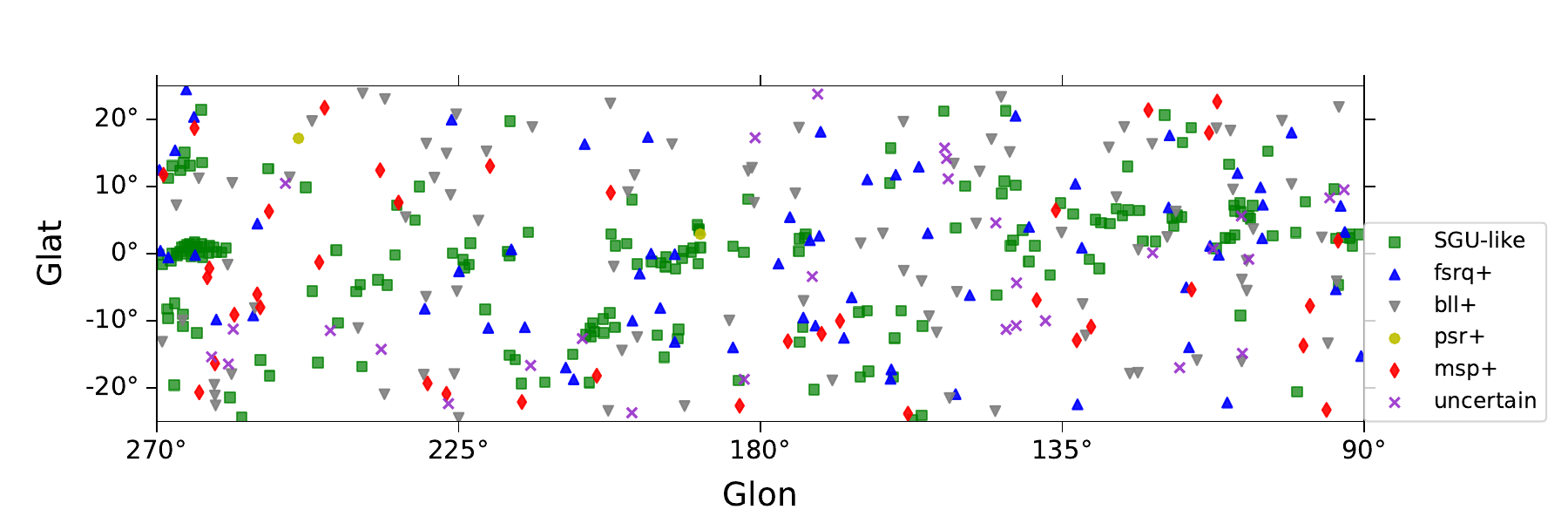}
\caption{Positions of unassociated sources with predicted classes in the prior shift model. Sources are attributed to a class if the corresponding class probability is larger than 0.5, otherwise the class is considered uncertain.}
\label{fig:sky_map}
\end{figure}

We show the positions of the class candidates in the prior shift model in Figure~\ref{fig:sky_map}.
Different symbols represent the sources with corresponding class probabilities $> 0.5$. 
If a source has all class probabilities smaller than 0.5, then it is labeled as uncertain. 

\begin{figure}
\centering
\includegraphics[width=0.49\textwidth]{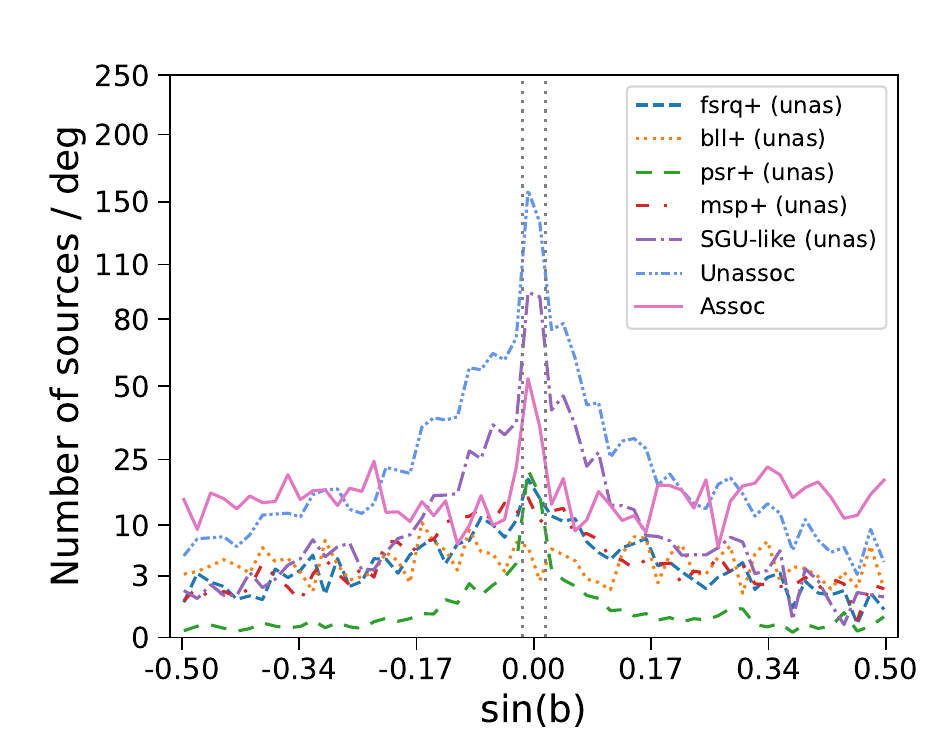}
\caption{Source distributions as a function of Galactic latitude. 
X-axis bins are equally spaced in $\sin(b)$ to give equal areas on the celestial sphere.
The bin size is $1^\circ$ near $b = 0^\circ$.
The vertical dashed lines show $b = \pm 1^\circ$.
The y-axis is a square-root scale.
}
\label{fig:cov_prior_glat}
\end{figure}

We compare the expected distributions of sources as a function of Galactic latitude in Figure~\ref{fig:cov_prior_glat}.
The distribution of bll+ sources is approximately uniform on the sky.
The distribution of fsrq+ sources has a slight increase close to the Galactic plane, which shows that there is likely a contamination from Galactic sources in this class.
This is not unexpected, provided that the fsrq+ sources have relatively low EPeak values (cf., the right panel of Figure~\ref{fig:prior_shift}).
As a result, one can expect a larger confusion between fsrq+ and SGU-like sources compared to bll+ sources.
More quantitatively, this mixing is estimated in the confusion matrix in Table~\ref{tab:cov_vs_prior_lowb} of Appendix~\ref{app:cov_prior_models}.
The distribution of SGU-like sources 
has spike and shoulder components both like the overall GU source distribution.

\section{Possible new classes 
\label{sec:new_classes}}
Conceivably, so many LAT sources are unidentified simply because one or more categories of gamma-ray bright celestial objects were absent from the association procedure.
Here, we investigate Galactic populations highlighted in the Literature, adding the catalogs cited below for each class to the procedure.



\subsection{Accreting Millisecond X-ray Pulsars (AMXPs)}
An AMXP is an accretion powered X-ray pulsar spinning at frequencies
$\nu\ge$ 100 Hz bound in a binary system with a donor companion of mass M $\sim 1 {\rm M}_\sun$ and with a weak surface magnetic field (B $\sim 10^{8-9}$ G) \citep{Pat21}. AMXPs show X-ray pulsations with millisecond periods during X-ray outbursts, likely caused by an accretion flow impacting the neutron-star surface. AMXPs are believed to be the direct predecessors of MSPs, and it was suggested that several AMXPs actually switch to being rotation powered pulsars during their quiescent state. AMXPs may thus be gamma-ray emitters.  We used the list of AMXPs from \cite{Sal20} to look for counterparts to 4FGL-DR4 sources. We obtained four high-confidence associations (based on spatial coincidence), already classified differently but with classes compatible with that of AMXP for three of them: MAXI J0911$-$655 (glc), NGC 6440 X$-$2 (glc), SAX J1808.4$-$3658 (psr). The fourth one, IGR J18245$-$2452, is located within the globular cluster M 28, whose gamma-ray emission is dominated by the MSP PSR J1824$-$2452A.   No subthreshold associations with probability $>$10\% have been found.

\subsection{Star Forming Regions (SFRs)}
\label{sec:SFRs}

In the standard paradigm, Galactic cosmic rays are accelerated mainly by diffusive shock acceleration (DSA) in supernova remnants.
SNRs, however, do not account for all observed features of cosmic-ray spectra and abundances \citep[for a recent review, see][]{GabiciSuperbubbles}.
The turbulent winds from massive stars and clusters of stars, as well as jets from protostars, can sweep up the dense material in the regions surrounding them, creating shocks. DSA can then locally accelerate particles, some of which will produce gamma rays. 
SFRs likely complement SNRs, and could create many {\Fermi}-LAT sources. 
We use ``SFR" as a catch-all name: a variety of specific objects within the regions can generate DSA in various configurations.

SFRs are not a new LAT source class: three 4FGL-DR4 sources are identified as SFRs, with extended GeV emission matching the SFR morphology. Three more are ``sfr" class. 
What \textit{is} new is the recognition that SFRs may be behind three or four times as many 4FGL sources, previously with no 4FGL association, or classified as ``unknown". Several GeV-bright SFRs have been reported in the literature \citep[e.g.][]{Tib21}. Most have radii from $0.3^\circ$ to  $1^\circ$. A few are larger, while others are point-like in LAT data. 
Here we consider three SFR catalogs: clusters of young stellar objects (RMS) ; H II regions (WISE) ; and stellar clusters detected in the optical band (Gaia).

We matched LAT sources with the 117 clusters of massive Young Stellar Objects (YSOs) that \citet{Urquhart2014_FIRclumps} built using the Red MSX Source survey catalog of YSOs \citep[RMS, ][]{RMSsurvey}. Bright 21 $\mu m$ emission flags massive YSOs otherwise obscured by surrounding dust and gas. 
RMS clusters are co-located with 38 individual LAT sources.


The WISE catalog \citep{WISE_HIIcatalog} targets HII regions, but also provides 21 $\mu m$ data with sensitivity probing deeper into the Galaxy than MSX.
\cite{Per24} found spatial coincidences between 138 4FGL-DR3 unassociated sources and WISE 22 $\mu m$ Galactic HII regions. Simulations predicted about 50 false-positive coincidences. To explore these sources' gamma-ray properties and their connection to SGUs, we performed a similar study. Unlike Peron et al., we excluded extended  4FGL-DR4 sources to limit the number of false positives, as these sources  can be quite wide (64 out of 82 have radii larger than 0.3\arcdeg).  We also  kept sources with low-confidence associations and included ``unknown" sources since they exhibit properties similar to the GUs. We found 157 gamma-ray sources lying within HII regions, with 90 random overlaps predicted by simulations. 
Most of the excess is in the spike of the Galactic latitude distribution, and is a significant fraction of it.  
Correcting for this leads to 98 true coincidences. Restricting the HII region radii to $<0.^\circ3$ leads to 71 coincidences with an estimated number of 33 true matches. Of these 71 WISE (38 RMS) associations, 50 (25) are SGU-like ($\sim 70$\%), consistent with statistical fluctuations ($1.7 \sigma$ and $0.4 \sigma$, respectively).

\begin{figure}
\centering
\includegraphics[width=13cm]{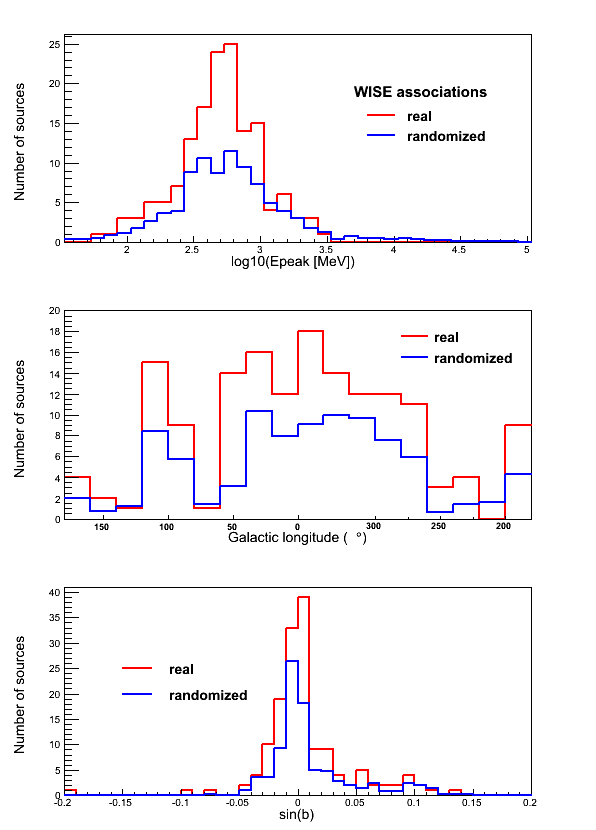}
\caption{From top to bottom: LP\_EPeak, Galactic longitude and latitude distributions of 4FGL-DR4 sources colocated with WISE HII regions. Distributions of sources with randomized locations are shown for comparison.} 
\label{fig:SFRs}
\end{figure}

Figure \ref{fig:SFRs} shows the LP\_EPeak, Galactic longitude and latitude distributions for the 157 gamma-ray sources lying within HII regions. The longitude and latitude distributions for the real and simulated skies are very similar. The real LP\_EPeak distribution is more peaked around 500 MeV than the simulated one.
LP\_EPeak values below 1 GeV agree with the measurements of RCW 38, RCW 36, RCW 41, and IRS 31 \citep[][ see Section \ref{sec:bright_SFRs}]{Ge24,Per24,Pan24}.
Those authors favor a hadronic scenario to understand the spectra, with locally
accelerated protons producing gamma rays via $\pi^0$ decay.


Gaia provides catalogs of stellar clusters in the optical band, and a selection path for gamma-ray emitting SFRs complementary to RMS and WISE. We used the open cluster catalog of \citet{CGGC} but, following \citet{CelliOpenClusters}, kept only the 387 younger than 30 Myr likely to still contain massive stars with violent winds. Sixty such clusters overlap $|b|<5^\circ$ LAT sources, shrinking to only 13 for LAT sources with position uncertainty or spatial extent $<1^\circ$, and keeping only unassociated or ``unknown" class sources. The number of random spatial coincidences is unknown, suggesting that the true number is smaller. However, particle acceleration and gamma-ray emission may occur outside of a stellar cluster \citep[e.g.,][]{2021MNRAS.504.6096M,2022MNRAS.512.1275V}. Such extended emission could fail our matching criteria, in which case the number would be larger.


 



This work aims to clarify the nature of a thousand unassociated low-latitude LAT sources. Of the hundreds of known massive SFRs, only a subset emit detectable gamma rays. The RMS and WISE catalogs place the number between $\sim 30$ and $\sim 100$ SFRs. More GUs than that will likely be identified, if they get grouped into extended sources in dedicated analyses, but not more than a few tens. While identifying $\sim 10$\% of the GUs is a significant step towards resolving the GU problem, SFRs are unlikely to explain a much larger fraction than that.

\subsection{Other relevant classes} 
The above remarks about objects and regions where DSA might efficiently lead to
detectable GeV photon fluxes are quite general. 
The crux of the matter is to find places where shocks are violent enough to 
accelerate particles to high energies; where the total mechanical energy suffices to
generate high fluxes; and where local densities are adequate for ample gamma-ray conversion.
Here follow some specific object classes cited in the literature, along with our attempts
to associate such objects with GUs. Note, however, that collective effects within SFRs
might make an ensemble more powerful than the simple sum of its components: neighboring
stars and protostars, at various evolutionary stages, can superpose turbulent winds and complex magnetic fields to create ion-accelerating hot spots.

\textbf{Wolf-Rayet (WR) stars.} The 4FGL association pipeline finds no significant matches with WR stars taken from the list of \cite{van01}. However, dust and gas obscure many Galactic WRs, and the list is
thus necessarily incomplete.  The possible association of  4FGL J1858.8+0354 with the open cluster Masgomas-6 is discussed by \citet{Wan22}.  They state that two WR stars contribute 60\% of the mechanical wind power in the system.

\begin{figure}
\centering
\includegraphics[width=13cm]{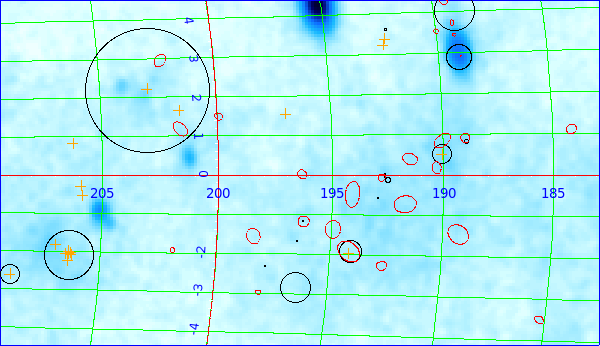} 
\caption{Gemini OB1 region. Black: Bright northern HII regions \citep{Sharpless-2}. Orange crosses: GOSC objects. Red: DR4 unassociated sources.\label{fig:GemOB1}}
\end{figure}

\textbf{OB and Be stars.} Milky Way O and B stars, like the WR stars, are born in dense regions that may
hide them from view. HII and 21 $\mu m$ observations can reveal their presence. 
The 4FGL association procedure yields no significant matches with OB stars in the \textit{Galactic O-star catalog} (GOSC). \footnote{\url{https://gosc.cab.inta-csic.es/}}
On the other hand, some GUs seem to be co-located with massive SFRs within the Gemini OB1 region (where source confusion is high). The GOSC catalog includes the Gemini OB1 region, see Figure \ref{fig:GemOB1}. 
Similarly, our procedure yielded no compelling matches with Be stars taken from the 
\textit{Be Star Spectra} (BeSS) list \footnote{\url{http://basebe.obspm.fr/basebe/}}.

\textbf{Colliding-wind binaries (CWB).}  Binary systems formed by early-type stars with strong winds may accelerate particles and thus shine in gamma-rays \citep{Mar21}. Only three objects in the list of CWBs established in \cite{deB13} are counterparts of 4FGL-DR4 sources:  Eta Carinae,  Kleinmann's star, and  $\gamma^2$ Velorum. They all show soft gamma-ray spectra, with power-law spectral indices between 2.3 and 2.6, or LP\_EPeak less than 610 MeV. Two more sources are associated with low confidence, namely  HD 318016 and BD 12 4988. 

\textbf{Bright stars.} Cosmic-ray electrons can Compton-scatter optical photons to gamma-ray energies in the Sun's atmosphere. In connection with EGRET's detection of the Sun \citep{Orl08}, \citet{Orl07} predicted that the {\Fermi}-LAT would be able to detect 9 nearby bright stars, mostly in Orion. These predictions motivated recent work \citep{deM21}, yielding only upper limits. Here we expand the search by looking at all bright stars from the Yale Bright Stars catalog
\footnote{\url{http://tdc-www.harvard.edu/catalogs/bsc5.html}}. 
No regular stars were found to be associated  with high confidence (P$>$0.8) with 4FGL-DR4 sources.

\textbf{Herbig-Haro (HH) and T Tauri objects.} 
These newborn stars are much less massive than WR or OB stars, and have weak winds in consequence.
However, they have collimated jets where strong shocks can occur, leading to predictions of particle
acceleration. 
\citet{Araya_HH219} argue that HH 219 contributes to 4FGL J0822.8–4207.
\citet{Yan_HH80-81} report a very soft LAT point source with TS $\approx 100$ at the position of the HH 80-81 system. The only nearby catalog source is 4FGL J1818.5-2036, $\sim 0.25^\circ$ away and with much lower significance, well co-located with a small RMS SFR and a Gaia open cluster.
We searched the \citet{Reipurth_HHcatalog} catalog for other HH stars near LAT sources. 
Only one has an angular separation less than the LAT 95\% semi-major localization axis: 
HH 408 is $\sim 0.09^\circ$ away from unidentified 4FGL J0441.8+2600c. 
Its latitude is $b = -13.2^\circ$ and source confusion is low. 

Nearly co-located with HH 408 is the T Tauri star Haro 6-33. A priori implausible as cosmic accelerators capable of generating LAT-detectable gamma-ray fluxes, \citet{GammaTTauri_delValle2011} predict that the LAT may see some, with a specific example in $\rho$ Oph. \citet{GammaTTauri_Filocomo2023} argue that a flaring T Tauri star made 3FGL J0546.4+0031c gamma-bright during the first two years of the Fermi mission, subsequently becoming invisible in 4FGL. \citet{gammaOrionB_Zeng2024} treat this same region, Orion B, more globally, as an ensemble of stellar clusters containing hundreds of OB stars and YSOs. They find an extended LAT source matching the cluster morphology, and discuss the possible gamma emission mechanisms.


\textbf{Gamma-ray binaries.}
Associations with 9 low- and 11 high-mass X-ray binaries have been reported in the 4FGL-DR4 catalog. These numbers represent small fractions of the known populations of X-ray binaries, amounting to 349 low- and 169 high-mass binaries as compiled in the recent XRBcats catalog \citep{Ava23,Neu23}. In addition, associations with 11 binary stellar systems (the brightest being Eta Carinae) have been found. The scarcity of the samples detected so far makes these classes unlikely to account for the bulk of the GUs.

To conclude this section, physically-motivated source categories beyond those used for the LAT catalog association pipeline shed light on about 10\% of the GUs, and constitute useful evolution but
not a revolution in understanding the bulk of the population.

\section{Correlation with eRASS1} 
\label{sec:erosita}

The recent release of the \textit{eRosita} (0.2-2.3 keV) eRASS1 catalog \citep{Mer24}
has paved the way to a deeper investigation of the X-ray - gamma-ray connection than allowed by the dated ROSAT all sky survey. Here we explore the above connection and try to assess to what extent it can shed light on the GU nature. The new  catalog includes over 900,000 sources in the Western ($\ell>180^\circ$) half of the sky, with a flux limit of about 5 $\times$ 10$^{-14}$ erg cm$^{-2}$ s$^{-1}$, and a typical positional accuracy of 5$''$. 

 The association procedure is performed via a likelihood-ratio method similar to that used for the FGL catalogs, described in Section 3.2 of \cite{2LAC} . 
 The likelihood ratio (LR) corresponds to the  ratio of  the probability to find the observed angular distance between the gamma-ray source and the counterpart if the association is real and the corresponding probability if the association is random. 
 The former probability depends on the sources' localization errors while the latter depends on the counterpart local spatial  density. 
 When computing this density, only sources with a flux greater than that of the considered counterpart are taken into account. 
 Comparing the LR distributions obtained for the real sky and for ``fake" ones, where the counterpart positions were shifted by a few degrees, enables us to estimate association probabilities. 
 The good uniformity of {\sl eRosita}'s sky coverage makes these probabilities much more reliable than what can be achieved for the Swift or XMM catalogs.  In contrast to the procedure followed by   \cite{May24} dedicated to searching for pulsars in the {\sl eRosita} sample, our approach is agnostic about the putative source class. 

\begin{figure}
\centering
\includegraphics[width=13cm]{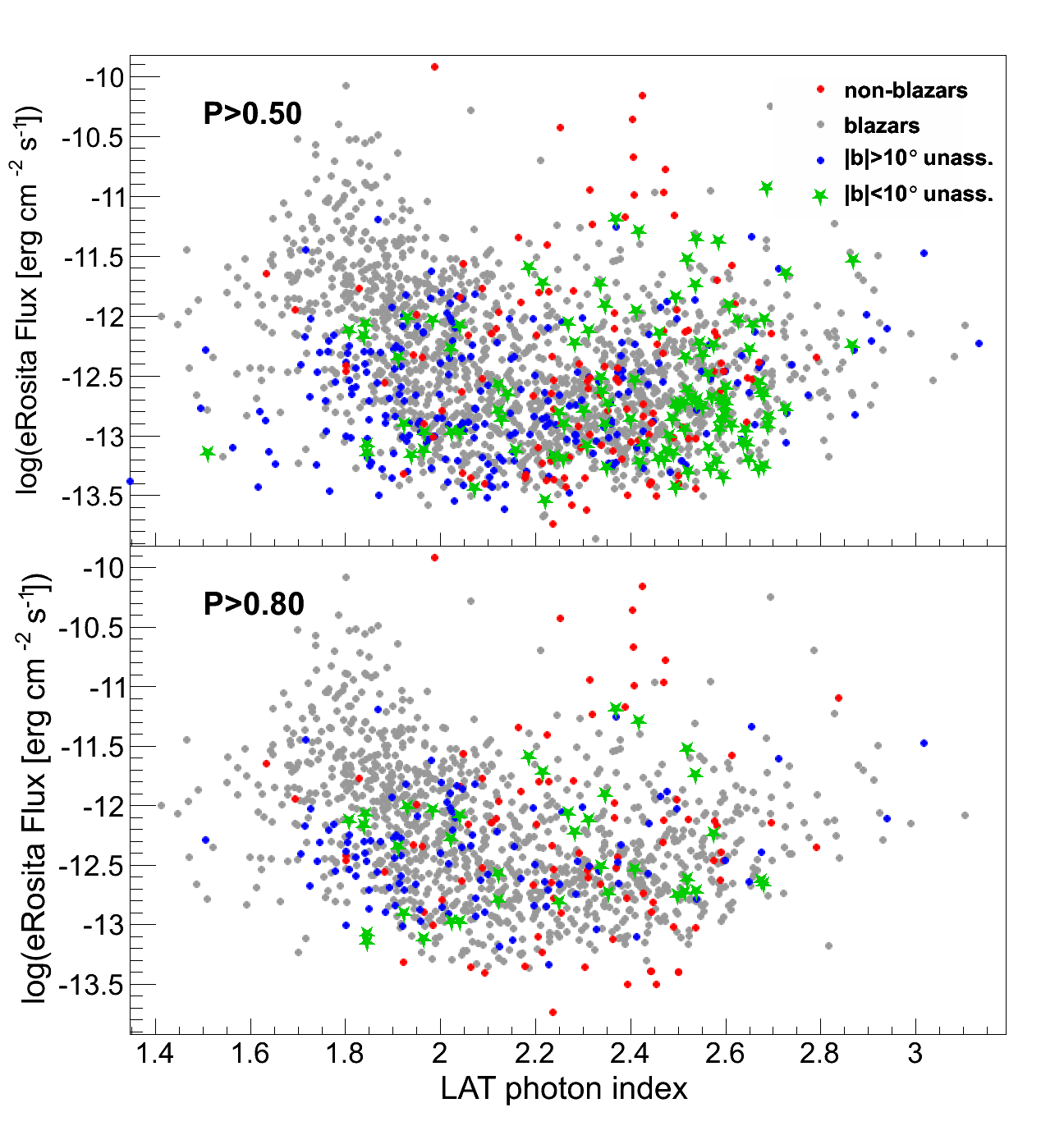} 
\caption{{\sl eRosita} flux plotted as a function of the LAT photon index for sources of different populations. The association probability threshold is 0.50 (top) and 0.80 (bottom). }
\label{fig:eRosita}
\end{figure}

Employing the same convention as that followed in the FGL catalogs, a probability $>0.8$  defines a high-confidence association. Out of 3626  $|b|>10^\circ$ 4FGL-DR4 sources, 1305 have a high-confidence {\sl  eRosita} counterpart. 
Not surprisingly, they comprise mostly active galactic nuclei, with a fraction of 4FGL-DR4 blazars with {\sl eRosita} counterparts of 54\%.
This fraction  drops to 25\% and 17\% for young and millisecond pulsars, respectively, but is quite high (60\%) for the 22 binary-star systems in 4FGL-DR4.   
As for the unassociated sources, the fractions of associations with {\sl eRosita} sources differ quite drastically as a function of the Galactic latitude, varying between 16\% (105/652) for $|b|>10^\circ$ and 5\% (30/650) for the GUs closer to the plane.  Figure \ref{fig:eRosita} displays the loci of the sources in the LAT photon index - {\sl eRosita} flux plane, for two different thresholds on the association probability.  In this Figure,  we are using the power law photon index (PL\_Index) instead of the LP\_EPeak parameter shown in the previous sections as about 50\% of the low-latitude sources with a high-confidence {\sl eRosita} association  are missing a meaningful (30 MeV $<$ LP\_EPeak $<$ 100 GeV) LP\_EPeak value. Let us recall that one of the defining criteria for SGUs in 4FGL-DR3 was a photon index greater than 2.4. High-confidence associations are displayed in Figure \ref{fig:eRosita} bottom. Blazars show an L-shaped distribution, with hard sources (low photon index) associated with bright X-ray counterpart. This is expected as most of these sources have a synchrotron hump in their spectral-energy distribution peaking near or in the X-ray band.  High-latitude unassociated sources are distributed quite similarly to blazars. Galactic associated sources are more uniformly distributed, as are the GUs. Out of the 30 GUs with high-confidence e-Rosita associations, 28 belong to the shoulder and only 7 are SGUs. Relaxing the probability association threshold from 0.80 to 0.50 (Figure \ref{fig:eRosita} top) leads to only 58 more soft GUs having eRASS1 counterparts. It can be concluded from the {low association rate with {\sl eRosita} counterparts} that SGUs are mainly faint in the soft X-ray band. 

\section{Mismodeled diffuse emission as a possible origin}
\label{sec:diffuse}

Diffuse emission is not fully understood and its modeling suffers many uncertainties. The possibility that the GU sources result from mismodeled Galactic emission has been investigated via three approaches: one explores whether allowing for more freedom to the fitted IEM at the region-of-interest scale reduces the number of detected GUs; the second tests whether underestimating the diffuse emission leads to spurious sources with spectral properties compatible with the GUs; the third checks whether GUs are slightly extended like diffuse clumps.

\subsection{Freeing more IEM components }

This test was among the first that we did, so it was carried out on the DR3 data (12 years) and catalog. It requires substantial resources, and the result does not depend qualitatively on adding two more years.
In the standard analysis performed in constructing the FGL catalogs \citep{LAT20_4FGL}, three parameters (the normalization and spectral index of the Galactic IEM component and the normalization of the isotropic one) are fitted for each considered region of interest (ROI) in addition to the individual-source parameters. We have explored the possibility that giving more freedom to the fit by allowing more of the original components of the iem\_v07 model\footnote{See \url{https://fermi.gsfc.nasa.gov/ssc/data/analysis/software/aux/4fgl/Galactic_Diffuse_Emission_Model_for_the_4FGL_Catalog_Analysis.pdf}.} to adapt independently for each ROI could affect the GUs and reduce their numbers. In order to leave the number of degrees of freedom (originally 35 because of the 10 Galactocentric rings) manageable only parameters of the inner rings (inside and including the local ring), contributing the most where the GU density is high, were set free. The two inner rings for HI are empty, and the three, fairly faint, rings CO\_1, CO\_2, CO\_3 were  merged together into CO\_123. The IC emission (very smooth so it is unlikely it can impact the GUs), the negative DNM (a negative component cannot be left free in a Poisson likelihood framework), the unresolved sources (a template obtained from Monte-Carlo simulations, so not smooth) and the outer rings were merged into a single component (the rest). Overall, this approach has at most 14 free parameters (normalizations of CO\_0, CO\_123, CO\_4 to CO\_6, HI\_2 to HI\_6, DNMp, patch, rest and isotropic) instead of three in the standard analysis\footnote{Model cubes of those components are available on request in FITS format (6 GB).}.
Parameters of very minor components in a given ROI (below a threshold $f_{\rm min}^{\rm Npred}$ on the fraction of model counts) are fixed to 1.
The spatial distributions of all components are very different, so they do not introduce large covariances between parameters.
We applied priors to ease convergence, setting the means to 1 (no change) and the widths to 0.2 (0.1 for the rest and the isotropic). We added the same fixed Sun and Moon components as in DR3.

We started from the same uw1216 seeds and extended sources as 4FGL-DR3 (not just from the final DR3 sources) and allowed removing DR1 and DR2 sources at TS $<$ 25 (contrary to the incremental DR3 approach). 
In the standard LAT catalog analysis pipeline \citep{LAT22_4FGL_DR3}, we applied this setting to the region defined as $|b|<10\arcdeg$ and $|l|<100\arcdeg$, where most GUs lie (background sources outside that region were taken from DR3). In a first attempt we set $f_{\rm min}^{\rm Npred}$ to 3\%.
Since the normalizations of minor components just above $f_{\rm min}^{\rm Npred}$ were well constrained, in a second attempt we reduced $f_{\rm min}^{\rm Npred}$ to 1\%.
In a third attempt, we added spectral freedom (a power law) to the 13 Galactic components, resulting in at most 27 free parameters. We applied priors to ease convergence, setting the mean indices to 0 (no change) and the widths to 0.1 (0.05 for the dominant ``rest'' component).

\begin{deluxetable*}{lrrrrrrr}
\tablecaption{SGU resilience to more {\bf freedom in the model of diffuse Galactic emission}
\label{tab:diffcomps}
}
\tablehead{
\colhead{Description} & \colhead{NPThresh} & \colhead{$\overline{{\rm NDFree}}$} & \colhead{$\overline{{\rm RatNpred}}$} & \colhead{$\sigma_{\rm RatNpred}$} &\colhead{$N_{\rm tot}$} & \colhead{$N_{\rm com}$} & \colhead{$N_{\rm SGU}$}
}
\startdata
4FGL-DR3     & 0\% & 3.0  & 0       & 0      & 1414 & 1414 & 527 (100\%) \\
Norms free   & 3\% & 7.6  & +0.34\% & 0.64\% & 1236 & 1157 & 379  (72\%) \\
Norms free   & 1\% & 9.6  & +1.30\% & 0.68\% & 1206 & 1141 & 371  (70\%) \\
Indices free & 1\% & 18.2 & +1.07\% & 0.77\% & 1169 & 1117 & 357  (68\%) \\
\enddata
\tablecomments{The results pertain to DR3 data in the area defined as $|b|<10\arcdeg$ and $|l|<100\arcdeg$. NPThresh is the minimum fraction of {\bf model} counts required for a component to be free. $\overline{{\rm NDFree}}$ is the average number of free diffuse parameters per ROI. $\overline{{\rm RatNpred}}$ is the average (over all ROIs) ratio of diffuse model counts to the DR3 reference, and $\sigma_{\rm RatNpred}$ its scatter. $N_{\rm tot}$ is the total number of point sources found at TS $>$ 25 in the entire area. $N_{\rm com}$ is the number of point sources common to DR3. $N_{\rm SGU}$ is the number of surviving DR3 sources flagged as SGUs (and their fraction in \%).
}
\end{deluxetable*}

\begin{deluxetable*}{llrrrrr}
\tablecaption{SGU resilience in the spike and shoulder
\label{tab:diffcompslat}
}
\tablehead{
\colhead{Description} & \colhead{Latitude} & \colhead{$N_{\rm tot}$} & \colhead{$N_{\rm com}$} & \colhead{$N_{\rm GU}$} & \colhead{$N_{\rm sppunk}$} & \colhead{$N_{\rm assoc}$}
}
\startdata
4FGL-DR3     & $|b| < 1\arcdeg$      &  397 &  397 (82.4) & 210 (64.1) &  88 (63.8) &  99 (338) \\
4FGL-DR3     & $1 < |b| < 10\arcdeg$ & 1017 & 1017 (57.2) & 620 (48.0) & 106 (47.4) & 291 (178) \\
\hline
Indices free & $|b| < 1\arcdeg$      &  342 &  332 (84\%) & 162 (77\%) &  73 (83\%) &  97 (98\%) \\
Indices free & $1 < |b| < 10\arcdeg$ &  827 &  785 (77\%) & 428 (69\%) &  80 (75\%) & 277 (95\%) \\
\enddata
\tablecomments{The results compare the DR3 data to the most flexible alternative within $100\arcdeg$ of the Galactic Center, separately for the spike ($|b| < 1\arcdeg$) and the shoulder ($1 < |b| < 10\arcdeg$). $N_{\rm tot}$ is the total number of point sources found at TS $>$ 25. $N_{\rm com}$ is the number of point sources common to DR3. $N_{\rm GU}$ is the number of surviving DR3 GUs. $N_{\rm assoc}$ is the number of surviving associated DR3 point sources, except SPP and UNK. $N_{\rm sppunk}$ is the number of surviving DR3 point sources classified as SPP or UNK. The value in parentheses for the 4FGL-DR3 rows is the median TS in each category. The value in parentheses for the ``Indices free'' rows is the surviving fraction in each category.
}
\end{deluxetable*}

The results are summarized in Table \ref{tab:diffcomps}. We discarded all sources at TS $<$ 25 in DR3 when comparing source numbers. We initially checked the sources bearing the SGU flag, and noted that their numbers decreased significantly (in proportion, much more than other sources), and continued decreasing when adding more freedom to the diffuse model. This is in part because the average TS of SGUs is much less than that of other sources (similar to that of GUs in Figure \ref{fig:unassoc_TS}).
Not surprisingly, the removed SGUs tend to be even closer to the TS threshold than the remaining ones. 

Encouraged by this, we went further with the third approach (with spectral freedom) and the results are reported in Table \ref{tab:diffcompslat}. In particular, we checked the fate of all GUs, whose numbers decreased similarly, and also of sources classified as SPP or UNK, both in the spike and in the shoulder. The SPP or UNK are a little more robust than the GUs, but 21\% did not survive. The reason why the surviving fraction is larger in the spike than the shoulder is that the median TS is larger in the spike. One may wonder whether the reason why associated sources survive more is the same (they have much larger median TS than the GU and SPP, which are similar). But even eliminating the brightest associated sources to force the same median TS as the GU and SPP, they remain more robust (90\% surviving fraction in the spike, 83\% in the shoulder).

The range of output diffuse parameters was typically a few tens of \% (except the dominant local HI ring that varied little), with some rising while others decreased. Table \ref{tab:diffcomps} shows that the overall diffuse level (measured by the model counts) increased by around 1\%, with a small scatter between ROIs. A larger value was expected because more degrees of freedom can fit more structures, but the overall increase was very modest, because the model must remain compatible with the data.
The log(Likelihood) reached with the alternative diffuse models was not better, even accounting approximately for the different numbers of free parameters. In other words, agreement with the data can be reached just as well by releasing constraints on the diffuse model or by adding more sources. In that sense, this test is not a proof that the GUs (even the fragile ones) are mostly due to imperfect modeling of the diffuse emission. Conversely, the fact that 71\% of the GUs survived the test does not prove that those are all real sources, because we cannot be sure that this test captures the true interstellar emission.
The conclusion is that leaving more freedom to the current model indicates that the GUs could be partly due to an imperfect diffuse model, so it does not rule out this option.  

\subsection{Simulations of underestimated diffuse emission}
\label{sec:simdiff}

The previous test justified going further in this direction.
If GUs arise from mismodelled diffuse emission, the energy spectra of these two components must be related. Indeed, summing all the GUs results in a spectrum peaking below 1 GeV, not so different from the spectrum of the IEM.
It is a little softer than the IEM, but this could be due to the fact that
GUs are fit as point sources, and if they are not (as expected for diffuse clumps) the fit misses part of the emission above the energy at which the instrumental PSF becomes smaller than the clump.

To test this idea we prepared simulations using the gtobssim tool of the Fermitools suite\footnote{\url{https://fermi.gsfc.nasa.gov/ssc/data/analysis/documentation/Cicerone/}} under different conditions. The simulations were carried out early in this study, so they were prepared for 12 years. However we reanalyzed the results in the DR4 software framework \citep[with priors on curvature,][]{LAT23_4FGL_DR4}, so that they can be compared with the DR4 GUs. We tested on one example that adding two more years does not change anything qualitatively. The simulations were carried out over 10 36$\arcdeg$x36$\arcdeg$ regions spanning the entire Galactic Plane. The standard IEM was the basis of the simulation in all cases, and we added different types of extra emission. In all cases we applied the standard catalog analysis in a pure point-source hypothesis and with the standard IEM (as is done in the real catalog), resulting in a fake catalog with TS and spectral parameters of all ``sources'' at TS $>$ 25.

In a first attempt we tested the possibility that the missing emission has the same spatial structure as the known one. So we added a simulated DNMp component (dark gas is the most uncertain part of the IEM, and we tested other components with similar results). This amounts to multiplying the DNMp by two relative to the standard IEM. The excess component was masked out above $|b|$=10$\arcdeg$ to avoid edge effects (the simulation covers $|b| < 18\arcdeg$). 
No point source was introduced. We then used the \textit{find\_sources} tool of the fermipy package\footnote{\url{https://github.com/fermiPy/fermipy}} to detect the source-like features arising from the resulting excess of photons. These features were then treated as seeds in the standard catalog analysis.

\begin{figure*}
\centering
   \begin{tabular}{cc}
   \includegraphics[width=0.5\textwidth]{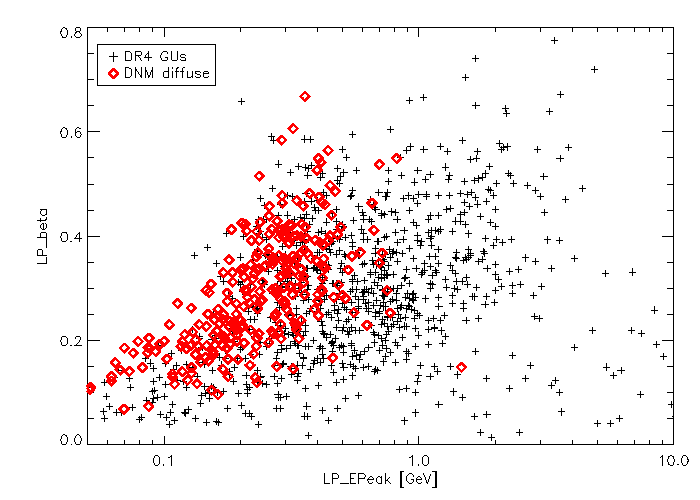} & 
   \includegraphics[width=0.5\textwidth]{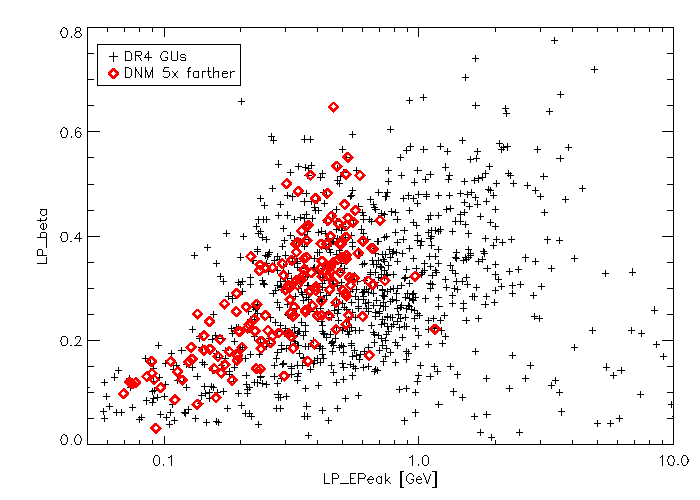} \\
   \includegraphics[width=0.5\textwidth]{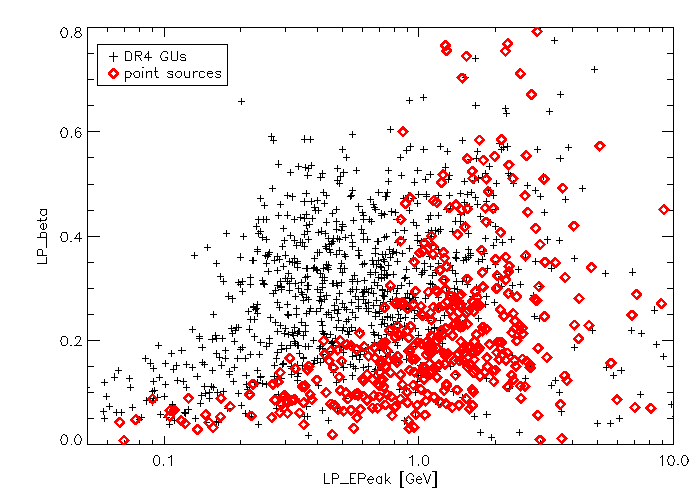} & 
   \includegraphics[width=0.5\textwidth]{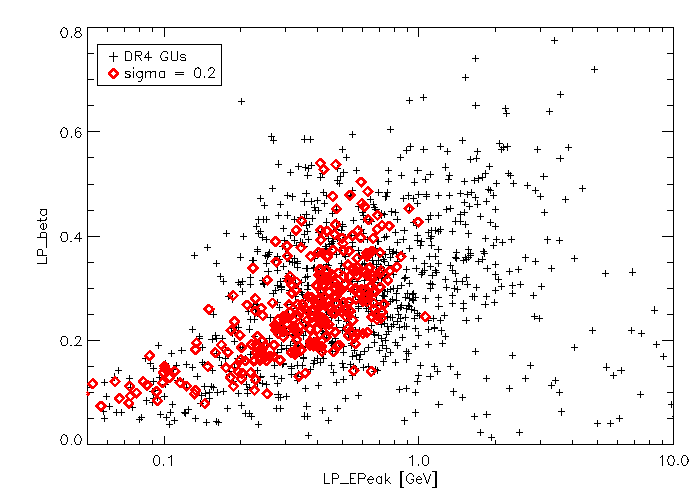} \\
   \includegraphics[width=0.5\textwidth]{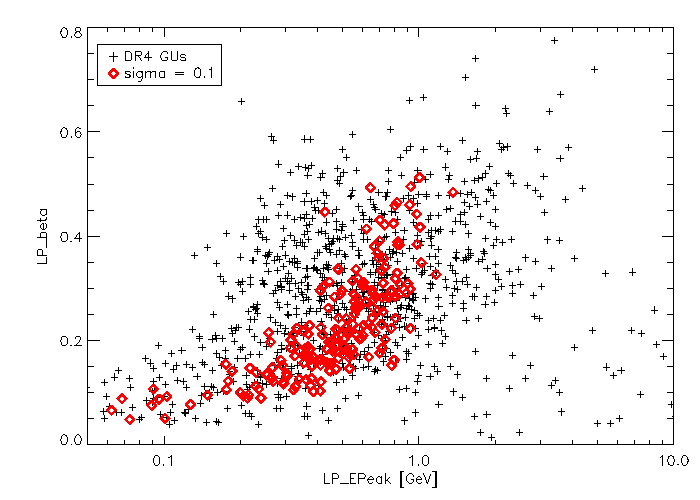} & 
   \includegraphics[width=0.5\textwidth]{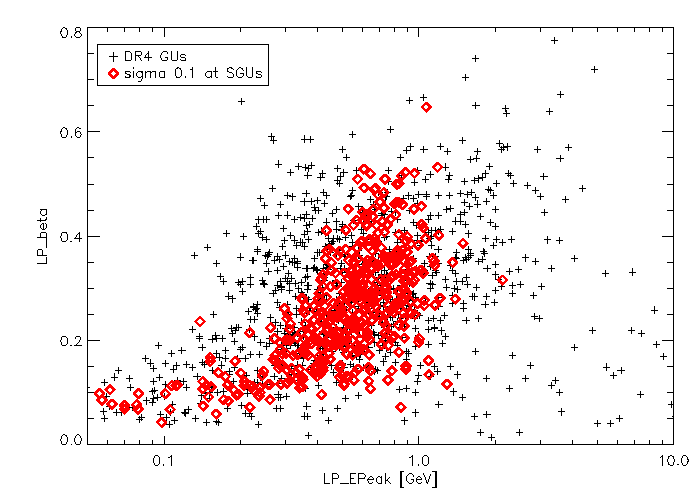} \\
   \end{tabular}
   \caption{LogParabola $\beta$ vs EPeak of fake sources resulting from simulations (bold red diamonds), compared to those observed for the DR4 GUs (small plus signs, same as Figure \ref{fig:epeak_beta}).
     Top left: Enhancing the DNMp diffuse component by a factor two.
     Top right: Reducing the spatial scale of the additional DNMp component by a factor five.
     Middle left: Point sources on a regular grid.
     Middle right: Gaussian extended sources on the same grid, with $\sigma = 0.2\arcdeg$.
     Bottom left: Gaussian extended sources on the same grid, with $\sigma = 0.1\arcdeg$.
     Bottom right: Gaussian extended sources at the SGU positions, with $\sigma = 0.1\arcdeg$.}
\label{fig:simDNMp}
\end{figure*}

It resulted in 336 point sources, with median TS = 48, similar to that of GUs.
Their spectral characteristics are illustrated in Figure \ref{fig:simDNMp}, top left. Their curvatures are similar to those of GUs, but their peak energies are nearly all below 500 MeV, clearly lower than those of GUs. The effect mentioned above (missing high-energy emission when a source is extended in truth but fit as a point source) is too strong, and the collective spectrum of those fake sources is softer than that of GUs. In other words, the known diffuse features are too broad to explain the GUs.

To go further we attempted to reduce the spatial scales in the DNMp, simply dividing by five the pixel scale in the template (as if the same structures were five times further) and repeating the template five times along longitude. This disconnects the simulated diffuse emission from the original DNMp component, but this difference does not affect the results (the DNMp component is not the major contributor to the IEM). It resulted in 196 point sources, with median TS = 45. Their spectral characteristics are illustrated in Figure \ref{fig:simDNMp}, top right. The peak energy is indeed found a little larger, but not enough.

In view of this, we abandoned the idea of simulating known diffuse features and tried simulating sources with the same spectrum as the IEM. We started with point sources (corresponding to the limit of very small clumps). They all had the same flux, chosen to be similar to the GUs. Since we were exploring their spectral (not spatial) characteristics, the sources were simulated on a regular grid every 2$\arcdeg$ in longitude and latitude (to avoid confusion effects), from $-$9 to 9$\arcdeg$ in latitude, and $-$98 to 98$\arcdeg$ in longitude, in the region where most GUs lie.
It resulted in 466 point sources, with median TS = 51. Their spectral characteristics are illustrated in Figure \ref{fig:simDNMp}, middle left. Their curvatures are somewhat lower, and their peak energies definitely larger, than those of GUs. This was expected, since we had noticed that the collective spectrum of the GUs is softer than that of the IEM.

We then moved to simulating extended sources (somewhat larger clumps) with the same IEM spectrum (while still fitting them as point sources). We chose a Gaussian distribution and varied its spatial scale $\sigma$.
We simulated only half as many as the point sources along longitude, but at about twice the flux (to compensate for the expected loss). 
Our first attempt was with $\sigma = 0.2\arcdeg$. It resulted in 312 point sources, with median TS = 93. Their spectral characteristics are illustrated in Figure \ref{fig:simDNMp}, middle right. This definitely goes in the right direction, but a little too far (peak energy is a little too small).

We then reduced $\sigma$ to 0.1$\arcdeg$. We simulated only 200 sources, but at a different flux depending on position on the grid, so that they should all be detected at TS $\sim$ 100. Their spectral characteristics are illustrated in Figure \ref{fig:simDNMp}, bottom left. The peak energy increased indeed, and became similar to the average peak energy of GUs.

In a final test we simulated sources with the same $\sigma=0.1\arcdeg$ at the positions of all SGUs in DR3, and somewhat fainter to have a median TS closer to that of GUs. This should be more realistic, putting the fake sources on top of the same diffuse features as the real GUs, and accounting for their clustering (confusion).
It resulted in 498 point sources, with median TS = 59. Their spectral characteristics are illustrated in Figure \ref{fig:simDNMp}, bottom right. Their peak energies remained similar to those of GUs.


It is clear from the final plots that the distribution of peak energies in GUs is broader than that of fake sources of identical width. This means that a distribution of widths would be required, which is not particularly surprising.
We conclude from this exercise that in theory it is possible to account for the spectral characteristics of GUs with unknown diffuse clumps with a distribution of widths around 0.1$\arcdeg$, just below the resolution of current surveys of the entire Galactic plane.
We discuss what they could be in Section \ref{sec:missing_gas}.

\subsection{Searching for extension in bright GUs}
\label{sec:extsearch}

The simulations described in Section \ref{sec:simdiff} concluded that, if the GUs are somehow diffuse clumps of interstellar gas, they must be small but appear slightly extended for the LAT, with Gaussian widths around 0.1$\arcdeg$.
The logical thing to do after this is to check on the real data whether there are indeed hints of extension in the observed GUs.
This is difficult because the GUs are soft, with peak energies mostly below 1 GeV, where the LAT PSF is broad. The Half-Width at Half-Maximum (HWHM, relevant for comparison with a Gaussian $\sigma$) of the LAT PSF at 1 GeV is 0.24$\arcdeg$, more than twice larger than the goal. There are however reasons to be more optimistic: first, the PSF3 event type (one fourth of all events) has HWHM = 0.21$\arcdeg$ at 1 GeV, slightly better; second, the spectra are strongly curved but extend to a few GeV, where the PSF improves strongly (HWHM = 0.15$\arcdeg$ at 2 GeV); third, with many photons it is possible to test extensions somewhat below the PSF width.
In any case, it is clear that such a small width can be measured only in bright enough sources. So we considered a sample restricted to the brightest GUs only (TS $>$ 230, containing 57 sources). We also tested four sources classified as SPP and five as UNK above the same TS threshold, since SPP and UNK were found similar to GUs in many respects.


Practically, we used the same 14-year data (split into PSF event types), ROIs, sources, diffuse background model and free parameters as in DR4 to test for extension using the \textit{extension} tool of the fermipy framework. We restricted the data to energies above 1 GeV (as argued above, energies below that bring little information for extension), except for nine GUs peaking far below 1 GeV, for which we reached down to 500 MeV, excluding PSF0 events and zenith angles $> 100\arcdeg$ up to 1 GeV. To save time we also limited the energy range to 10 GeV, since GUs emit very little above that, except for four sources with peak energy above 3 GeV, which were analyzed up to 100 GeV. We used small pixels of 0.05$\arcdeg$ to maximize the resolution over 8$\arcdeg$ square ROIs (enough above 1 GeV), and the DR4 weights. The same background sources as in DR4 were left free in the process (up to 9 for the most complex ROIs). We relocalized the GUs first (since DR4 is an incremental catalog, source positions were left at their first appearance, so the localizations of bright sources are often based on 8 years). Then we attempted to fit a Gaussian extension to each source, recording TS$_{\rm ext}$ (twice the likelihood ratio between the extended and point-source options), the best fit and the upper limit to the Gaussian $\sigma$, the offset between the DR4 and best-fit localizations and the localization error.

\begin{figure}
\centering
\includegraphics[]{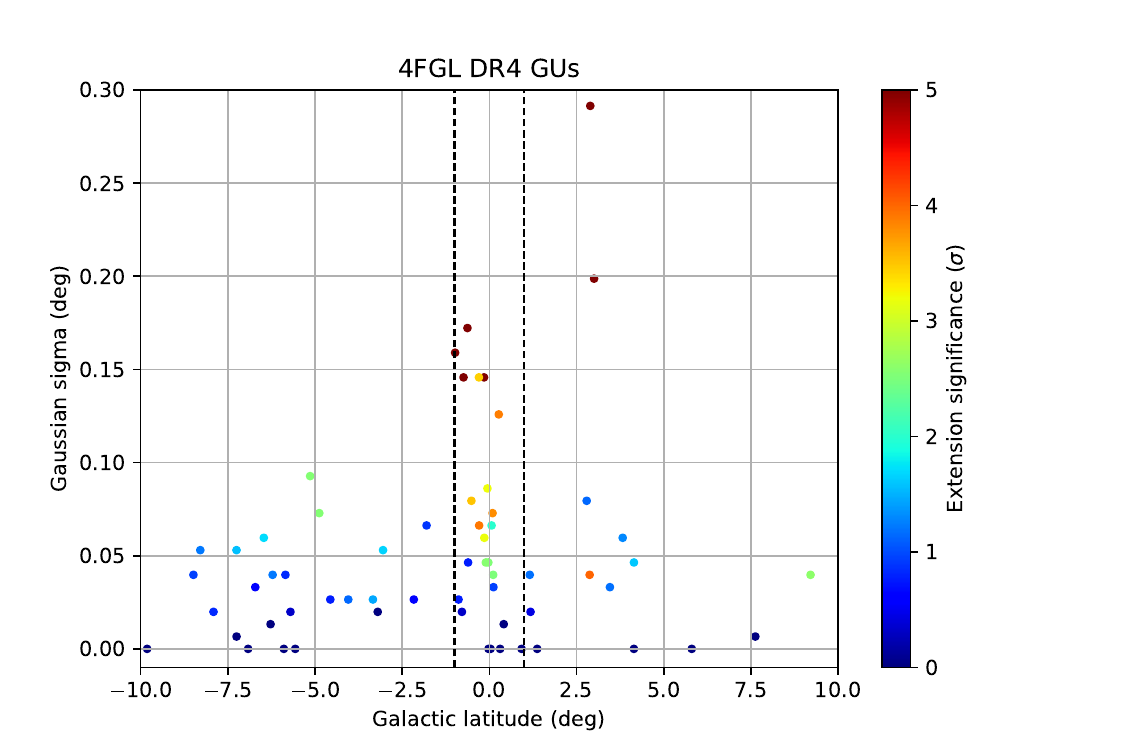}
\caption{Best-fit Gaussian extension of 62 GUs, SPP and UNK sources at TS $>$ 230 in DR4, as a function of their Galactic latitude. The color denotes the extension significance $\sqrt{{\rm TS}_{\rm ext}}$. Blue is not significant ($< 2\sigma$). The vertical dashed lines materialize $|b| < 1\arcdeg$. Four GUs at TS $>$ 230 converge to $\sigma > 0.3\arcdeg$ so they do not appear on this plot.}
\label{fig:GUExtentSig}
\end{figure}

The results are shown in Figure \ref{fig:GUExtentSig}.
Three GUs (4FGL J1046.7$-$6010, J1048.5$-$5923 and J2038.4+4212) are very special. They are inside large known extended sources in Carina and Cygnus, and were added to limit local residuals, but were known in advance to be diffuse features. They come out indeed with $\sigma \sim 0.4\arcdeg$ (outside Figure \ref{fig:GUExtentSig}).
Three sources (4FGL J0616.5+2235, J0650.6+2055 and J2217.5+6346, the last two being UNK) are much harder than the bulk of GUs, with peak energies around 10 GeV, to the right of the plots in Figure \ref{fig:simDNMp}. The first two are found slightly extended at $\sigma = 0.04\arcdeg$ (small extensions are measurable in hard sources). They appear respectively as the orange point at $b = 2.9\arcdeg$ and the green point at $b = 9.2\arcdeg$ in Figure \ref{fig:GUExtentSig}. They might be small pulsar wind nebulae, but they are largely unrelated to the main GU population.

Among the 60 other sources, 20 show indications of extension, at TS$_{\rm ext} > 4$ (above $2\sigma$ significance). Most widths range from 0.04 to 0.2$\arcdeg$. Two (4FGL J1929.0+1729 and J2108.0+5155) are found to have even larger widths (0.46 and 0.29$\arcdeg$, respectively). The former (outside Figure \ref{fig:GUExtentSig}) is close to PSR J1928+1746, a high $\dot{E}$ but faint pulsar detected in timing only. The latter (the largest $\sigma$ visible in Figure \ref{fig:GUExtentSig}) is an interesting special case because it is quite far from the Galactic plane ($b = 2.9\arcdeg$), in a small cluster of sources on top of strong diffuse emission. 
The smallest significant extensions tend to be found in the brightest sources, but this is just a selection effect (such small extensions cannot be significant in fainter sources).

Among the 60 sources considered above, 16 moved far enough that their original 4FGL position was outside the 95\% error circle from $fermipy$ over 14 years (not accounting for any systematic effect), indicating that their position was unstable. Most of those (11) were also found extended. Among the five point sources that moved far, one was inside FHES J1741.6$-$3917 and two (4FGL J2004.3+3339 and J1759.7$-$2141) were UNK, which actually moved away from their putative counterparts, confirming that the UNK associations are not very solid.

The most striking feature in Figure \ref{fig:GUExtentSig} is the concentration of extended GUs (defined from TS$_{\rm ext} > 4$) very close to the Galactic plane (in the spike rather than the shoulder).
This can be quantified by comparing the latitude distributions of extended GUs and of point-like GUs.
Among the 60 GUs mentioned above excluding the six special cases, 20 are probably extended and 40 are not. Only 4 extended GUs are at $|b| > 1\arcdeg$ vs 16 at $|b| < 1\arcdeg$. Conversely, 31 point-like GUs are at $|b| > 1\arcdeg$ vs 9 at $|b| < 1\arcdeg$.
The binomial probability to get 4 or fewer among 20 and 31 or more among 40 in two draws assuming the same $p$ in both is only $5 \times 10^{-6}$ (for $p$ = 0.589), a $4.6\sigma$ effect. We conclude that extended GUs are definitely concentrated in the spike within $1\arcdeg$ of the Galactic plane.

Corroborating the same conclusion, most GUs at TS$_{\rm ext} < 4$ (and therefore most GUs in the shoulder) have 95\% upper limits to the Gaussian $\sigma < 0.1\arcdeg$, whereas Section \ref{sec:simdiff} would require at least that width, if they are interstellar clumps illuminated by cosmic rays.
Besides 4FGL J2108.0+5155 mentioned above at $\sigma \sim 0.3\arcdeg$, the other broad GU outside the spike (at $\sigma = 0.2\arcdeg$ in Figure \ref{fig:GUExtentSig}) is J1714.9$-$3324 in the Galactic Bulge.

The combined conclusion of Sections \ref{sec:simdiff} and \ref{sec:extsearch} is that the ``diffuse clump'' hypothesis is plausible for GUs close to the Galactic plane, but not a very likely explanation further off, in the shoulder.


\section{Missing Gas}
\label{sec:missing_gas}

The previous section simulations indicate that a distribution of
clump-like structures may be a possible origin for some of the low flux
SGUs.
Line intensities for the $^{12}$CO emissions become optically thick for
higher densities, and so the gas surveys \citep{2001ApJ...547..792D}
tracing the molecular gas for the 4FGL IEM under estimate the high
column densities coming from probing the full depth of molecular cloud
profiles.
Because of this the `missing gas' may lead to a distribution of slightly
extended excesses following the lower density regions traced by the
optically thinner $^{12}$CO emissions -- as possibly evidenced by the SGUs.
An avenue to investigate this possibility is to use the emissions of the
line transitions for rarer CO isotopologues, e.g., $^{13}$CO, which
allow for a more complete tracing of the molecular gas to higher column
densities.

Correlation of unassociated 4FGL sources with these was explored by
\citet{2023PhRvD.107l3032K}. They used the Mopra Southern Galactic Plane
CO survey data release 3 \citep{2013PASA...30...44B,2018PASA...35...29B}
to investigate whether unaccounted molecular gas could explain some of
the unassociated sources in the 4FGL-DR2.
Mopra is a single-dish radio telescope located approximately 450~km
northwest of Sydney, Australia.
The survey covers Galactic longitudes $l=300^\circ$--$350^\circ$ for
$|b|<0.5^\circ$ at high spatial (0.01$^\circ$) and spectral
(0.1~km~s$^{-1}$) resolution of the emissions for four CO isotopologue
$J=1\rightarrow0$ transition lines: $^{12}$CO, $^{13}$CO, C$^{18}$O, and
C$^{17}$O.
\citet{2023PhRvD.107l3032K} generated molecular gas maps based on Mopra
$^{12}$CO and $^{13}$CO data (other species had insufficient coverage or
were not detected), and compared the $\gamma$-ray emissions predicted
with the GALPROP code using these to those based on the
\citet{2001ApJ...547..792D}.
Under the assumption that all other diffuse emissions processes
accurately model the $\gamma$-ray sky (electron Compton scattering of
the interstellar radiation field, along with electron bremsstrahlung and
nuclei pion production on the atomic and ionised gas), they developed a
list of potential sources spatially distributed as the higher density
regions traced by the Mopra data.
For those that were determined to have a test statistic $TS>16$, they
found 23 sources.
Following correlation using 95\% containment radii with the 4FGL-DR2
unassociated sources they concluded that it was likely that some
fraction of their newly identified sources could already be apparent in
the catalogue.
But whether the entirety of their sample has been detected is unclear.
Examining their Figure~1 shows overlap between some of their sources and
4FGL-DR2 unassociated ones, but there are many over the Mopra region
that are not near any catalogue source.

To quantitatively test the hypothesis that SGUs in this region
preferentially trace high-density molecular gas, we performed a
comparative analysis using the Mopra data. The Mopra survey region
(covering approximately region `6' in Figure~\ref{fig:map_ROIs}) contains
72 SGUs, the majority of which are faint sources with a test statistic
$TS < 100$. For each SGU, we integrated the Mopra $^{12}$CO and
$^{13}$CO emissions within its 95\% location uncertainty radius
($\sim$0.05$^\circ$), after smoothing the spectra with a 1~km~s$^{-1}$
Gaussian kernel to reduce noise. To establish a control sample,
we defined a background population by applying an identical
integration method to all non-overlapping, comparably sized (0.1$^\circ$
diameter) regions across the survey area, integrating above the
2$\sigma$ noise limit.

\begin{figure}
   \centering
   \includegraphics[]{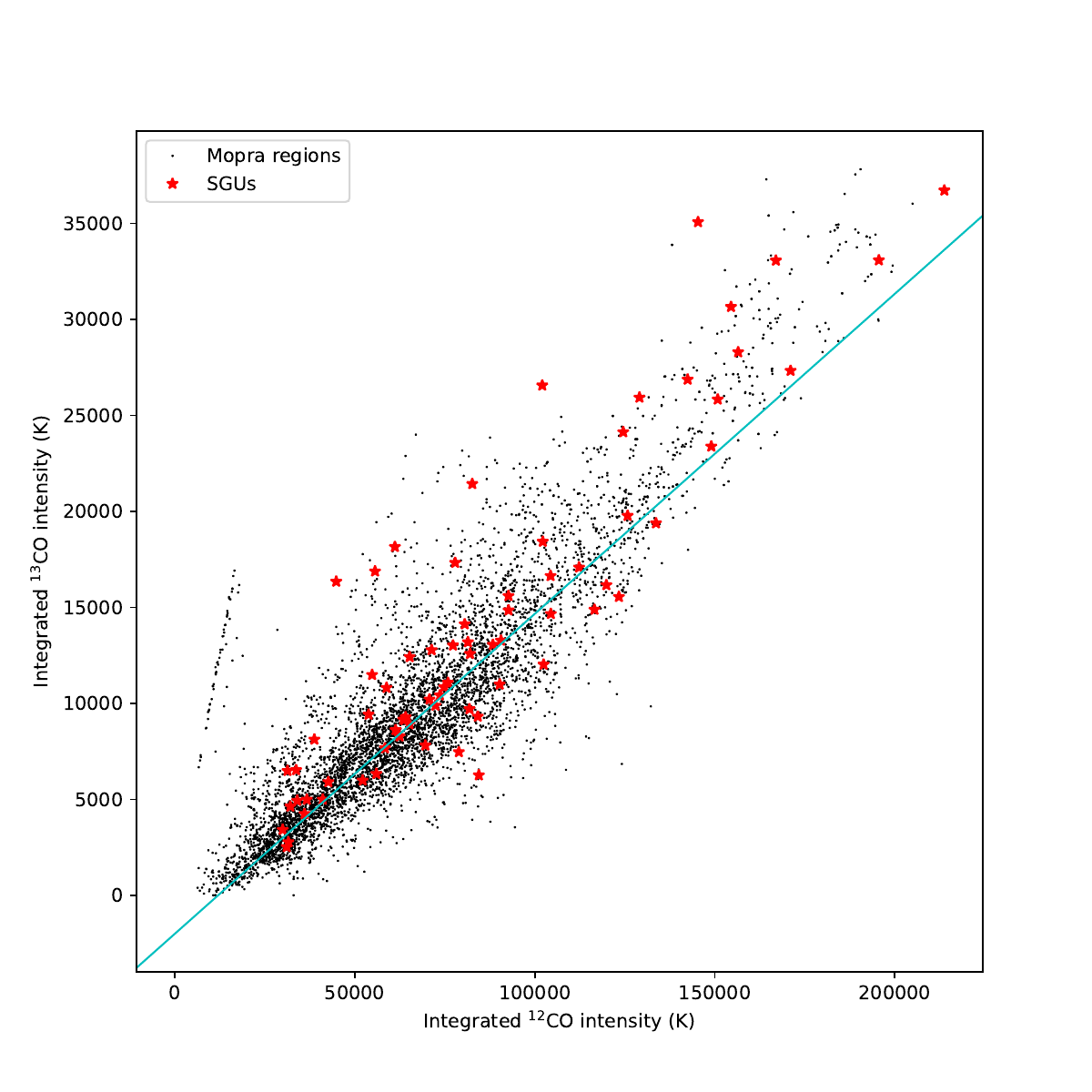}
   \caption{Correlation of integrated CO line intensities over the Mopra
survey region $l=300^\circ$--$350^\circ$ for $|b|<0.5^\circ$ for
overlapping SGUs (red stars) and all other $\sim$0.1$^\circ$ regions
(black dots). The cyan line has slope 1/6.}
   \label{fig:fig_mopra_sgu}
\end{figure}

Figure~\ref{fig:fig_mopra_sgu} presents the integrated $^{13}$CO versus
$^{12}$CO intensities for both the SGU-coincident regions (red stars)
and the background population (black dots). If SGUs were
predominantly associated with optically thick gas, their distribution
would be biased towards higher $^{13}$CO/$^{12}$CO intensity ratios
compared to the control sample. However, the two populations are
observed to be statistically indistinguishable. The
SGU-coincident regions occupy the same parameter space as the background
regions, following the same correlation trend, including the upward
curvature at high intensities (beyond $^{13/12}$CO intensities
$\sim$15000/100000~K). We find no evidence for a distinct
sub-population of SGUs that would support an association with the
"missing gas" traced by $^{13}$CO. The small cluster of points at low
$^{12}$CO intensity, likely an artifact of our noise thresholding, does
not impact this primary conclusion.

As a supplementary check, a systematic visual inspection of the
Mopra spectra was performed at the location of each SGU to search for
kinematic alignment with prominent $^{13}$CO spectral features. This
qualitative analysis revealed no systematic correlation; while a small,
incidental number of cases showed some spatial and velocity alignment
with a gas peak, the vast majority of SGUs showed no corresponding
features that would distinguish their sightlines from random locations
in the survey. Given the null result from both the integrated
intensity analysis (Figure~\ref{fig:fig_mopra_sgu}) and this spectral
inspection, a more computationally intensive line-fitting analysis for
the full sample was not pursued as it was unlikely to yield a
statistically significant association.

\section{Multiwavelength investigation of the brightest GUs}
\label{sec:bright_GUs}







\begin{figure}
\centering
\includegraphics[width=18cm]{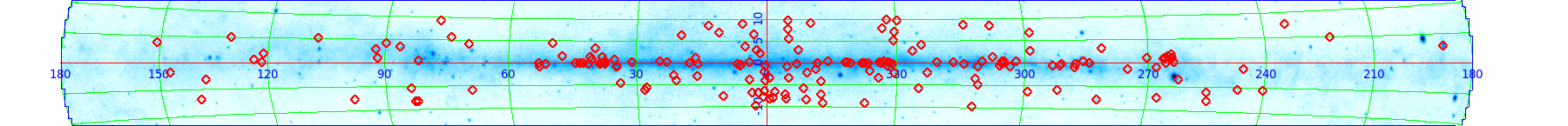} 
\caption{ Sky map (Galactic coordinates, Hammer-Aitoff projection) displaying the positions of the sources in the bright sample. The background is the 14-year Fermi intensity map above 1 GeV.}
\label{fig:sky_bright_GUs}
\end{figure}

\begin{figure}
\centering
\includegraphics[width=13cm]{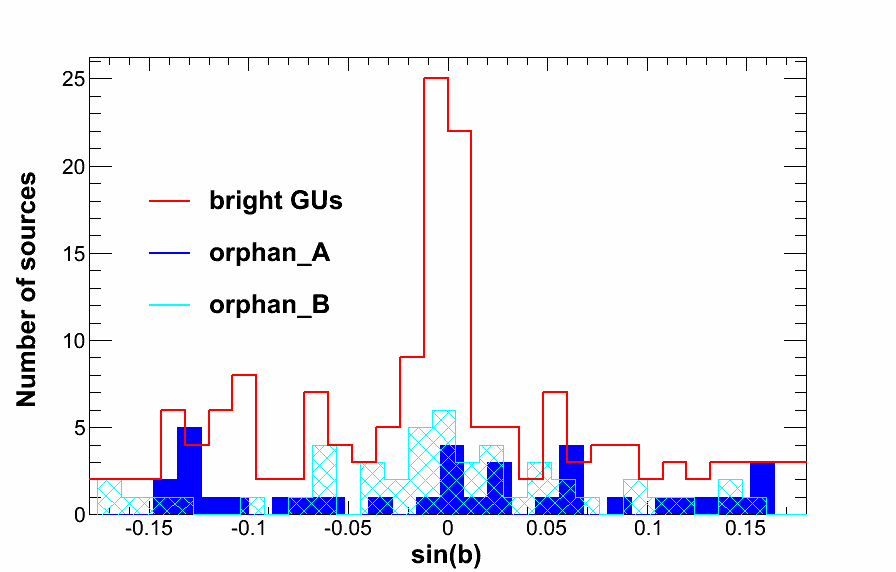} 
\caption{ Galactic-latitude distribution of the sources in the bright sample and in the ``orphan" one.  }
\label{fig:lat_bright_GUs}
\end{figure}

\begin{figure}
\centering
\includegraphics[width=13cm]{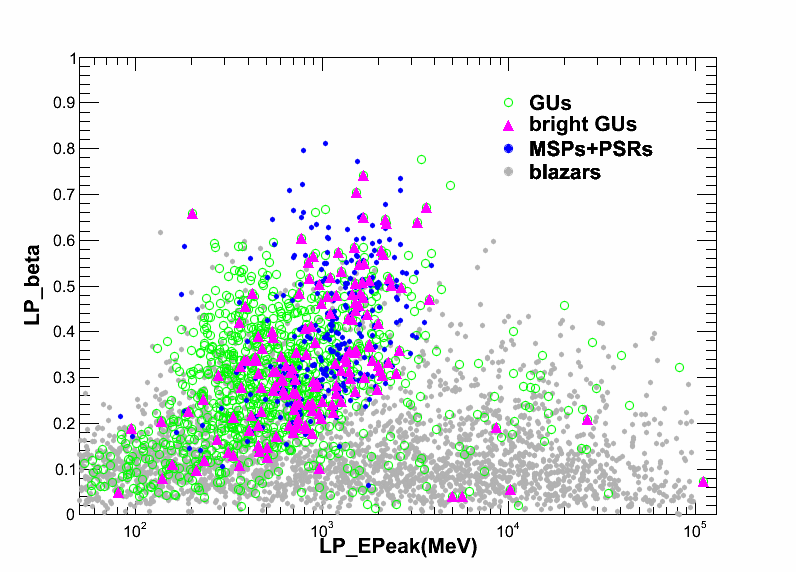} 
\caption{ Same as Figure \ref{fig:epeak_beta}, with the locii of the bright GUs highlighted. }
\label{fig:epeak_beta_GUs}
\end{figure}

In this section, we discuss the possible nature of the 175  brightest GUs reported in 4FGL-DR4 (Test Statistic greater than about 100)\footnote{The selection criterion is TS$>$ 100 in 4FGL-DR3.}, which may help to shed light on the whole GU population.  
 A notable fraction of these sources has already been the focus of dedicated investigations. In addition to these works, improved source localizations based on 16 yr of data and the use of more extended counterpart catalogs have enabled solid associations to be found in FL16Y for about {25\% of these bright GUs \citep{FL16Y}. This population is notably less sensitive to the diffuse emission model than the less significant one, as reflected by the different fractions of sources with at least one 4FGL-DR4 analysis flag set (0.43 vs. 0.79, see caption of Table \ref{tab:bright_SGUs_1} for the definitions of these flags).  Given the above and their larger significance, these sources are more likely to be real point sources, with a better localization than for the rest of the GUs.  Table \ref{tab:bright_SGUs_1} provides the 4FGL-DR4 gamma-ray properties of the sample. 
Table \ref{tab:bright_SGUs_2} lists the {\sl eRosita} and {\sl Swift} counterparts to the  sample sources.

Figure \ref{fig:sky_bright_GUs} shows the sky bright-GU loci, populating both the spike and shoulder components.
The Galactic latitude distribution displayed in Figure \ref{fig:lat_bright_GUs} supports the idea
that the bright GUs sample the whole GU population reasonably well (compare to Figure \ref{fig:latitude_1}).

 Figure \ref{fig:epeak_beta_GUs} compares the positions of the bright GUs to that of the whole population in the LP\_beta vs. LP\_EPeak plane. The bright sources span a large range in LP\_EPeak, straddling the 1 GeV mark. The bright GUs overpopulate the region with  LP\_beta$>0.2$ and LP\_EPeak$>$1 GeV, where the pulsars dominantly lie, as compared to the bulk of the GUs. This supports the idea that pulsars (probably mostly young ones
 given the latitude selection) with thus-far undetected pulsations are present in the sample.  On the other hand, sources with low spectral curvature and high LP\_EPeak could be either blazars, supernova remnants or pulsar-wind nebulae.

Table \ref{tab:bright_SGUs_3} lists the features reported by other groups or noted in the present work concerning some bright GUs.   Appendix~\ref{app:MW_GU_2} gives more details about specific sources.\footnote{The data presented in Section 8 and Appendix C is available in Zenodo at \dataset[doi: 10.5281/zenodo.20342245]{https://doi.org/10.5281/zenodo.20342245}. In addition to the machine readable version of tables 5 - 7, there is also an mrt file that aggregates the information from these tables and Appendix C.} 


\textbf{Active Galactic Nuclei Candidates.} 
Five sources are now associated with  AGNs thanks to the improvements described above:
4FGL J0057.9+6326, 4FGL J0725.7$-$0549, 4FGL 1812.8$-$3144, 4FGL J1819.9$-$1530, and 4FGL J2109.6+3954.  Another source, 4FGL J0928.4$-$5256, presents a hard spectrum with low curvature significance, making its nature as blazar of the BL Lac subclass probable. When lying in the hemisphere covered by the eRASS1 catalog, a high-confidence counterpart is found for these sources. Distinguishing blazars of the flat-spectrum radio quasar subclass is more difficult since their spectra are significantly softer than those of BL Lacs and more comparable to the GU ones. Variability is of little help in the present context because it is often not significant for  the low to moderate overall detection significance  considered here. Nevertheless, two sources are found variable in 4FGL-DR4.  Three sources have subthreshold associations with AGNs in 4FGL-DR4, albeit with low probabilities (P$<$0.29).

\textbf{Pulsars.}  Pulsations have recently been detected in 4FGL J1208.0$-$6900, 4FGL J1415.4$-$6458 \citep{Ker25}, 4FGL J1603.3$-$6010, 4FGL J1752.8$-$4449  \citep[TRAPUM collaboration ]{TRA23}, and 4FGL J2051.7+5051 \citep{Den24},  which are all MSPs, and in 4FGL J1953.5+3841, 4FGL J2116.2+3701 \citep{Don23}, 4FGL J0235.3+5650, 4FGL J0736.9$-$3231, 4FGL J1616.6$-$5341 \citep{Ein10}, five PSRs.

The 3PC catalog lists 49 ATNF pulsars colocated with LAT sources (Table 5). Two sources show pulsar-like SEDs, making the association plausible. Despite the colocation, two other associations  are considered unlikely given the low  pulsar $\dot E$. 

Five sources were considered plausible pulsar candidates by \cite{Bru23}. This selection was  based on the presence of at least one radio counterpart in their error box with L-band spectral index and flux compatible with those expected for pulsars. 

Eight other sources were searched for pulsation (prior to 2018)  by the Fermi Pulsar Consortium at positions within their 95\% error ellipse, with no positive results thus far. 4FGL J1808.5$-$3701 is possibly associated with an accreting millisecond pulsar \citep{deO16}.

\textbf{Pulsar Wind Nebulae.}
Two sources, 4FGL J1427.8$-$6051 and 4FGL J1929.0+1729, have TeV counterparts and have been proposed as possible PWNe \citep{Dev21}. 

\textbf{Supernova Remnants - Molecular cloud Candidates.} Eight sources lie close to known (sometimes recently discovered) shell supernova remnants and three other to nearby molecular clouds. Seven have extended TeV counterparts. Nine sources  present a significant spectral break, possibly manifesting a ``pion bump" \citep{Abd22}.

\textbf{Star-forming regions - young clusters candidates\label{sec:bright_SFRs}.}
Out of the 71 sources associated with a WISE HII region with a radius $<$0.3\arcdeg (with an estimated false-positive rate of about 50\%, see Section \ref{sec:SFRs} ), 32 belong to the bright sample. 
Five sources belong to the Vela Molecular Ridge and have been associated with SFRs in different works. Two others are part of the Carina Nebula Complex, one of the most active and nearest massive star-forming regions in our Galaxy. One (4FGL J2038.4+4212)  lies in a complex region at the edge of the Cygnus Cocoon.

 We note that several gamma-ray sources having fairly robust associations with  SFRs show spectral properties typical of the SGU population, with LP\_EPeak $< 700$ MeV. 

\textbf{Globular clusters.}
The source 4FGL J1836.8$-$2354  lies within the extent of the globular cluster M 22. 

\textbf{Sources coincident with structures in the diffuse emission.} We visually inspected the regions surrounding the sources in the bright sample via the maps available from radio to X-ray surveys or archival X-ray observations. The list is given in the  Appendix~\ref{app:MW_GU}.
The map inspection revealed that eleven sources are sitting on top or near extended IR or radio structures, although any firm associations cannot be claimed.

\textbf{``Orphan" sources.} Sources without any prominent counterparts at other wavelengths, dubbed orphans in the following,  are of special interest.  
In the association procedure, the individual probabilities of false positives were quantitatively assessed when considering the radio (SUMSS and NVSS) and X-ray (ROSAT) surveys via the likelihood-ratio method,  when putting together the 4FGL catalog and subsequent releases. So a spatial coincidence between a gamma-ray source and a source from these surveys  does not warrant excluding the former from the list of orphan sources considered here. 

We define two sets of orphan sources, one in which no bright radio, optical or X-ray counterpart is present in the error ellipse (orphan\_A\footnote{These sources are indicated  by a $^\bigstar$ symbol after theirs names in Table \ref{tab:bright_SGUs_1}}, 35 sources) and the other one where at least one counterpart exists in these bands (for example as listed in Table \ref{tab:bright_SGUs_2}) but there is no indication that it belongs to a known class of gamma-ray emitters (orphan\_B\footnote{These sources are indicated  by a $^\lozenge$ symbol after their names in Table \ref{tab:bright_SGUs_1}}, 48 sources).  Together, these two sets represent 47\%  of the bright GUs.  These sources have similar significance and LP\_EPeak medians as the whole sample but the spike component in the Galactic-latitude distribution  is strongly suppressed (Figure \ref{fig:lat_bright_GUs}). This may simply be due, at least in part, to the larger density of counterparts in the plane enhancing false positives.   We encourage further investigation of this sample, which may reveal novel classes of gamma-ray sources.

\section{Discussion and summary}
\label{sec:conclusions}
The GUs represent a large fraction (54\%) of the sources detected within 10$^\circ$ of the Galactic plane. Their number rises sharply below TS=100, explaining why the peculiarities of these sources have only been recognized fairly recently. The majority of the GUs exhibit three main properties not commonly found in  known classes of gamma-ray emitters: i)  a soft spectrum (hence their denomination as SGUs),  with a pronounced curvature leading to a SED peaking below 1 GeV (ii)  two distinct Galactic-latitude components with different extensions, the spike and the shoulder  iii) some evident source clusters, 
sometimes co-located with prominent regions of the Galaxy (e.g., the Vela Molecular Ridge).  
Altogether, on the basis of these properties, young and millisecond pulsars, by far the largest Galactic classes  detected  by the LAT,  cannot explain the bulk of the GU population. The presence of SGU sources high above and below the Galactic plane suggests that the SGU sources in the shoulder population have a local origin, while sources in the spike are distributed in the disk of the Milky Way Galaxy at large distances.

Several source clusters have been highlighted, both in the spike and the shoulder. Their LP\_EPeak distributions all peak below 1 GeV, but the strength of their high-energy wings, extending above 1 GeV and possibly populated by pulsars, varies significantly across the sample.  This may illustrate a variety of situations giving rise to these clusters. 

In a machine learning approach, we modeled the differences between LP\_beta and LP\_EPeak distributions for GUs and for the sources with the known classes discussed in Section~\ref{sec:GU_prop} by introducing a new class of sources with a Gaussian distribution in parameter space.  
Most of the distribution for this new class (right panel of Figure~\ref{fig:prior_shift})  has 0.2 GeV $<$ LP\_EPeak $<1$ GeV, which is smaller than LP\_EPeak for pulsars (generally $> 1$ GeV) but larger than for FSRQs (generally $\lesssim$ 100 MeV). This subpopulation, called SGU-like as it presents a large overlap with SGUs determined by selecting sources with power-law index above 2.4, displays a 
spatial distribution (Figure~\ref{fig:cov_prior_glat}) 
resembling the general distribution of GU sources (Figure~\ref{fig:latitude_1}), i.e., it consists of a thin population close to the Galactic plane (the spike) and a population that extends up to $10^\circ$ and more away from the Galactic plane (the shoulder).   This approach enables us to assess the SGU tally to about 630 sources.


Some general remarks about the number of as-yet undiscovered pulsars among the GUs are in order.  The number of known gamma-ray pulsars accumulated since Fermi's launch increases with a rate near 13 yr$^{-1}$ over the last decade \citep[3PC, Figure 1, ][]{3PC}. Over half are radio MSPs far from the plane, but the rest used to be GUs. These GUs were ``orphans'' until pulses were found. Most are not among the $>$few mJy L-band radio sources collated by \cite{Bru23}, having much lower radio flux densities (see 3PC Figure 6). Neutron stars are barely visible at optical wavelengths (mag $> 26$), and only a subset of gamma-ray pulsars are seen in X-rays. Hundreds of LAT sources have been searched for radio and/or gamma-ray pulsations \citep[see][and references therein]{TRA23,EatHome2025}, but failed searches do not mean that there is no pulsar -- glitches, binary acceleration, scintillation, radio-frequency light pollution, as well as faint signal strength all conspire to foil detection until more sensitive searches, or comparable searches at more favorable epochs, occur. New instruments \citep[such as FAST, see][]{FAST_J0447} and methods \citep{PulsarTOAlattice,GibbsSampling} continually yield fresh pulsar harvests.
Population syntheses typically predict that about 200 more gamma-ray pulsars over the whole sky await pulsation detection \citep{gammaPopSynth,gammaMSPpopSynth}. Table \ref{tab:cov_vs_prior_lowb} in Appendix A predicts 21 young and 126 millisecond pulsars among the GUs. In conclusion, pulsars do not explain the bulk of the GUs, but tens of GUs will likely be identified as pulsars in the coming years.

Possible classes not fully considered so far when establishing the LAT general source catalogs have been explored.   Three extended star-forming regions (Westerlund 2, rho Ophiuchi, Cygnus X) are included in the 4FGL-DR4 catalog, along with three candidates  Sh 2-148,  Sh 2-152, and  NGC 346 (this one in the SMC). There is growing evidence that more LAT sources  could be associated with star-forming regions \citep[][]{Tib21,Per24}.  We have confirmed the results of \cite{Per24} indicating that several tens of LAT sources are likely to be associated with objects in  the WISE catalog of HII regions \citep{WISE_HIIcatalog}. However, the rate of false positives remains high due to the large spatial densities of WISE sources, impeding the identification of high-confidence counterparts. The SFR candidates found in our analysis belong dominantly to the spike component. We conclude that though SFRs represent a so-far somewhat overlooked  (at least in the FGL catalogs) real class of gamma-ray emitters and may underlie some notable SGU clusters  (co-located with e.g., the Vela Molecular Ridge or the Gemini OB1 complex), they probably constitute a small fraction ($<10\%$) of the unassociated population.  Although some individual associations have been claimed by different groups,  none of the alternative classes we have explored (including Wolf-Rayet, OB, Be,  T Tauri stars, bright nearby stars, Herbig-Haro objects, AMXPs, and CWBs) have been found to make up a sizeable part of the GUs.  

We have taken advantage of the recent release of the {\sl eRosita }eRASS1 catalog to search for soft X-ray counterparts to GUs, using a likelihood-ratio approach.  This search has turned out mostly unsuccessful, only 30  high-confidence ($P>0.8$) associations being found. Out of the 30, 28 belong to the shoulder, a result which can be partly due to the high-detection limit of {\sl eRosita} along the Galactic plane, and only 7 qualify as SGUs.  

The possibility that SGUs are merely residuals related to mismodeled diffuse emission has been explored in detail. Leaving more degrees of freedom to the IEM components in the fitting procedure reduces the number of SGUs above the detection threshold by 20-30\%, but does not improve the fit quality substantially. This analysis does not rule out the possible SGU origin as unaccounted-for clumps of diffuse emission. We have explored this scenario further by looking for the expected spectral connection between the SGUs and the diffuse emission. To this end, we have performed Monte Carlo simulations under different conditions.  Doubling the DNMp contribution relative to the standard model or lowering the DNMp clump spatial scales  have led to only small improvements in the reproduction of the peak energies.  


Taking another approach, sources have been simulated with the same spectrum as the IEM and varying spatial extensions and have been again analyzed assuming point sources.  Under ad hoc extension of $\sigma=0.1^\circ$  and realistic (median TS and positions similar to those of the SGUs) conditions, the SGU spectral properties (curvature and peak energies) can be recovered reasonably well.  This indication of extension on a small spatial scale, which naturally fits in the {mismodeled diffuse emission} scenario,  has triggered a search for confirmation in the data. Only the brightest sources lend themselves to such a search. Of the 60 sources considered, significant extension was found for 20 of them, 16 belonging to the spike. This observation confirms the viability of the clump scenario, at least for the spike component.   We note that the prevalently soft spectrum for the missing clumps may be due to an analysis bias since systematic searches for extended emission in LAT data have focused on the energy range $>10$~GeV \citep[][]{2FGES}.      

Historically, some gas missed by the usual tracers, now known as ``dark gas",  was put in evidence by the gamma-rays in the EGRET data. The SGU population may be indicative of a similar situation of ``missing gas" in our IEM. We have explored the possibility that some unaccounted-for gas could be due to the saturation of the $^{12}$CO line. To this end, we have examined the $^{13}$CO line, less prone to a saturation effect, using the MOPRA data. The results show no evidence that SGUs are preferentially detected at locations where saturation effects are most probable, as determined from the comparison between the intensities of the $^{12}$CO and $^{13}$CO lines. 

An alternative hypothesis yet to be explored is that clumps of mismodeled diffuse emission trace regions of enhanced cosmic-ray densities during the first steps of their propagation around injection sites. The ansatz used in the construction of LAT catalogs is that sources coincide with active sites of particle acceleration and the diffuse background model accounts for the interactions of the large-scale population of cosmic rays with the interstellar medium. However, the process by which locally-injected particles merge into the large-scale cosmic-ray population is poorly understood. Recent observations point to the existence of extended halos or bubbles of gamma rays tracing localized cosmic-ray overdensities surrounding pulsars, supernova remnants and star-forming regions \citep[e.g.,][and references therein]{Tib21}.
  
Considering that highly significant sources are  more likely to pinpoint active particle accelerators, we have searched for counterparts at other wavelengths to the brightest 175 GUs, looking for trends that could shed light on the nature of the whole population. Some of these sources have been the subject of detailed MW investigations published by different groups, pointing out plausible associations with sources of various classes. About 32 sources show such associations or lie in regions where extended gamma-ray emission has already been reported.  Five more display pulsar-like spectra. Overall, no clue on the existence of a new significant class emerges from this analysis.  A comprehensive set of MW maps have been  inspected for conspicuous radio to X-ray counterparts  within the GU error boxes or small-scale underlying structures. A total of 83 sources, mostly in the shoulder, are found to be ``orphan", i.e., show no plausible candidates of gamma-ray emitters in their error boxes. 

Overall, no decisive evidence has been found that excludes the possibility that SGUs are real point sources or, conversely, residuals associated with mismodeled diffuse emission.  
However, different situations appear to prevail for the spike and shoulder components. Indications have been found in favor of the  ``diffuse clump" scenario for the spike component, in particular the significant spatial extension for some of the brightest sources. 
Moreover, tens of spike GUs can plausibly be associated with WISE-detected  star-forming regions and account for a small but significant part of this population.

The origin of the shoulder component, where similar indications are lacking,  remains more elusive.  The Galactic-longitude distributions are markedly different for the spike and shoulder, which is another feature supporting separate origins,  despite the similarity in spectral properties.  
If the shoulder is not related to diffuse emission, then the apparent correlation between the Galactic-longitude distributions of the CO  or patch components and that of the GUs (dominated by the shoulder) seen in Figure \ref{fig:longitude_patch_CO_npred} would reflect a more complex scenario. Among the unexpected features revealed by the shoulder Galactic-longitude distribution is the strong enhancement in the direction of the Galactic Center (Figure \ref{fig:longitude_1}),  corresponding to region 4 in Figure \ref{fig:map_ROIs}.  If the photons attributed to the shoulder originate in the local ring, then there are no reasons why the particular direction of the Galactic Center should be favored.  If, on the other hand, these photons come from the innermost rings, as their directions would support, then their latitude spread is surprising. This spread (10$^\circ$) roughly corresponds to the radius of the Galactic bulge \citep[1.5 kpc, ][]{Weg13} seen from Earth. What source/structure would emit these photons in the bulge remains an open question, as well as why similar, although less pronounced, features appear at other places along the Galactic plane.   

The link to the GeV excess \citep[e.g.,][]{2011PhLB..697..412H, Cal14, Ack17} is worth considering. Besides the possibility of a dark-matter origin, its spectrum favors a mostly unresolved population of  millisecond pulsars as its origin, but few GUs match the properties of this source class (see the LP\_EPeak distribution in Figure \ref{fig:Epeak}). Moreover, if these sources were located at the Galactic-Center distance, their luminosities, derived from their energy-flux distributions, would typically exceed those of the 3PC MSPs by one order of magnitude. 

The possible connection with the Fermi bubbles  also deserves attention. Two features disfavor a common origin with the SGUs. The Fermi bubble spectrum peaks above 10 GeV, so the connection to the very soft sources that make up the bulk of the GUs seems unrealistic. Second, the Galactic-longitude profile of the base of the bubbles around the Galactic Center \citep{Her19} is very different from that observed for the GUs. 

The connection of the shoulder to other components like the base of the North Polar Spur/Loop I (possibly related to the {\sl eRosita} bubbles seen in the X-rays) whose contributions to the IEM are encapsulated in the patch seems more compelling. We recall that, by design, these components cannot absorb spatial structure on scales less than $\sim 4^\circ$, leaving room for the emergence of point-like sources spawn from smaller structures.  However, given our poor knowledge about the exact nature of objects like the North Polar Spur \citep[see for example ][]{Lal22}, pinning down the above connection may remain elusive for the years to come.       

The development of an improved diffuse emission model, which is deemed a prerequisite for producing the next Fermi-LAT catalog (5FGL), is currently underway. It will hopefully help shedding light on the link between SGUs and the diffuse emission. If a direct link can eventually be ruled out, a new class of gamma-ray emitters is the most likely alternative scenario. To identify this class, we encourage multiwavelength follow-up analysis of the brightest sources, especially those for which no viable counterparts have been found so far (the ``orphan" sample). A gamma-ray telescope with improved PSF in the MeV-GeV energy range would also be extremely valuable towards clarifying the nature of the SGUs. 
In conclusion, a minority of GUs exhibit properties compatible with established classes of gamma-ray emitters. This fraction is estimated to about 30\% from the machine-learning analysis. Pulsars could represent about 15\% of the GUs.  Looking for new classes beyond those considered so far in the FGL catalogs that could underlie the SGUs,  a  maximum of 10\% of them can be attributed to SFRs, almost exclusively populating the spike component. Only a few tens of sources potentially belong to the other "minor" classes considered in this work. Even for the brightest sample of 175 sources, 47\% of them are found to be "orphan" of some kind. Some ad hoc conditions have been found to successfully reproduce the SGU spectral properties assuming small extended clumps of diffuse emission as their origin. Varying the clump spatial sizes can potentially account for the properties of the whole SGU population.  


\begin{acknowledgments}
The \textit{Fermi} LAT Collaboration acknowledges generous ongoing support
from a number of agencies and institutes that have supported both the
development and the operation of the LAT as well as scientific data analysis.
These include the National Aeronautics and Space Administration and the
Department of Energy in the United States, the Commissariat \`a l'Energie Atomique
and the Centre National de la Recherche Scientifique / Institut National de Physique
Nucl\'eaire et de Physique des Particules in France, the Agenzia Spaziale Italiana
and the Istituto Nazionale di Fisica Nucleare in Italy, the Ministry of Education,
Culture, Sports, Science and Technology (MEXT), High Energy Accelerator Research
Organization (KEK) and Japan Aerospace Exploration Agency (JAXA) in Japan, and
the K.~A.~Wallenberg Foundation, the Swedish Research Council and the
Swedish National Space Board in Sweden.
Work at NRL is supported by NASA.
 
Additional support for science analysis during the operations phase is gratefully
acknowledged from the Istituto Nazionale di Astrofisica in Italy and the Centre
National d'\'Etudes Spatiales in France. This work performed in part under DOE Contract DE- AC02-76SF00515.

\end{acknowledgments}

\software{fermipy\footnote{\url{https://fermipy.readthedocs.io/en/master}}   \citep{Wood17}, HEALPix\footnote{\url{http://healpix.jpl.nasa.gov/}} \citep{2005ApJ...622..759G}, 
Astropy\footnote{\url{https://www.astropy.org/}} \citep{Ast22}, Matplotlib\footnote{\url{https://matplotlib.org/}} \citep{Hunter:2007}, pandas\footnote{\url{https://pandas.pydata.org/}} \citep{mckinney-proc-scipy-2010}, 
scikit-learn\footnote{\url{https://scikit-learn.org/stable/}} \citep{scikit-learn}}.

\newpage
\appendix 

\section{Machine learning classification of gamma-ray sources with covariate and prior shift models}
\label{app:cov_prior_models}

Classification of objects with ML is based on the assumption that the joint distributions of input features $x$ and output features, i.e., classes $k$, are the same for the training and target datasets:
\be
p_{\rm train} (x, k) = p_{\rm target} (x, k).
\ee
In presence of a dataset shift the training and target distributions are different
$p_{\rm train} (x, k) \neq p_{\rm target} (x, k)$.
In particular, the distributions of associated \Fermi-LAT sources (training dataset) and unassociated sources (target dataset) are different, \citep[e.g.][or Figure \ref{fig:unassoc_TS}]{LAT22_4FGL_DR3, 2023RASTI...2..735M}.
The joint distribution can be written as a product of conditional probability times a prior distribution in two different ways:
\be
p(x, k) = p(k|x) p(x) = p(x|k) p(k).
\ee
Correspondingly, there are two special cases of the dataset shift \citep{MorenoTorres2012AUV}:
\vspace{-1mm}
\ben
\item
Covariate shift: $p_{\rm train}(k|x) = p_{\rm target}(k|x)$, but $p_{\rm train}(x) \neq p_{\rm target}(x)$;
\vspace{-1mm}
\item
Prior shift: $p_{\rm train}(x|k) = p_{\rm target}(x|k)$, but $p_{\rm train}(k) \neq p_{\rm target}(k)$.
\een
\vspace{-1mm}

Typical ML classification algorithms with supervised learning, such as decision trees (boosted decision trees, random forest), logistic regression, multi-layer perceptrons, support vector machines, give the conditional probability $p_{\rm train}(k|x)$ as a result of the training.
If this conditional probability is then used, e.g., for the classification of unassociated sources, then differences in the distributions of associated and unassociated sources are attributed solely to overall distribution of sources as a function of input features $p_{\rm train}(x) \neq p_{\rm target}(x)$.
Thus, the covariate shift assumption is implicitly used in this case~\citep{2023RASTI...2..735M}.
Classification of 4FGL-DR4 sources with the covariate shift assumption using the four classes specified above with seven input features, log10(Energy\_Flux100), log10(Unc\_Energy\_Flux100), log10(Signif\_Avg), LP\_index1GeV, LP\_beta, LP\_SigCurv, log10(Variability\_Index), where LP\_index1GeV is the index of the LogParabola spectrum fit at 1 GeV, has been performed in~\cite{Mal24CovPrior, MalyshevGCEcov}. 

In case of the prior probability shift, the distributions of sources in each class as a function of the input features are assumed to be the same for the associated and unassociated sources, $p_{\rm unas}(x|k) = p_{\rm assoc}(x|k)$, while the overall fractions of associated and unassociated sources can be different 
$p_{\rm unas}(k) \neq p_{\rm assoc}(k)$. 
In general, also the distributions of associated and unassociated sources in the different classes can be different due to an association bias: it is easier to find associations for bright sources than for faint ones, i.e., the fraction of associated and unassociated sources may depend on the flux.
A prior probability shift model 
with a covariate shift that modulates the distributions of all classes in the same way for high latitude sources has been constructed by~\cite{2025arXiv250314584A}.
A more general model that takes into account the possible dependence of the fractions of associated and unassociated sources as a function of flux for different classes has been presented in~\cite{Mal24CovPrior}. In this model the constant priors are replaced by priors that depend on flux $p_{\rm unas}(k, F)$. These functions have been modeled as a sigmoid plus a constant as a function of the logarithm of energy flux above 100 MeV.
It has been observed in~\cite{Mal24CovPrior} that even this more general model cannot account for all unassociated sources, i.e., there is an excess of sources with high spectral curvature and low LP\_EPeak values.
This excess has been modeled as a Gaussian distribution in the input features, log10(Energy\_Flux100), LP\_beta, log10(LP\_EPeak).

The model for the distribution of unassociated sources is
\be
\label{eq:prior_model}
p_{\rm unas} (x) = \sum_k p_{\rm assoc} (x|k) \pi_k(x) + G(x),
\ee
where $x$ are the three input features, $k$ labels the four classes of associated sources (fsrq+, bll+, psr+, and msp+ specified above), $p_{\rm assoc} (x|k)$ are the PDFs of the four classes determined with Gaussian mixture models (GMMs).
We split each of the four classes of associated sources into 50\% training and 50\% validation datasets.
The GMMs are obtained with the training datasets. 
In order to avoid overfitting, we determine the number of Gaussians in the mixture by applying the model to the validation dataset and by taking the model with the minimal Bayesian information criterion. This results in one Gaussian kernel for the msp+ class and in two kernels for the other three classes. Note that we use the GMM here to determine the overall probability density function of each of the classes, i.e., we do not use the individual components of the GMMs.
The scaling functions $\pi_k(x)$ in the standard prior shift model are constants $\pi_k$, which encode class prevalence, i.e., the fraction of class members relative to the total number of samples.
In the standard prior shift model the coefficients $\pi_k$ are adjusted to obtain the best-fit of the model to the target data.
Nevertheless, this adjustment of $\pi_k$ is not sufficient to obtain a good fit of the unassociated {\Fermi}-LAT sources. As a result, two further improvements of the model were introduced in~\cite{Mal24CovPrior}: a new class, modeled as a Gaussian distribution in the input features $G(x)$, and 
flux-dependent modulations
\be
\label{eq:pshift}
\pi_k(x) = \frac{a}{1 + e^{(x_1 - b) / c}} + d,
\ee
where $x_1 =$ Log10(Energy\_Flux100). 
The motivation for the introduction of the flux-dependent modulations is that the relative fraction of unassociated sources is higher at low fluxes (most of high-flux sources are associated). As a result, in a model determined from PDFs of associated sources, we need to decrease the relative contribution from high-flux sources, which is achieved with the flux-dependent modulation of the corresponding PDFs.
Parameters $(a,\ b,\ c,\ d)$ are obtained from the fit to the data.
The function $G(x)$ is a product of three Gaussians in the three input variables, i.e., it has 7 free parameters.
The parameters are determined by maximizing the log likelihood of the model in Eq.~(\ref{eq:prior_model})
given the distribution of unassociated sources
\be
\log L = 
\sum_{i \in {\rm unas}} \log(p_{\rm unas} (x_i))
- N_{\rm unas} \int p_{\rm unas} (x) dx.
\ee
We note that the model performance characteristics, such as accuracy, precision, and recall, used in supervised learning are not suitable for estimation of model performance in presence of dataset shift.
For instance, if class prevalence is different in the training and the validation datasets, then the predictions based on the prevalence determined for the training dataset will give biased estimates for the validation dataset and vice versa, prevalence optimized for the validation dataset, would result in worse performance if applied to the training dataset.
The main measure of performance, in this case, is the likelihood itself, i.e., how well the model actually describes the data~\citep[cf. Figure 2 of][]{Mal24CovPrior}.


The presence of a possible new component of sources among the unassociated sources can be also observed in the covariate shift classification of sources presented in~\cite{MalyshevGCEcov}.
In the covariate shift model all unassociated sources are classified into the known classes used for training of the ML algorithm.
If there exists a new class, then in the covariate shift model
the existing classes are scaled proportionally in the domain of the new class to account for the sources in this class.
In other words, expectations for the number of sources in all classes inside the domain of the new class should be higher than the relative expectations outside of the domain.
We illustrate this in Figure~\ref{fig:cov_excess}.
The only assumption necessary to estimate the possible excess
is the domain in the input parameters, where we can expect the presence of a new component.
In this work, we choose the ranges of LP\_beta and log10(LP\_EPeak) parameters that approximately cover the Gaussian distribution used to model the SGU-like sources, i.e., from 0.1 to 0.5 in LP\_beta and from 150 MeV to 1.5 GeV in LP\_EPeak.
These ranges are shown as grey dotted vertical lines in Figure~\ref{fig:cov_excess}.
The Gaussian model for the SGU-like sources is shown by the dash-dotted purple line.
The differences between the covariate shift model distribution of expected contributions of different classes and the scaled distributions of associated sources are shown by 
the blue dashed, orange dotted, green sparse dashed, and red sparse dash-dotted lines for the fsrq+, bll+, psr+, and msp+ classes respectively.
The scalings are obtained from the fit of the distributions of associated sources to the expected contributions to the unassociated ones in the areas outside of the dotted grey vertical lines.
We note that for all classes the corresponding differences are generally positive inside the grey lines, with the exception of LP\_EPeak distribution for msp+ and psr+ classes above about 1 GeV (right panel in Figure~\ref{fig:cov_excess}).
The sum of the differences, shown by the solid grey lines, generally agrees with the expectation for the SGU-like sources modeled by the Gaussian distribution.
This is the expected behavior of classification in the covariate shift model, if there existed a new component of sources inside the vertical grey lines.

\begin{figure}
\centering
\includegraphics[width=0.47\textwidth]{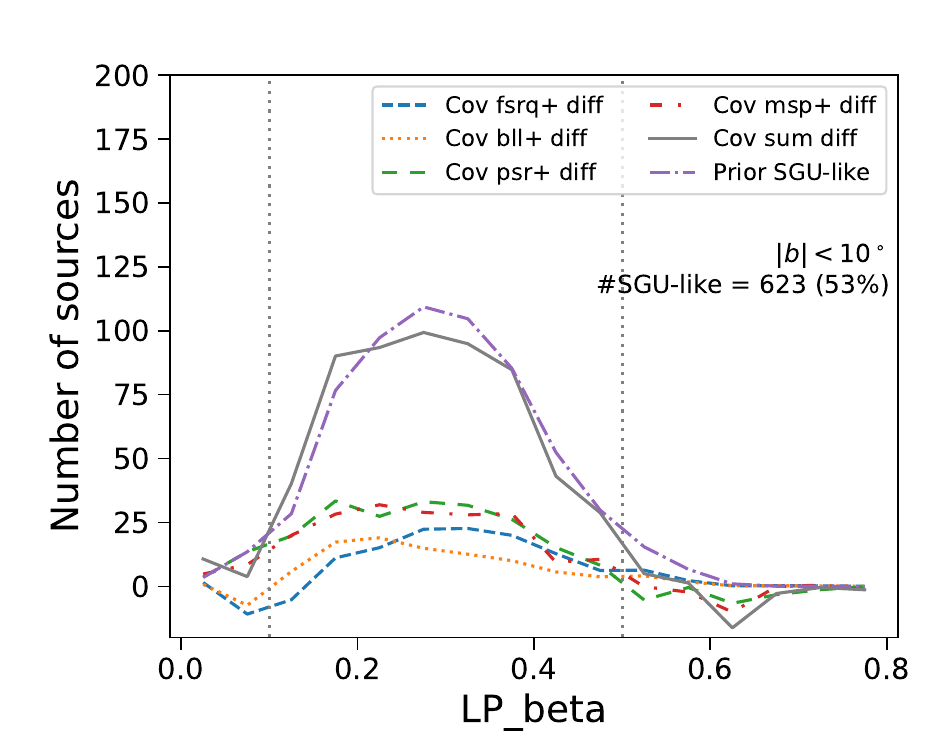}
\includegraphics[width=0.47\textwidth]{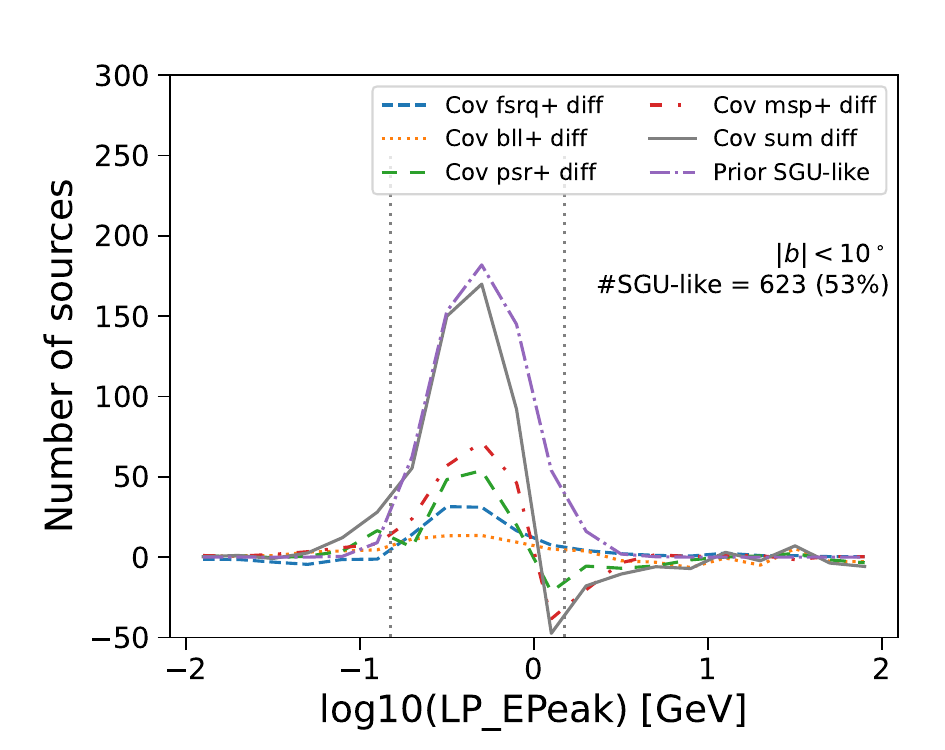}
\caption{Covariate shift model excesses vs SGU-like component of the prior shift model for sources within $|b| < 10^\circ$.
See text for the details on the derivation of the lines.
}
\label{fig:cov_excess}
\end{figure}

We compare the predictions for the numbers of high-probability class candidates ($p > 0.5$) in the prior and covariate shift models for the low-latitude ($|b| < 10^\circ$) sources in Table~\ref{tab:cov_vs_prior_lowb}.
The sources with all class probabilities below 0.5 fall into the uncertain category.
There are many more uncertain sources in the covariate shift model at low latitudes relative to the prior shift model (510 vs 87).
Most of the uncertain sources in the covariate shift model correspond to the Gaussian SGU-like component in the prior shift model:
out of 510 uncertain classifications in the covariate shift model, 338 sources are classified to be SGU-like in the prior shift model (``uncert'' row in Table~\ref{tab:cov_vs_prior_lowb}).
There is also a significant contribution to the SGU-like component from psr+ and msp+ classes in the covariate shift model (cf. ``psr+'' and ``msp+'' rows in Table~\ref{tab:cov_vs_prior_lowb}).
Generally, the prior shift model predicts very few psr+ candidates.
Most of the psr+ candidates in the covariate shift model are attributed to the SGU-like sources in the prior shift case.
There are also 37 psr+ prior shift candidates classified as fsrq+ sources in the covariate shift model.
Since the prior shift model uses only spectral parameters of the sources, there is a large overlap between pulsars and SGU-like sources and FSRQs with curved spectra, which may lead to misclassification of some pulsars as SGU-like of fsrq+ sources in the prior shift case.
We also note that the number of sources classified as SGU-like in Table~\ref{tab:cov_vs_prior_lowb} in the prior shift model above the probability threshold of 0.5 (710 sources) is different from the expected number of SGU-like sources (623 sources), which is derived by summing the SGU-like probabilities for all unassociated and unk sources. 

\begin{table*}
\centering
\footnotesize
\begin{tabular}{llllllll}
\hline
Covar$\backslash$Prior & fsrq+ & bll+ & psr+ & msp+ & SGU-like & uncert & total \\
\hline
fsrq+ & 9 & 0 & 0 & 0 & 21 & 2 & 32 \\
bll+ & 5 & 75 & 0 & 0 & 0 & 13 & 93 \\
psr+ & 37 & 5 & 19 & 9 & 175 & 21 & 266 \\
msp+ & 0 & 0 & 2 & 80 & 176 & 16 & 274 \\
uncert & 79 & 21 & 0 & 37 & 338 & 35 & 510 \\
total & 130 & 101 & 21 & 126 & 710 & 87 & 1175 \\
\hline
\end{tabular}
\caption{Numbers of unassociated sources with $|b| < 10^\circ$ classified by the covariate shift model (rows) and by the prior shift model (columns) with class probabilities $> 0.5$. If all class probabilities are below 0.5, then the class of the source is considered uncertain (``uncert'' column and row).}
\label{tab:cov_vs_prior_lowb}
\end{table*}

\section{Multi-wavelength data used for bright-GU inspection}
\label{app:MW_GU}

 We inspected the maps provided by the following surveys or facilities (accessed with SkyView\footnote{http://skyview.gsfc.nasa.gov}) on a source-by-source basis  to look for counterparts in the 95\% error ellipses and underlying diffuse structures:  
\begin{itemize}
\item the HI4PI full sky survey of the 1.4 GHz (21 cm) HI line, with all velocity being integrated and used as a proxy of all atomic gas along a given line of sight;
\item the Dame et al. survey of the 2.6 mm line of the 1-0 rotational transition of 12CO, used as a proxy for molecular gas;
\item the radio and  IR surveys: SUMSS \citep{SUMSS}, VLSS \citep{VLSS} , PMN \citep{PMN}, MIPS24, GLIMPSE4, WISE \footnote{https://irsa.ipac.caltech.edu/data/WISE/docs/release/All-Sky/index.html}, iris100 \footnote{https:/irsa.ipac.caltech.edu/data/IRIS/};
\item the RASS ROSAT survey and dedicated pointings in the 0.2$-$2.5 keV band;
\item mosaics of pointed observations from XMM, Chandra, and Swift in the 0.5$-$10 keV band;
\item the HESS Galactic plane survey (HGPS) at energies starting above $\simeq$250 GeV up to a few tens of TeV.
\end{itemize}

We also made use of the Swift-XRT web tool dedicated to the MW Fermi unassociated sources \citep{Str13}. 

\section{Details on individual bright GUs}
\label{app:MW_GU_2} 
				
This Appendix examines high-significance unassociated sources near the Galactic plane ($|b|<$10 deg.) and pursues identifications for that subset.  Setting a threshold of TS $>$ 100 selects 175 GU sources, of which 113 are SGUs and 62 are non-SGUs.  The working hypothesis is that this threshold gives sources so high in gamma-ray flux, and also in statistical significance in fitting, that they cannot be mismodeled diffuse.  To be very clear, this hypothesis is not proven.  There might be ways to attempt such proof by studying fluctuations and gradients of the diffuse near GU sources, but that is out of scope for this paper.  There is, in fact, no known boundary in any parameter space such that sources to one side of the boundary are immune to the possibility of their being mismodeled diffuse.  
To make the working hypothesis quantitative, if we assume that TS $>$ 100 sources, being approximately 8 $\sigma$, have probability $\ll$0.0001 of being spurious then a list of 175 such sources has an expectation $\ll$0.018 of containing even one spurious source.   We then pursue identification possibilities for that list.  Furthermore, the goal is not to establish these identifications by some procedure but merely to establish candidacies for further work.  It is permitted to entertain more than one candidate per source, although it will be seen that such an outcome is comparatively rare.  
The more problematic outcome occurs when no respectable candidate is found.   It leads toward these possible conclusions, (1) that TS $>$ 100 does not exclude mismodeled diffuse; (2) that there is a gamma-ray class so distinct in its properties as to elude identification of candidates (such as compact zones of dark matter annihilation that produce gammas and nothing else); or (3) that more work needs to be done, such as deeper searching in other wavelengths.  Sometimes the LAT R95 for the source is filled only with undistinguished stellar sources, with the exact density of such stars varying from case to case.  Cases of this kind are called “orphans.”
Results of this TS$>$100 exercise are tabulated in Tables 5, 6, and 7.  This appendix gives further detail on  particular cases.  
Two special cases of counterpart candidacy need mention.  The first is the case where, rather than the counterpart candidate being inside r95, it is the other way around: r95 is contained within some recognized object of large angular extent.  The second is the case where the candidate counterpart is such as to preclude any hope of detecting any more distant candidate.  It is possible for a counterpart to meet both of these conditions, as, for example, when the 95\% confidence error circle falls entirely inside portions of a globular cluster.  
Many of the candidate counterparts are “active” in the sense that one would anticipate an acceleration site inside the counterpart, but some seem to be plausibly “passive,” that is, concentrations of target material that might become sites of enhanced gamma ray emission if suitably bombarded with cosmic rays.  A good example of this is 4FGL J1906.9+0712, which has 4 molecular clouds and 7 sub-millimeter sources all in a small region within R95.   For this purpose, star formation regions (SFR) are considered “active” but they are usually cases where the LAT error region is inside the SFR and not the other way around. 

The following conclusions can be drawn: 
(1) Many of these TS $>$ 100 GUs have promising leads for follow up if there are resources to pursue that follow up.  In some cases what is missing is only the investment in theory and detailed analysis to model data already in the archives.  That investment is out of scope for this paper.
(2) The remainder, which we designate as “orphans” should not automatically be envisioned as blank sky regions where nothing of interest is known.  Often, they are in dense star fields with hundreds of catalogued stars within r95, creating a complex foreground that makes it harder to pursue identification of objects relevant to high-energy astrophysics.  It is premature to conclude that the orphans present a case for a new class of sources seen only in GeV photons.  Sometimes surveys deliberately avoid these regions.   The entries below indicate when this situation of dense stellar foreground is encountered.  Often the entry gives the source Galactic latitude.  Conversely, if a nearly blank r95, without a high stellar foreground, is encountered, this is also noted, but it is a rare situation. 

\noindent{\bf 4FGL J0340.4+5302} One of five bright Orphan B sources with TS $>$ 500 and $|$b$| > $1$\arcdeg$.  These are good candidates for pulsar searches. \\
{\bf 4FGL J0426.5+5434} One of four bright Orphan A sources with TS $>$500 and $|$b$|>$1$\arcdeg$.  These are good candidates for pulsar searches. \\
 {\bf 4FGL J0722.4$-$2650} lies 150" away from a galaxy, LEDA 77115. This galaxy seems unrelated to the source eRosita counterpart. \\
 {\bf 4FGL J0744.9$-$4028} has a low-probability (P=0.13) association with an AGN. A 8th magnitude star HD 63006  is offset from the LAT position by only 37$\arcsec$. \\ 
{\bf 4FGL J0758.8-1450} One of five bright Orphan A sources with TS $>$ 500 and $|$b$| > $1$\arcdeg$.  These are good candidates for pulsar searches. \\
{\bf 4FGL J0826.1$-$5053}  A dark nebula G267.29-7.32 is near this source. 
The offset is 4$\arcmin$, comparable to the size of the gamma-ray source. This is a case of a passive source candidate. \\ 
 {\bf 4FGL J0848.8$-$4328} lies in the Vela Molecular Ridge, with many young stellar objects. \\
 {\bf 4FGL J0853.6$-$4306} is colocated with a huge number of young stellar, with molecular clouds and sub-millimeter sources. \\ 
 {\bf 4FGL J0900.5$-$4434c}  This source is colocated with three submillimeter and two dense-core sources, all with $\simeq1\arcmin$ offset, but present no X-ray or radio candidates. \\ 
 {\bf 4FGL J0901.1$-$4456c} This source is colocated with two submillimeter sources and one dense core source about 1$\arcmin$ offset, and several more of each slightly more offset. The region hosts an unusual number of dense core sources. \\
 {\bf 4FGL J0933.8$-$6232} One of five bright Orphan B sources with TS $>$ 500 and $|$b$| >$1$\arcdeg$.  These are good candidates for pulsar searches. \\
 {\bf 4FGL J1020.4$-$5314}  The eRosita counterpart is not obviously connected to any particular object in the vicinity. \\
 {\bf 4FGL J1026.2$-$5731}  The region of this source hosts multiple young stellar objects. The source presents no X-ray or  radio counterparts. \\
 {\bf 4FGL J1046.7$-$6010}  The eRosita counterpart corresponds to the NGC 3372 H II region. \\ 
 {\bf 4FGL J1112.2$-$6055} There are two sub-millimeter sources close to the LAT position. The closer one is AGAL G291.159$-$00.337, which is 45" off the LAT position. This is a candidate for a passive source.  \\
 {\bf 4FGL J1127.9$-$6158}   There is an emission line star spectral type OBe at the eRosita position, at a distance of 3.4 kpc. The star is CPD-61 2338. \\ 
 {\bf 4FGL J1216.8$-$5955}  The closest thing of interest seems to be the bright 8.9m star HD 106671, with spectral type A0 V and parallax distance 558 pc.  The eRosita counterpart does not correspond to anything. \\
 {\bf 4FGL J1220.1$-$5558}  is colocated with two galaxies, the closer being LEDA 3079414. \\ 
 {\bf 4FGL J1244.3$-$6233}  lies near a cluster of NIR, sub-millimeter, objects, so it has passive source possibilities.\\ 
 {\bf 4FGL J1312.3$-$6257}  The Swift source position corresponds to two sub-millimeter sources.  This is a good candidate for a passive source. \\ 
 {\bf 4FGL J1325.3$-$5413}  The eRosita counterpart is not found to  correspond to anything else nearby. \\ 
 {\bf 4FGL J1329.9$-$6108} One of five bright Orphan B sources with TS $>$ 500 and $|$b$| >$1$\arcdeg$.  These are good candidates for pulsar searches. \\
 {\bf 4FGL J1404.8$-$5237}  There is a peculiar object in the error ellipse, the star HD 122558 at 690 pc, which is not only spectral class F3/F5 V (i.e., luminosity class V) but is also thought to be a double or multiple system. \\ 
 {\bf 4FGL J1517.9$-$5233} One of five bright Orphan B sources with TS $>$ 500 and $|$b$| >$1$\arcdeg$.  These are good candidates for pulsar searches. \\
 {\bf 4FGL J1534.7$-$5842}  A radio pulsar, with angular offset of about 7$\arcmin$ from the LAT position, is PSR J1535-5848, which has period $\simeq$ 0.3 s and \.{E}= $3.7\times10^{33}$ erg s$^{-1}$. \\ 
 {\bf 4FGL J1538.0$-$4638}  The radio source PMN J1539$-$4643 at is only 8$\arcsec$ off the LAT position. \\ 
 {\bf 4FGL J1547.4$-$4802}  The eRosita conterpart does not correspond to anything, the field has only stars.\\
 {\bf 4FGL J1616.6$-$5341}  The X-ray counterpart does not corresponding to anything except HD 146018, a double or multiple star at Vmag = 8.53 (distance=250 pc) and spectral class A0 V. \\
 {\bf 4FGL J1622.7$-$4934c}  lies within a cluster of 12 molecular clouds near the  HII region G334.1+0.0986. \\ 
 {\bf 4FGL J1634.0$-$4742c}  is near a luminous B star, LS 2832, a large number of molecular clouds and the SFR G336.7+0.0. \\ 
 {\bf 4FGL J1636.9$-$4710c} is coincident with the submillimeter source HIGALBM G337.4066+00.0237 and two molecular clouds at nearby position. It has an AGILE counterpart.\\ 
 {\bf 4FGL J1639.3$-$5146} One of five bright Orphan B sources with TS $>$ 500 and $|$b$| >$1$\arcdeg$.  These are good candidates for pulsar searches. \\
 {\bf 4FGL J1649.3$-$4441}  The eRosita counterpart corresponds to a LEDA galaxy, while the Swift one matches a molecular cloud. The source is spatially coincident with the young stellar cluster NGC 6216. \\  
 {\bf 4FGL J1653.2$-$4349} is located in an IRAS HII region, with bubble /molecular clouds. \\ 
 {\bf 4FGL J1656.5$-$2733.}  A radio source is a possible lead: NVSS J165650-273438. The offset is 220$\arcsec$. \\ 
 {\bf 4FGL J1703.6$-$2850}  A radio source is a possible lead: NVSS J170341-285343. The offset is 221$\arcsec$. \\ 
 {\bf 4FGL J1704.8$-$4030} A LAT refit of this region associated the gamma-ray sources with SNR CTB 37B \citep{Xin16}. Their fit identifies gamma-ray sources in the region that are not in 4FGL, but one of them corresponds to 4FGL J1704.8-4030. \\ 
 {\bf 4FGL J1711.0$-$3002}  The eRosita counterpart is CD-29 13265, a double or multiple star system about 534 pc away. A radio source NVSS J171112$-$300328 lies nearby.\\
 {\bf 4FGL J1729.1$-$3503}  The re-analysis of LAT sources near HESS J1731$-$347 finds two LAT sources near this position. The HESS source is associated with SNR G353.6$-$00.7. There are submilllimeter, dark nebula and molecular cloud sources in the vicinity but those are not necessarily unrelated to the SNR. \\
 {\bf 4FGL J1729.9$-$4148} A radio source is the possible lead, PMN J1729$-$4148, with an offset of only 102". \\
 {\bf 4FGL J1737.3$-$3332} lies in field of stars, with no likely associations. \\ 
 {\bf 4FGL J1747.0$-$3505} There is a radio detection in a survey of Southern Hemisphere LAT positions with ATCA. \\ 
 {\bf 4FGL J1754.6$-$2933}  The Chandra Galactic Bulge Survey finds an X-ray source at an offset of 86". There is no other X-ray, no radio, and no passive source candidate. \\ 
 {\bf 4FGL J1757.4$-$3125} This source has an eRosita counterpart, but no other X-ray, radio, or passive candidates. \\ 
 {\bf 4FGL J1804.4$-$0852}  has a Swift counterpart and lies close to a WISE mid-IR source. \\
 {\bf 4FGL J1805.1$-$3618} This source presents no radio, no X-ray candidates.  One galaxy of interest, LEDA 2055844 is  offset by 83". \\ 
 {\bf 4FGL J1808.4$-$3358} is near a luminous B star LS 2832. Also near a large number of molecular clouds and the SFR G336.7+0.0. \\
 {\bf 4FGL J1815.8$-$1416}  This region hosts a combination of a sub-millimeter source, dark nebula, and molecular cloud. In a different direction there is a PMN radio source.  \\ 
 {\bf 4FGL J1830.8$-$3132} This source is spatially coincident with the radio source NVSS J183102$-$312544. \\
 {\bf 4FGL J1832.4$-$0847} This region hosts 4 sub millimeter sources, close to each other. \\ 
 {\bf 4FGL J1834.7$-$0724c.}  This region has a great host of submillimeter sources and molecular clouds ranging in offsets from 26$\arcsec$ to about 3$\arcmin$.\\ 
 {\bf 4FGL J1852.6+0203} The error box hosts a sub-millimeter source, a DNe, and a molecular cloud in mutual proximity.\\  
 {\bf 4FGL J1852.8+0156c}   The only thing of moderate interest here is ZTF J185236.94+020423.1, a BY Draconis-type flare star. These stars and also RS Cn Ven stars can be outstanding flaring X-ray sources. \\
 {\bf 4FGL J1857.1+0056} This source is colocated with  two dark nebulae and three sub-millimeter sources and is a passive source possibility. \\
 {\bf 4FGL J1857.4+0106} This is an interesting passive source possibility  due to the colocation with a dark nebula, three sub-millimeter sources, and one molecular cloud, mutually close to each other but about 3$\arcmin$ off the LAT position. \\ 
 {\bf 4FGL J1900.7+0426}  The error box hosts 7 submillimeter sources, one HII and one molecular cloud all within 1$\arcmin$. A radio source from the 5 GHz VLA survey \citep{Beck94}   is also colocated at 7$\arcsec$. It in turn agrees  with two of the sub-millimeter sources and the HII region. \\ 
 {\bf 4FGL J1902.2+0448}  lies close to five sub-millimeter sources and a nebula. \\
 {\bf 4FGL J1906.9+0712}   There are 4 molecular clouds and 7 sub millimeter sources catalogued here, an unusual total count in a small region. This could be the prototype for a passive source. \\
 \noindent{\bf 4FGL J1908.8$-$0131 } One of four bright Orphan A sources with TS $>$ 500 and $|$b$| >$1$\arcdeg$.  These are good candidates for pulsar searches. \\
 {\bf 4FGL J1912.5+1320} This source lies near a collection of sub-millimeter, bubbles and young stellar objects. \\
 {\bf 4FGL J1944.8+4301} One may consider the eruptive variable, UCAC4 666$-$079929 a good positional coincidence with a rare kind of object.  \\
 {\bf 4FGL J1948.9+3414.}  This source is colocated with the cataclysmic variable V1449 Cyg. \\
 \noindent{\bf 4FGL 2041.1+4736 } One of four bright Orphan A sources with TS $>$ 500 and $|$b$| >$1$\arcdeg$.  These are good candidates for pulsar searches. \\
 {\bf 4FGL J2218.8+4736}  This source presents no X-ray, radio, or passive candidates. One galaxy, WISE J221917.91+473633.4, is located just outside the r95 region. \\
{\bf 4FGL J2220.8+6319} The source is located near two HII regions, two IR sources, and two young stellar objects in mutual proximity. \\

\bibliography{GU_papers}

\begin{longrotatetable}
\startlongtable
\begin{deluxetable*}{lcccccccccc}
\label{tab:bright_SGUs_1} 
\tablecaption{Gamma-ray properties of the bright GU sample}
\tablehead{\colhead{{Source Name\tablenotemark{$\ddag$}}} & \colhead{{L }} & \colhead{{B }} & \colhead{{TS $^\diamond$}} & \colhead{{PL\_Index}} &
\colhead{{$\sigma_{Curv}$}} & 
\colhead{{LP\_beta}\tablenotemark{$\dagger$}} &
\colhead{{LP\_EPeak{\tablenotemark{$\dagger$}} }} & \colhead{Var.}&
\colhead{Flags\tablenotemark{$\vartriangle$}} & ML class \\
 &  \colhead{{($^\circ$)}} & \colhead{{($^\circ$)}} & & &  & &\colhead{{(MeV)}} & Index\tablenotemark{$\star$} & & }
\startdata
J0034.6+6438$^\bigstar$ & 121.13 & 1.83 & 114 & 2.45$\pm$0.08& 2.62 & 0.15$\pm$0.08 & 508$\pm$488&16.9& 3 & SGU \\ 
J0039.1+6257$^\bigstar$ & 121.54 & 0.12 & 845 & 2.36$\pm$0.03& 9.31 & 0.46$\pm$0.08 & 1585$\pm$175&11.3&  & SGU \\ 
J0057.9+6326 & 123.66 & 0.58 & 125 & 1.79$\pm$0.12& 0.6 & \nodata  &\nodata  &16.0&  & bll+ \\ 
J0204.7+6656 & 130.04 & 5.08 & 147 & 2.59$\pm$0.07& 4.74 & 0.28$\pm$0.08 & 475$\pm$195&8.8&  & SGU \\ 
J0235.3+5650 & 136.82 & -3.19 & 296 & 2.6$\pm$0.05& 5.64 & 0.34$\pm$0.09 & 443$\pm$119&17.8&  & SGU \\ 
J0237.8+5238 & 138.83 & -6.92 & 749 & 2.29$\pm$0.04& 8.73 & 0.48$\pm$0.08 & 1612$\pm$162&9.0&  & msp+ \\ 
J0340.4+5302$^\lozenge$ & 146.79 & -1.82 & 1056 & 2.95$\pm$0.03& 15.33 & 0.66$\pm$0.07 & 202$\pm$18&9.9&  & \nodata \\ 
J0426.5+5434$^\bigstar$ & 150.88 & 3.82 & 953 & 2.64$\pm$0.03& 13.59 & 0.45$\pm$0.05 & 390$\pm$41&11.4&  & SGU \\ 
J0616.5+2235 & 188.96 & 2.91 & 243 & 1.98$\pm$0.05& 2.62 & 0.05$\pm$0.02 & 10197$\pm$6605&13.7& 2,5 & psr+ \\ 
J0722.4$-$2650 & 240.26 & -5.67 & 138 & 2.34$\pm$0.08& 2.54 & 0.25$\pm$0.12 & 1001$\pm$478&12.3&  & SGU \\ 
J0725.7$-$0549 & 222.06 & 4.93 & 202 & 1.93$\pm$0.08& 2.76 & 0.19$\pm$0.09 & 8568$\pm$2949&16.8&  & bll+ \\ 
J0736.9$-$3231 & 246.79 & -5.57 & 279 & 2.54$\pm$0.05& 5.53 & 0.34$\pm$0.08 & 385$\pm$98&21.0&  & SGU \\ 
J0744.9$-$4028 & 254.57 & -8.0 & 212 & 2.24$\pm$0.06& 4.57 & 0.3$\pm$0.09 & 1679$\pm$385&15.8&  & msp+ \\ 
J0752.0$-$2931$^\lozenge$ & 245.78 & -1.28 & 149 & 2.5$\pm$0.07& 6.29 & 0.57$\pm$0.14 & 897$\pm$164&8.8&  & msp+ \\ 
J0754.9$-$3953 & 255.02 & -6.06 & 189 & 2.24$\pm$0.07& 6.74 & 0.64$\pm$0.14 & 2212$\pm$286&6.7&  & msp+ \\ 
J0758.8$-$1450$^\bigstar$ & 233.96 & 7.61 & 680 & 2.22$\pm$0.04& 8.38 & 0.47$\pm$0.09 & 1672$\pm$187&15.0&  & msp+\\ 
J0826.1$-$5053$^\bigstar$ & 267.31 & -7.39 & 228 & 2.37$\pm$0.06& 4.06 & 0.26$\pm$0.09 & 1198$\pm$402&10.7&  & SGU \\ 
J0828.4$-$4444 & 262.47 & -3.52 & 168 & 2.52$\pm$0.08& 4.37 & 0.42$\pm$0.14 & 2005$\pm$548&15.1& 2,5 & msp+ \\ 
J0848.8$-$4328$^\lozenge$ & 263.69 & 0.16 & 209 & 2.73$\pm$0.06& 4.72 & 0.25$\pm$0.08 & 232$\pm$119&7.4& 14 & SGU \\ 
J0853.6$-$4306$^\lozenge$ & 263.98 & 1.07 & 116 & 2.76$\pm$0.07& 5.62 & 0.48$\pm$0.13 & 421$\pm$122&3.6& 5,14 & SGU \\ 
J0854.8$-$4504$^\lozenge$ & 265.61 & -0.03 & 1507 & 2.36$\pm$0.03& 10.53 & 0.36$\pm$0.05 &1525$\pm$159&11.0&  & SGU\\ 
J0857.7$-$4256c & 264.33 & 1.75 & 127 & 2.57$\pm$0.07& 3.1 & 0.18$\pm$0.08 & 301$\pm$227&10.4& 3,6,14 & SGU \\ 
J0859.2$-$4729 & 267.94 & -1.03 & 341 & 2.54$\pm$0.05& 4.41 & 0.16$\pm$0.05 & 275$\pm$154&7.1& 14 & SGU \\ 
J0859.3$-$4342 & 265.1 & 1.46 & 93 & 2.55$\pm$0.08& 4.18 & 0.31$\pm$0.11 & 672$\pm$331&14.5& 14 & SGU \\ 
J0900.5$-$4434c$^\lozenge$ & 265.9 & 1.06 & 73 & 2.55$\pm$0.09& 3.41 & 0.32$\pm$0.13 & 425$\pm$183&13.0& 2,5,6,14 & SGU \\ 
J0901.1$-$4456c$^\lozenge$ & 266.24 & 0.89 & 138 & 2.46$\pm$0.07& 4.33 & 0.28$\pm$0.1 & 944$\pm$407&9.9& 3,6,14 & SGU \\ 
J0917.9$-$4755 & 270.39 & 0.99 & 113 & 2.66$\pm$0.07& 5.25 & 0.42$\pm$0.12 & 361$\pm$106&16.5&  & SGU \\ 
J0928.4$-$5256 & 275.1 & -1.4 & 171 & 2.04$\pm$0.08& 0.8 & \nodata  &\nodata  &27.8&  & bll+ \\ 
J0933.8$-$6232$^\lozenge$ & 282.24 & -7.91 & 2780 & 2.11$\pm$0.02& 17.29 & 0.57$\pm$0.05 &2144$\pm$92&10.1&  & msp+\\ 
J1020.4$-$5314$^\lozenge$ & 281.41 & 3.2 & 160 & 2.46$\pm$0.07& 4.61 & 0.41$\pm$0.12 & 874$\pm$193&11.3&  & SGU \\ 
J1026.2$-$5731 & 284.41 & 0.02 & 301 & 2.6$\pm$0.05& 7.73 & 0.57$\pm$0.12 & 1217$\pm$208&6.0& 5 & psr+ \\ 
J1037.8$-$5810c & 286.06 & 0.24 & 154 & 2.62$\pm$0.07& 3.86 & 0.26$\pm$0.1 & 468$\pm$269&10.6& 2,5,6 & SGU \\ 
J1046.7$-$6010 & 288.01 & -0.97 & 326 & 2.6$\pm$0.04& 7.34 & 0.35$\pm$0.08 & 603$\pm$165&16.1& 2 & SGU \\ 
J1048.5$-$5923 & 287.85 & -0.17 & 396 & 2.64$\pm$0.04& 8.4 & 0.48$\pm$0.1 & 754$\pm$159&5.5&  & SGU \\ 
J1058.4$-$6625 & 291.96 & -5.99 & 123 & 2.02$\pm$0.09& 0.74 & \nodata  &\nodata  &21.4& 3 & bll+ \\ 
J1112.2$-$6055 & 291.17 & -0.33 & 173 & 2.39$\pm$0.06& 6.15 & 0.28$\pm$0.06 & 919$\pm$276&6.4&  & SGU \\ 
J1115.1$-$6118 & 291.63 & -0.57 & 301 & 2.34$\pm$0.06& 1.7 & \nodata  &\nodata  &7.3& 3 & fsrq+ \\ 
J1127.9$-$6158$^\lozenge$ & 293.3 & -0.68 & 132 & 2.49$\pm$0.06& 5.41 & 0.38$\pm$0.11 & 915$\pm$269&13.3&  & SGU \\ 
J1208.0$-$6900 & 299.04 & -6.46 & 309 & 2.42$\pm$0.05& 5.27 & 0.27$\pm$0.07 & 705$\pm$193&14.0&  & SGU \\ 
J1216.8$-$5955$^\lozenge$ & 298.61 & 2.66 & 110 & 2.54$\pm$0.08& 3.62 & 0.22$\pm$0.09 & 474$\pm$280&11.0& 1,2,14 & SGU \\ 
J1220.1$-$5558$^\lozenge$ & 298.53 & 6.64 & 148 & 2.44$\pm$0.08& 2.23 & 0.13$\pm$0.07 & 333$\pm$348&21.2&  & SGU \\ 
J1244.3$-$6233$^\lozenge$ & 302.11 & 0.31 & 239 & 2.47$\pm$0.05& 4.05 & 0.2$\pm$0.07 & 810$\pm$451&4.9&  & SGU \\ 
J1257.0$-$6339 & 303.55 & -0.79 & 327 & 2.48$\pm$0.05& 4.07 & 0.22$\pm$0.08 & 970$\pm$463&16.7&  & SGU \\ 
J1309.1$-$6223 & 304.98 & 0.41 & 213 & 2.61$\pm$0.05& 6.97 & 0.52$\pm$0.12 & 852$\pm$183&8.5& 5,14 & SGU \\ 
J1312.3$-$6257 & 305.31 & -0.19 & 140 & 2.54$\pm$0.07& 2.32 & 0.11$\pm$0.06 & 157$\pm$243&19.1& 5,14 & fsrq+ \\ 
J1312.6$-$6231c & 305.37 & 0.24 & 130 & 2.48$\pm$0.07& 3.9 & 0.4$\pm$0.16 & 1828$\pm$624&7.8& 2,6,14 & SGU \\ 
J1317.5$-$6316 & 305.86 & -0.56 & 492 & 2.52$\pm$0.04& 8.79 & 0.58$\pm$0.11 & 1487$\pm$179&16.8& 14 & psr+ \\ 
J1325.3$-$5413$^\lozenge$ & 307.93 & 8.31 & 227 & 2.55$\pm$0.07& 4.08 & 0.27$\pm$0.09 & 619$\pm$252&12.2& 2 & SGU \\ 
J1329.9$-$6108$^\lozenge$ & 307.56 & 1.38 & 792 & 2.35$\pm$0.04& 8.19 & 0.37$\pm$0.06 & 1793$\pm$220&21.1&  & SGU\\ 
J1351.6$-$6142 & 310.0 & 0.34 & 217 & 2.53$\pm$0.05& 6.54 & 0.43$\pm$0.13 & 1390$\pm$437&19.7& 14 & SGU \\ 
J1403.5$-$6236$^\lozenge$ & 311.12 & -0.88 & 276 & 2.43$\pm$0.05& 7.6 & 0.64$\pm$0.14 & 2191$\pm$276&23.6& 3,14 & msp+ \\ 
J1404.8$-$5237$^\lozenge$ & 314.07 & 8.65 & 181 & 2.41$\pm$0.07& 3.71 & 0.24$\pm$0.09 & 846$\pm$348&11.6&  & SGU \\ 
J1412.1$-$6631 & 310.86 & -4.9 & 355 & 2.3$\pm$0.05& 7.07 & 0.44$\pm$0.09 & 1748$\pm$254&12.7&  & msp+ \\ 
J1415.4$-$6458 & 311.67 & -3.55 & 120 & 2.27$\pm$0.09& 3.93 & 0.36$\pm$0.14 & 2611$\pm$750&16.6&  & msp+ \\ 
J1427.8$-$6051 & 314.4 & -0.15 & 339 & 2.47$\pm$0.05& 3.49 & 0.12$\pm$0.04 & 233$\pm$183&8.1& 3,10,14 & fsrq+ \\ 
J1443.7$-$7037$^\lozenge$ & 312.05 & -9.77 & 127 & 2.44$\pm$0.08& 1.77 & \nodata  &\nodata  &20.1&  & fsrq+ \\ 
J1444.9$-$5939$^\lozenge$ & 316.83 & 0.11 & 135 & 2.61$\pm$0.06& 6.23 & 0.31$\pm$0.07 & 456$\pm$166&11.4& 2,3,5,14 & SGU \\ 
J1510.1$-$5750$^\lozenge$ & 320.56 & 0.2 & 99 & 2.53$\pm$0.09& 1.69 & \nodata  &\nodata  &7.4& 2,14 & SGU \\ 
J1517.9$-$5233$^\lozenge$ & 324.26 & 4.14 & 658 & 2.33$\pm$0.04& 8.64 & 0.51$\pm$0.09 & 1783$\pm$181&11.4&  & msp+\\ 
J1526.3$-$4501 & 329.59 & 9.67 & 141 & 2.25$\pm$0.09& 0.01 & \nodata  &\nodata  &8.7&  & \nodata \\ 
J1534.0$-$5232$^\lozenge$ & 326.29 & 2.79 & 241 & 2.43$\pm$0.06& 3.33 & 0.19$\pm$0.08 & 823$\pm$424&6.4&  & SGU \\ 
J1534.7$-$5842$^\lozenge$ & 322.79 & -2.29 & 114 & 2.26$\pm$0.09& 4.89 & 0.64$\pm$0.19 & 3256$\pm$502&8.4&  & msp+ \\ 
J1536.8$-$4327 & 332.06 & 9.86 & 75 & 2.77$\pm$0.1& 2.22 & 0.19$\pm$0.11 & 95$\pm$112&33.3&  & fsrq+ \\ 
J1538.0$-$4638 & 330.32 & 7.18 & 178 & 2.73$\pm$0.07& 4.75 & 0.3$\pm$0.09 & 278$\pm$110&9.0&  & SGU \\ 
J1547.4$-$4802$^\lozenge$ & 330.72 & 5.11 & 174 & 2.59$\pm$0.07& 3.76 & 0.21$\pm$0.08 & 333$\pm$181&13.3&  & SGU \\ 
J1548.1$-$4416$^\lozenge$ & 333.18 & 7.97 & 110 & 2.67$\pm$0.08& 4.15 & 0.39$\pm$0.13 & 456$\pm$162&12.2&  & SGU \\ 
J1603.3$-$6010 & 324.73 & -5.7 & 476 & 2.2$\pm$0.05& 6.67 & 0.33$\pm$0.07 & 2269$\pm$330&8.8&  & msp+ \\ 
J1610.3$-$5154c$^\lozenge$ & 331.01 & -0.23 & 138 & 2.55$\pm$0.06& 4.52 & 0.26$\pm$0.1 & 738$\pm$400&8.8& 1,2,6,14 & SGU \\ 
J1611.9$-$5125c & 331.52 & -0.04 & 136 & 2.58$\pm$0.06& 4.76 & 0.25$\pm$0.08 & 443$\pm$233&13.7& 3,5,6,14 & SGU \\ 
J1613.0$-$5102 & 331.91 & 0.11 & 212 & 2.58$\pm$0.05& 6.67 & 0.29$\pm$0.08 & 610$\pm$255&16.4& 14 & SGU \\ 
J1616.6$-$5341$^\lozenge$ & 330.48 & -2.17 & 557 & 2.47$\pm$0.04& 9.76 & 0.74$\pm$0.12 & 1661$\pm$140&12.4&  & psr+ \\ 
J1616.6$-$5009 & 332.94 & 0.37 & 100 & 2.67$\pm$0.07& 5.51 & 0.4$\pm$0.12 & 542$\pm$231&4.7& 3,4,5,14 & SGU \\ 
J1618.0$-$5119 & 332.28 & -0.63 & 135 & 2.42$\pm$0.07& 2.81 & 0.18$\pm$0.09 & 892$\pm$707&15.6&  & SGU \\ 
J1620.8$-$4958c & 333.55 & 0.03 & 158 & 2.52$\pm$0.06& 4.8 & 0.28$\pm$0.12 & 952$\pm$596&8.5& 2,5,6,14 & SGU \\ 
J1622.7$-$4934c & 334.04 & 0.09 & 107 & 2.57$\pm$0.06& 5.67 & 0.3$\pm$0.09 & 624$\pm$344&8.5& 5,6,14 & SGU \\ 
J1634.0$-$4742c & 336.69 & 0.03 & 167 & 2.55$\pm$0.05& 8.02 & 0.57$\pm$0.12 & 1046$\pm$212&12.8& 2,5,6,14 & psr+ \\ 
J1636.9$-$4710c & 337.41 & 0.03 & 170 & 2.53$\pm$0.05& 5.08 & 0.21$\pm$0.06 & 451$\pm$222&10.1& 2,5,6,14 & SGU \\ 
J1639.3$-$5146$^\lozenge$ & 334.25 & -3.33 & 1856 & 2.25$\pm$0.03& 9.6 & 0.3$\pm$0.04 & 2103$\pm$205&18.1&  & SGU\\ 
J1639.8$-$4642c & 338.08 & -0.02 & 134 & 2.52$\pm$0.06& 3.94 & 0.15$\pm$0.05 & 317$\pm$261&10.4& 5,6,14 & fsrq+ \\ 
J1640.3$-$4917$^\bigstar$ & 336.21 & -1.8 & 125 & 2.5$\pm$0.08& 3.48 & 0.24$\pm$0.1 & 885$\pm$463&21.7& 14 & SGU \\ 
J1643.3$-$3148$^\bigstar$ & 349.82 & 9.24 & 124 & 2.34$\pm$0.08& 3.34 & 0.24$\pm$0.1 & 1136$\pm$467&15.3&  & SGU \\ 
J1649.3$-$4441 & 340.69 & 0.03 & 172 & 2.52$\pm$0.06& 5.01 & 0.35$\pm$0.13 & 1499$\pm$606&6.3& 14 & SGU \\ 
J1651.7$-$4359$^\bigstar$ & 341.51 & 0.15 & 145 & 2.6$\pm$0.06& 6.13 & 0.44$\pm$0.15 & 1110$\pm$452&15.0& 5,14 & SGU \\ 
J1653.2$-$4349 & 341.8 & 0.06 & 299 & 2.59$\pm$0.04& 7.03 & 0.27$\pm$0.06 & 546$\pm$198&16.8& 14 & SGU \\ 
J1656.5$-$2733 & 355.0 & 9.68 & 132 & 2.36$\pm$0.08& 5.92 & 0.65$\pm$0.16 & 1674$\pm$258&10.7&  & msp+ \\ 
J1703.6$-$2850 & 354.93 & 7.66 & 205 & 2.42$\pm$0.07& 3.52 & 0.21$\pm$0.08 & 744$\pm$342&21.0& 3 & SGU \\ 
J1704.8$-$4030$^\lozenge$ & 345.74 & 0.43 & 267 & 2.43$\pm$0.05& 5.14 & 0.27$\pm$0.08 & 1501$\pm$546&18.1& 14 & SGU \\ 
J1711.0$-$3002$^\lozenge$ & 354.91 & 5.64 & 189 & 2.49$\pm$0.06& 6.02 & 0.52$\pm$0.13 & 1105$\pm$202&5.7&  & msp+ \\ 
J1714.8$-$3849 & 348.25 & -0.13 & 190 & 2.47$\pm$0.06& 4.3 & 0.23$\pm$0.08 & 891$\pm$469&4.4& 5,14 & SGU \\ 
J1714.9$-$3324$^\lozenge$ & 352.65 & 3.01 & 260 & 2.48$\pm$0.06& 5.13 & 0.33$\pm$0.1 & 1173$\pm$339&8.2& 3,14 & SGU \\ 
J1721.3$-$5257$^\lozenge$ & 337.24 & -9.13 & 171 & 2.48$\pm$0.07& 5.53 & 0.5$\pm$0.14 & 964$\pm$187&17.6& 14 & msp+ \\ 
J1721.7$-$3917 & 348.62 & -1.49 & 108 & 2.49$\pm$0.08& 3.74 & 0.24$\pm$0.09 & 751$\pm$376&11.8& 14 & SGU \\ 
J1725.1$-$1924$^\bigstar$ & 5.58 & 9.03 & 174 & 2.4$\pm$0.07& 2.71 & 0.15$\pm$0.07 & 454$\pm$319&12.9&  & SGU \\ 
J1727.6$-$2304$^\bigstar$ & 2.79 & 6.55 & 158 & 2.28$\pm$0.07& 3.95 & 0.27$\pm$0.1 & 1984$\pm$652&13.3&  & SGU \\ 
J1729.1$-$3503 & 352.96 & -0.33 & 102 & 2.48$\pm$0.08& 4.43 & 0.52$\pm$0.19 & 2285$\pm$642&7.7& 3,14 & msp+ \\ 
J1729.9$-$4148 & 347.41 & -4.18 & 134 & 2.35$\pm$0.08& 1.86 & \nodata  &\nodata  &18.1&  & SGU \\ 
J1730.8$-$3806 & 350.6 & -2.31 & 116 & 2.38$\pm$0.08& 3.66 & 0.31$\pm$0.12 & 1407$\pm$504&15.4& 2,3 & SGU \\ 
J1737.3$-$3332$^\lozenge$ & 355.17 & -0.93 & 108 & 2.65$\pm$0.07& 3.79 & 0.28$\pm$0.1 & 368$\pm$209&13.2& 3,4,14 & SGU \\ 
J1739.3$-$2531$^\bigstar$ & 2.17 & 2.98 & 149 & 2.48$\pm$0.07& 3.6 & 0.23$\pm$0.09 & 714$\pm$369&10.9& 14 & SGU \\ 
J1740.7$-$2640$^\lozenge$ & 1.36 & 2.1 & 152 & 2.49$\pm$0.07& 4.88 & 0.43$\pm$0.13 & 1458$\pm$379&5.8& 14 & SGU \\ 
J1742.8$-$2246$^\bigstar$ & 4.94 & 3.74 & 87 & 2.55$\pm$0.09& 3.78 & 0.36$\pm$0.14 & 725$\pm$300&14.1& 14 & SGU \\ 
J1743.7$-$4321$^\bigstar$ & 347.45 & -7.17 & 257 & 2.41$\pm$0.06& 6.47 & 0.48$\pm$0.11 & 1069$\pm$162&21.3& 2,14 & msp+ \\ 
J1744.0$-$1311$^\bigstar$ & 13.34 & 8.45 & 223 & 2.31$\pm$0.07& 5.15 & 0.37$\pm$0.1 & 1812$\pm$382&11.5&  & msp+ \\ 
J1744.7$-$1557$^\bigstar$ & 11.01 & 6.91 & 182 & 2.44$\pm$0.07& 4.18 & 0.24$\pm$0.08 & 719$\pm$284&6.0& 14 & SGU \\ 
J1747.0$-$3505$^\lozenge$ & 354.89 & -3.46 & 131 & 2.53$\pm$0.07& 3.96 & 0.3$\pm$0.11 & 721$\pm$291&15.0& 3,14 & SGU \\ 
J1748.3$-$2906 & 0.17 & -0.6 & 236 & 2.55$\pm$0.05& 5.92 & 0.34$\pm$0.1 & 1327$\pm$467&12.4& 3,5,14 & SGU \\ 
J1748.8$-$3915$^\lozenge$ & 351.49 & -5.89 & 274 & 2.39$\pm$0.05& 7.11 & 0.53$\pm$0.12 & 1266$\pm$186&25.2&  & msp+ \\ 
J1752.8$-$4449 & 346.99 & -9.32 & 164 & 2.34$\pm$0.07& 3.91 & 0.34$\pm$0.11 & 1850$\pm$476&14.0&  & msp+ \\ 
J1753.8$-$2538$^\bigstar$ & 3.77 & 0.13 & 1462 & 2.35$\pm$0.03& 10.2 & 0.32$\pm$0.05 & 2051$\pm$244&6.5&  & psr+\\ 
J1754.6$-$2933$^\lozenge$ & 0.48 & -2.01 & 169 & 2.51$\pm$0.07& 5.33 & 0.48$\pm$0.14 & 1459$\pm$325&16.5& 14 & msp+ \\ 
J1757.4$-$3125$^\lozenge$ & 359.16 & -3.46 & 127 & 2.48$\pm$0.08& 3.88 & 0.29$\pm$0.1 & 943$\pm$382&17.6& 14 & SGU \\ 
J1758.7$-$4109$^\bigstar$ & 350.77 & -8.48 & 406 & 2.36$\pm$0.05& 3.04 & 0.11$\pm$0.04 & 358$\pm$251&21.6&  & fsrq+ \\ 
J1800.9$-$2407 & 5.88 & -0.51 & 131 & 2.29$\pm$0.08& 2.36 & 0.1$\pm$0.06 & 969$\pm$1064&12.2& 5 & SGU \\ 
J1801.6$-$2326 & 6.57 & -0.32 & 327 & 2.44$\pm$0.04& 9.61 & 0.3$\pm$0.04 & 610$\pm$112&5.0& 3,5,10,14 & \nodata \\ 
J1801.8$-$2358 & 6.11 & -0.61 & 202 & 2.27$\pm$0.07& 0.01 & \nodata  &\nodata  &14.1& 3,5 & \nodata \\ 
J1802.4$-$3041$^\lozenge$ & 0.33 & -4.04 & 278 & 2.14$\pm$0.06& 6.23 & 0.47$\pm$0.11 & 3754$\pm$483&9.4&  & msp+ \\ 
J1804.4$-$0852$^\lozenge$ & 19.62 & 6.24 & 161 & 2.51$\pm$0.07& 2.63 & 0.14$\pm$0.06 & 316$\pm$299&8.5&  & fsrq+ \\ 
J1805.1$-$3618$^\bigstar$ & 355.66 & -7.24 & 284 & 2.33$\pm$0.06& 7.24 & 0.55$\pm$0.11 & 1592$\pm$204&6.9&  & msp+ \\ 
J1808.4$-$3358 & 358.05 & -6.73 & 399 & 2.51$\pm$0.05& 8.68 & 0.55$\pm$0.1 & 848$\pm$111&6.9& 14 & SGU \\ 
J1808.5$-$3701 & 355.34 & -8.18 & 161 & 2.49$\pm$0.07& 4.35 & 0.33$\pm$0.11 & 656$\pm$192&15.8& 14 & SGU \\ 
J1809.2$-$2726$^\lozenge$ & 3.9 & -3.76 & 113 & 2.45$\pm$0.08& 2.91 & 0.17$\pm$0.07 & 566$\pm$394&11.3& 2,14 & SGU \\ 
J1812.8$-$3144 & 0.48 & -6.5 & 174 & 2.33$\pm$0.07& 4.05 & 0.3$\pm$0.1 & 1418$\pm$419&10.2&  & SGU \\ 
J1814.7$-$3420$^\lozenge$ & 358.35 & -8.06 & 162 & 2.5$\pm$0.07& 4.22 & 0.31$\pm$0.11 & 551$\pm$178&7.6& 3,14 & SGU \\ 
J1815.8$-$1416 & 16.21 & 1.19 & 105 & 2.63$\pm$0.08& 3.0 & 0.2$\pm$0.09 & 457$\pm$401&6.1& 14 & SGU \\ 
J1817.2$-$3035 & 1.95 & -6.78 & 207 & 2.41$\pm$0.07& 0.86 & \nodata  &\nodata  &17.9&  & \nodata \\ 
J1817.9$-$3334 & 359.34 & -8.28 & 220 & 2.37$\pm$0.06& 6.23 & 0.48$\pm$0.11 & 1196$\pm$198&8.0&  & msp+ \\ 
J1818.6$-$3206$^\bigstar$ & 0.73 & -7.76 & 125 & 2.45$\pm$0.07& 5.34 & 0.46$\pm$0.13 & 988$\pm$235&25.6& 2,5,14 & msp+ \\ 
J1819.9$-$2926$^\bigstar$ & 3.25 & -6.78 & 106 & 2.46$\pm$0.09& 3.33 & 0.26$\pm$0.11 & 727$\pm$329&19.6& 3,14 & SGU \\ 
J1819.9$-$1530 & 15.59 & -0.26 & 141 & 2.67$\pm$0.06& 6.28 & 0.7$\pm$0.18 & 1528$\pm$297&7.5& 2,5,14 & psr+ \\ 
J1823.3$-$1340$^\bigstar$ & 17.6 & -0.11 & 1138 & 2.46$\pm$0.03& 7.73 & 0.22$\pm$0.04 & 1149$\pm$245&7.4&  & psr+\\ 
J1830.7$-$1634$^\bigstar$ & 15.85 & -3.04 & 470 & 2.36$\pm$0.05& 4.94 & 0.25$\pm$0.06 & 1262$\pm$301&18.8&  & SGU \\ 
J1830.8$-$3132$^\lozenge$ & 2.41 & -9.81 & 257 & 2.2$\pm$0.06& 6.22 & 0.5$\pm$0.12 & 2356$\pm$318&9.2&  & msp+ \\ 
J1832.4$-$0847$^\lozenge$ & 22.94 & 0.19 & 179 & 2.5$\pm$0.06& 6.54 & 0.58$\pm$0.15 & 2070$\pm$398&10.1& 14 & psr+ \\ 
J1834.7$-$0724c & 24.43 & 0.31 & 113 & 2.55$\pm$0.06& 5.54 & 0.37$\pm$0.1 & 815$\pm$268&15.8& 5,6,14 & SGU \\ 
J1836.8$-$2354 & 9.94 & -7.64 & 162 & 2.59$\pm$0.07& 4.54 & 0.39$\pm$0.12 & 721$\pm$207&11.2& 2 & SGU \\ 
J1840.4$-$1139 & 21.31 & -2.88 & 114 & 2.41$\pm$0.09& 4.93 & 0.55$\pm$0.15 & 1685$\pm$315&7.9&  & msp+ \\ 
J1843.3$-$1242$^\bigstar$ & 20.69 & -4.0 & 145 & 2.41$\pm$0.08& 3.74 & 0.28$\pm$0.1 & 1193$\pm$416&24.3&  & SGU \\ 
J1847.2$-$0141 & 30.95 & 0.16 & 72 & 2.54$\pm$0.07& 3.85 & 0.22$\pm$0.08 & 419$\pm$268&14.2& 5,14 & SGU \\ 
J1851.5+0718$^\bigstar$ & 39.46 & 3.3 & 119 & 2.5$\pm$0.08& 4.04 & 0.36$\pm$0.15 & 1401$\pm$598&9.2& 14 & SGU \\ 
J1852.6+0203$^\bigstar$ & 34.9 & 0.67 & 111 & 2.5$\pm$0.08& 3.02 & 0.19$\pm$0.09 & 669$\pm$524&15.5& 3,14 & SGU \\ 
J1855.2+0456 & 37.77 & 1.4 & 138 & 2.74$\pm$0.06& 4.12 & 0.2$\pm$0.07 & 137$\pm$89&13.6& 3,14 & SGU \\ 
J1857.1+0056$^\lozenge$ & 34.43 & -0.85 & 165 & 2.4$\pm$0.06& 6.0 & 0.5$\pm$0.14 & 2676$\pm$579&9.6& 1,3 & psr+ \\ 
J1857.4+0106$^\lozenge$ & 34.61 & -0.83 & 114 & 2.4$\pm$0.07& 3.89 & 0.33$\pm$0.16 & 2002$\pm$998&10.9& 2,5 & SGU \\ 
J1858.0+0354 & 37.17 & 0.31 & 346 & 2.58$\pm$0.04& 8.51 & 0.41$\pm$0.09 & 809$\pm$189&18.3& 2,14 & psr+ \\ 
J1900.4+0339 & 37.22 & -0.34 & 253 & 2.56$\pm$0.05& 7.07 & 0.35$\pm$0.09 & 821$\pm$238&18.8& 14 & SGU \\ 
J1900.7+0426 & 37.95 & -0.05 & 67 & 2.58$\pm$0.08& 5.03 & 0.36$\pm$0.1 & 476$\pm$200&8.2& 1,4,5,14 & SGU \\ 
J1901.1+0730$^\bigstar$ & 40.72 & 1.26 & 206 & 2.46$\pm$0.06& 4.29 & 0.19$\pm$0.06 & 717$\pm$381&16.2& 3,14 & SGU \\ 
J1901.8$-$0718 & 27.59 & -5.62 & 122 & 2.37$\pm$0.09& 3.77 & 0.33$\pm$0.12 & 1351$\pm$423&18.0&  & msp+ \\ 
J1902.2+0448 & 38.45 & -0.22 & 167 & 2.6$\pm$0.05& 6.95 & 0.39$\pm$0.09 & 550$\pm$164&8.7& 2,3,14 & SGU \\ 
J1902.5+0654 & 40.35 & 0.68 & 181 & 2.45$\pm$0.06& 3.87 & 0.18$\pm$0.07 & 698$\pm$418&11.8& 3,14 & SGU \\ 
J1904.7$-$0708 & 28.07 & -6.2 & 641 & 2.45$\pm$0.04& 6.79 & 0.26$\pm$0.05 & 633$\pm$136 &22.7&  & SGU\\ 
J1906.1+1651 & 49.62 & 4.45 & 127 & 2.18$\pm$0.09& 5.54 & 0.67$\pm$0.16 & 3605$\pm$503&11.4&  & msp+ \\ 
J1906.9+0712 & 41.12 & -0.15 & 392 & 2.39$\pm$0.04& 7.33 & 0.21$\pm$0.04 & 994$\pm$324&11.2&  & psr+ \\ 
J1908.7+0812 & 42.2 & -0.08 & 225 & 2.58$\pm$0.05& 7.31 & 0.34$\pm$0.07 & 440$\pm$106&14.7& 3,14 & SGU \\
J1908.8$-$0131$^\bigstar$& 33.55 & -4.57 & 695 & 2.32$\pm$0.04& 9.11 & 0.45$\pm$0.07 & 1510$\pm$162&12.5&  & SGU\\  
J1910.2+0904c & 43.14 & 0.01 & 125 & 2.29$\pm$0.07& 1.53 & \nodata  &\nodata  &16.8& 1,5,6,10 & fsrq+ \\ 
J1912.5+1320$^\bigstar$ & 47.19 & 1.47 & 100 & 2.72$\pm$0.07& 3.58 & 0.23$\pm$0.08 & 192$\pm$130&9.1& 14 & SGU \\ 
J1912.7+0957 & 44.21 & -0.15 & 119 & 2.47$\pm$0.06& 5.46 & 0.33$\pm$0.09 & 645$\pm$212&9.7& 3,5,14 & SGU \\ 
J1926.4+1602$^\bigstar$ & 51.15 & -0.25 & 84 & 2.43$\pm$0.09& 1.79 & \nodata  &\nodata  &11.5& 5,14 & fsrq+ \\ 
J1929.0+1729 & 52.72 & -0.09 & 302 & 2.56$\pm$0.04& 7.91 & 0.31$\pm$0.06 & 572$\pm$162&23.5& 3,14 & SGU \\ 
J1931.1+1656 & 52.48 & -0.79 & 469 & 2.54$\pm$0.04& 7.73 & 0.28$\pm$0.06 & 563$\pm$148&8.0& 3,14 & SGU \\ 
J1944.8+4301$^\bigstar$ & 76.82 & 9.27 & 209 & 2.18$\pm$0.07& 3.96 & 0.31$\pm$0.11 & 2511$\pm$604&5.7&  & msp+ \\ 
J1948.9+3414 & 69.51 & 4.24 & 147 & 2.37$\pm$0.08& 5.27 & 0.51$\pm$0.14 & 1651$\pm$318&18.4&  & msp+ \\ 
J1953.5+3841 & 73.84 & 5.69 & 171 & 2.5$\pm$0.07& 2.47 & 0.18$\pm$0.09 & 406$\pm$306&14.5&  & SGU \\ 
J2027.0+2811$^\bigstar$ & 68.83 & -5.88 & 402 & 2.52$\pm$0.05& 9.48 & 0.6$\pm$0.1 & 773$\pm$94&21.1&  & psr+ \\ 
J2038.4+4212 & 81.53 & 0.54 & 375 & 2.63$\pm$0.04& 8.25 & 0.32$\pm$0.06 & 367$\pm$86&13.0& 3 & SGU \\ 
J2041.1+4736$^\bigstar$ & 86.1 & 3.45 & 1335 & 2.43$\pm$0.03& 9.75 & 0.29$\pm$0.04 & 901$\pm$132&10.4&  & SGU\\ 
J2051.7+5051 & 89.72 & 4.13 & 419 & 2.42$\pm$0.05& 7.49 & 0.46$\pm$0.1 & 1505$\pm$237&9.4&  & msp+ \\ 
J2108.0+5155$^\lozenge$ & 92.21 & 2.91 & 258 & 2.55$\pm$0.05& 5.74 & 0.32$\pm$0.09 & 696$\pm$229&18.7&  & SGU \\ 
J2109.6+3954 & 83.54 & -5.43 & 130 & 1.74$\pm$0.09& 2.6 & 0.21$\pm$0.1 & 26391$\pm$13200&19.1& 3 & bll+ \\ 
J2114.3+5023$^\lozenge$ & 91.76 & 1.15 & 230 & 2.48$\pm$0.06& 4.06 & 0.2$\pm$0.07 & 738$\pm$395&20.0&  & SGU \\ 
J2116.2+3701 & 82.3 & -8.34 & 290 & 2.54$\pm$0.05& 7.16 & 0.56$\pm$0.12 & 698$\pm$110&10.5&  & SGU \\ 
J2117.9+3729$^\bigstar$ & 82.88 & -8.27 & 231 & 2.34$\pm$0.07& 4.25 & 0.32$\pm$0.12 & 1274$\pm$349&3.7&  & SGU \\ 
J2218.8+4736$^\bigstar$ & 98.08 & -7.78 & 153 & 2.31$\pm$0.08& 5.35 & 0.5$\pm$0.13 & 1504$\pm$276&8.4&  & msp+ \\ 
J2220.8+6319$^\bigstar$ & 106.95 & 5.22 & 197 & 2.6$\pm$0.05& 6.26 & 0.34$\pm$0.08 & 396$\pm$111&8.5& 5 & SGU

\enddata

\tablenotetext{\diamond}{ 4FGL-DR4 value}
\tablenotetext{\ddag}{$^\bigstar$ and  $^\lozenge$ symbols indicate that the source belongs to the ``orphan\_A"  and  "orphan\_B" sample, respectively.}
\tablenotetext{\dagger}{omitted if SigCurv is less than 2 $\sigma$ or Epeak is outside the 30--10$^{5}$ MeV range.}
\tablenotetext{\star}{A value greater than 27.68 indicates that the source is variable at a $>$99\% confidence level.}
\tablenotetext{\vartriangle}{
Flags from 4FGL-DR4.
Their meanings are briefly summarized here \citep[see ][ for details]{LAT23_4FGL_DR4}.
1: TS$<$25 with other model or analysis; 2: Moved beyond 95\% error ellipse; 3: Flux changed with other model or analysis; 4: Source/background ratio$<$10\%; 5: Confused; 6: Interstellar gas clump (c sources);  10: Bad spectral fit quality; 14: Located in a high-density region of SGUs.}
\vspace{-7mm}
%
\end{deluxetable*}
\end{longrotatetable}
\tabletypesize{\normalsize}

\startlongtable
\begin{deluxetable*}{|c|cc|c|}
\label{tab:bright_SGUs_2} 
\tablecaption{{\sl eRosita} and {\sl Swift} counterparts of sources in the bright GU sample}
\tablehead{\colhead{Source Name} & \colhead{Probability$^\dagger$} & \colhead{ {\sl eRosita} name} & \colhead{ {\sl Swift} name} }
\startdata
J0057.9+6326 &  &  not visible & 2SXPS J005758.3+632640  \\ 
J0616.5+2235 & 0.98 & J061637.6+223630 & \nodata  \\ 
J0722.4$-$2650 & 0.75 & J072219.3$-$264734 & \nodata  \\ 
J0725.7$-$0549 & 1.0 & J072547.6$-$054826 & 2SXPS J072547.5$-$054830  \\ 
J0752.0$-$2931 & 0.82 & J075210.2$-$293018 & \nodata  \\ 
J0754.9$-$3953 & 0.07 & J075450.4$-$395029 & \nodata  \\ 
J0828.4$-$4444 & 0.06 & J082844.1$-$444756 & \nodata  \\ 
J0848.8$-$4328 & 0.32 & J084833.4$-$432701 & 2SXPS J084819.2$-$432919  \\ 
J0853.6$-$4306 & 0.09 & J085313.9$-$431141 & \nodata  \\ 
J0854.8$-$4504 & 0.13 & J085452.6$-$450623 & \nodata  \\
J0859.2$-$4729 & 0.77 & J085905.4$-$473042 & 2SXPS J085857.5$-$472724  \\ 
J0859.3$-$4342 & 0.57 & J085925.9$-$434514 & 2SXPS J085903.4$-$434834  \\ 
J0917.9$-$4755 & 0.07 & J091859.9$-$475800 & \nodata  \\ 
J0928.4$-$5256 & 0.82 & J092818.4$-$525706 & 2SXPS J092818.6$-$525703  \\ 
J0933.8$-$6232 & \nodata & \nodata & 2SXPS J093400.9$-$623350 \\
J1020.4$-$5314 & 0.24 & J102001.5$-$531341 & \nodata  \\ 
J1046.7$-$6010 & 0.23 & J104726.8$-$600529 & \nodata  \\ 
J1058.4$-$6625 & 0.98 & J105831.8$-$662600 & 2SXPS J105832.0$-$662602  \\ 
J1115.1$-$6118 & 0.7 & J111507.3$-$611538 & 2SXPS J111510.7$-$611639  \\ 
J1127.9$-$6158 & 0.71 & J112753.9$-$620125 & \nodata  \\ 
J1216.8$-$5955 & 0.25 & J121631.9$-$595430 & \nodata  \\ 
J1257.0$-$6339 & 0.38 & J125719.7$-$633845 & \nodata  \\ 
J1309.1$-$6223 & 0.8 & J130855.3$-$622442 & \nodata  \\ 
J1312.3$-$6257 & \nodata & \nodata & 2SXPS J131238.0$-$625328  \\ 
J1325.3$-$5413 & 0.32 & J132528.1$-$541137 & \nodata  \\ 
J1329.9$-$6108 & 0.57 & J132939.7$-$610746 & \nodata \\
J1412.1$-$6631 & 0.09 & J141221.4$-$662742 & \nodata  \\ 
J1415.4$-$6458 & 0.19 & J141514.8$-$650207 & \nodata  \\ 
J1443.7$-$7037 & 0.24 & J144410.0$-$704008 & 2SXPS J144405.4$-$703957  \\ 
J1444.9$-$5939 & 0.28 & J144518.4$-$593813 & \nodata  \\ 
J1526.3$-$4501 & 0.89 & J152618.5$-$450244 & 2SXPS J152618.3$-$450242  \\ 
J1536.8$-$4327 & 0.27 & J153635.6$-$432554 & \nodata  \\ 
J1547.4$-$4802 & 0.71 & J154715.8$-$480012 & \nodata  \\ 
J1613.0$-$5102 & \nodata & \nodata & 2SXPS J161314.9$-$510002  \\ 
J1616.6$-$5341 & 0.64 & J161648.6$-$534141 & 2SXPS J161648.7$-$534141  \\ 
J1616.6$-$5009 & 0.43 & J161637.9$-$495844 & 2SXPS J161702.4$-$500350  \\ 
J1618.0$-$5119 & 0.45 & J161818.0$-$511555 & \nodata  \\ 
J1649.3$-$4441 & 0.74 & J164921.7$-$444358 & 2SXPS J164906.8$-$444217  \\ 
J1653.2$-$4349 & 0.2 & J165323.7$-$435144 & \nodata  \\ 
J1703.6$-$2850 & \nodata & \nodata & 2SXPS J170341.9$-$284746  \\ 
J1711.0$-$3002 & 0.42 & J171103.0$-$295839 & \nodata  \\ 
J1714.8$-$3849 & 0.21 & J171502.4$-$384645 & \nodata  \\ 
J1714.9$-$3324 & \nodata & \nodata & 2SXPS J171452.2$-$332601  \\ 
J1721.3$-$5257 & 0.33 & J172135.8$-$525437 & \nodata  \\ 
J1721.7$-$3917 & 0.13 & J172143.0$-$392201 & \nodata  \\ 
J1729.9$-$4148 & 0.56 & J172945.8$-$414831 & 2SXPS J172946.3$-$414826  \\ 
J1737.3$-$3332 & 0.51 & J173733.5$-$333236 & 2SXPS J173732.2$-$333539  \\ 
J1740.7$-$2640 & \nodata & \nodata & 2SXPS J174047.9$-$263923  \\ 
J1748.8$-$3915 & \nodata & \nodata & 2SXPS J174854.0$-$391739  \\ 
J1752.8$-$4449 & \nodata & \nodata & 2SXPS J175246.4$-$444845  \\ 
J1757.4$-$3125 & 0.4 & J175718.3$-$312250 & \nodata  \\ 
J1804.4$-$0852 & &  not visible & 2SXPS J180425.0$-$085003  \\ 
J1808.4$-$3358 & \nodata & \nodata & 2SXPS J180825.1$-$335615  \\ 
J1808.5$-$3701 & \nodata & \nodata & 2SXPS J180827.5$-$365842  \\ 
J1809.2$-$2726 &  & not visible & 2SXPS J180926.9$-$272838  \\ 
J1817.2$-$3035 &  & not visible & 2SXPS J181720.3$-$303256  \\ 
J1832.4$-$0847 &  & not visible & 2SXPS J183222.5$-$084545  \\ 
J1836.8$-$2354 &  & not visible & 2SXPS J183658.0$-$234453  \\
J1904.7$-$0708 &  & not visible & 2SXPS J190444.5$-$070739  \\
J1855.2+0456 &  & not visible & 2SXPS J185502.9+045946  \\ 
J2038.4+4212 &  & not visible & 2SXPS J203815.9+421201  \\ 
J2109.6+3954 &  & not visible & 2SXPS J210936.2+395514

\enddata
\tablenotetext{^\dagger}{
Probability computed as described in Section \ref{sec:erosita}
}
\end{deluxetable*}

\startlongtable
\begin{deluxetable*}{|l|l|l|}
\label{tab:bright_SGUs_3} 
\tablecaption{Noteworthy bright GUs}
\tablehead{\colhead{Source Name} & \colhead{Comment$^\dagger$}  & \colhead{References$^\star$} }
\startdata
J0057.9+6326 & now assoc. with 3HSP J005758.4+632639 & \\
J0204.7+6656 & pulsar-like in the radio & Bru23 \\
J0235.3+5650 & pulsations detected & E@H\\
J0237.8+5238 &  ass. with the pulsar PSR J0237+5238   & Liu25\\
J0340.4+5302 & spectral break, possibly related to LHAASO J0341+5258 & Abd22 \\  & searched for pulsations & E@H \\
J0426.5+5434 & spectral break, searched for pulsations  & Abd22, E@H \\  
J0616.5+2235 & spatially coinc. with IC 443, extended (16.1) &  \\
J0725.7$-$0549 & now associated with the BL Lac 3HSP J072547.9$-$054832 &  \\
J0736.9$-$3231 & pulsations detected & E@H\\
J0744.9$-$4028 & low-probability association with an AGN &  \\
 & searched for pulsations &  PSC\\
J0752.0$-$2931 & pulsar-like in the radio & Bru23 \\
J0754.9$-$3953 & pulsar-like in the radio, searched for pulsations & Bru23, PSC \\
 &  now ass. with the UNK NVSS J075452$-$395317 &  \\
J0758.8-1450 & searched for pulsations & Ker25, E@H\\
J0826.1$-$5053 & searched for pulsations & PSC \\  
J0828.4$-$4444 & now SPP, assoc. with SNR G263.9-03.3 & \\
J0848.8$-$4328 & in the  Vela Molecular Ridge & \\
J0854.8$-$4504 & extended (6.4), in the  Vela Molecular Ridge & \\ &  searched for pulsations  &  E@H\\
J0857.7$-$4256c & possibly ass. with IRS 31 &  Per24 \\
J0859.2$-$4729 & possibly ass.  with RCW 38, extended (25.8), & Ge24, Pan24 \\
               &   coinc. with WISE G267.935-01.075 & \\
J0859.3$-$4342 & possibly ass. with RCW 36, searched for pulsations &  Per24, E@H \\
J0917.9$-$4755 & possibly ass. with RCW 41  &  Per24\\
J0928.4$-$5256 & spectrum compatible with an AGN, variable  & \\
J0933.8$-$6232 & searched for pulsations & Ker25, E@H\\
J1026.2$-$5731 & coinc. with WISE G284.362+00.025, searched for pulsations  & E@H \\
J1037.8$-$5810 & coinc. with WISE G285.808+00.106, in RCW 51 (NGC 3293)  &  \\
 & searched for pulsations & E@H\\
J1046.7$-$6010 & in the  Carina Nebula Complex, extended (206.6) & Ge24 \\
& searched for pulsations & E@H \\    
J1048.5$-$5923 & in the  Carina Nebula Complex, extended (176.3) & Ge24 \\
& searched for pulsations & E@H \\ 
J1058.4$-$6625 & now ass. with the UNK  1eRASS J105831.8$-$662600 &\\
               & searched for pulsations & PSC \\ 
J1112.2$-$6055 & extended (43.8), coinc. with WISE G291.154$-$00.321  &  2FGES \\
J1115.1$-$6118 & spatially coinc. with the  young star cluster NGC 3603, &  \\
&  extended (64.4) & \\
J1208.0$-$6900 & pulsations detected, MSP PSR J1207$-$6900 &  Cla24\\
J1257.0$-$6339 & spatially coinc. with a radio structure,   &  \\
               & coinc. with WISE G303.445$-$00.745 & \\
J1309.1$-$6223 & coinc. with WISE G305.056+00.372  &  \\
J1312.3$-$6257 & coinc. with the open cluster Danks 1, extended (34.7), & Liu04, 2FGES \\
               & coinc. with WISE G305.322$-$00.255 & \\
J1312.6$-$6231 & coinc. with WISE G305.503+00.214, in the open cluster Danks 1 &  \\
J1317.5$-$6316 & coinc. with the open cluster Danks 1, extended (12.3) &  Liu04 \\
J1329.9$-$6108 & searched for pulsations & E@H \\ 
J1325.3$-$5413 & searched for radio pulsations & PSC \\
 & searched for pulsations & E@H \\
J1351.6$-$6142 & spectral break, coinc. with WISE G309.917+00.342  &  Abd22 \\
               & and PSR J1352$-$6141, searched for pulsations & E@H \\
 J1403.5$-$6236 & searched for pulsations & E@H \\  
 J1404.8$-$5237 & searched for pulsations & PSC \\  
J1412.1$-$6631 & extended (6.5), searched for pulsations & PSC  \\
J1415.4$-$6458 & pulsations detected, MSP PSR J1415.4$-$6458 & Ker25 \\
J1427.8$-$6051 & HESS J1427$-$608, PSR/PWN?, extended (38.4)& Guo17, Dev21 \\
& searched for pulsations & E@H \\
J1444.9$-$5939 & association with  PSR J1444$-$5941 deemed unlikely & \\
& searched for pulsations & E@H \\ 
J1510.1$-$5750 & coinc. with WISE G320.590+00.190  &  \\
J1517.9$-$5233 & coinc. with an optical variable, searched for pulsations & Tur26, E@H\\
J1534.0$-$5232 & spectral break, searched for pulsations  & Abd22, PSC \\  
J1536.8$-$4327 & variable & \\
J1603.3$-$6010  &  pulsations detected, MSP PSR J1603$-$6011 & Ker25  \\
J1610.3$-$5154c & spatially coinc. with the large molecular cloud Clump 12 & \\
J1611.9$-$5125c & spatially coinc. with a radio structure,  &  \\
                & coinc. with WISE G331.580$-$00.022  &  \\
J1613.0$-$5102 & now ass. as SPP with SNR G332.0+00.2 & \\
J1616.6$-$5009 & association with PSR J1616$-$5017 deemed unlikely,  &  \\ 
 & searched for pulsations & E@H\\
J1616.6$-$5341 & pulsation detected & E@H \\
J1618.0$-$5119 & coinc. with WISE G332.394$-$00.668  &  \\
J1620.8$-$4958 & coinc. with WISE G333.580+00.058  &  \\
J1622.7$-$4934c & within a cluster of molecular clouds near an HII region, & \\
                & coinc. with WISE G334.022+00.106 &  \\
J1634.0$-$4742c & coinc. with WISE G336.753+00.097 & \\
J1636.9$-$4710 & coinc. with WISE G337.415+00.034, searched for pulsations & E@H \\
J1639.3$-$5146 & coinc. with an optical variable, searched for pulsations  & Tur26, E@H\\
J1639.8$-$4642 & coinc. with WISE G338.080$-$00.028  &  \\
& searched for pulsations & PSC, E@H \\
J1643.3$-$3148 & searched for pulsations & PSC \\
J1649.3$-$4441 & spatially coinc. with the young stellar cluster NGC 6216 & \\
& searched for pulsations & E@H \\
J1653.2$-$4349 & located in an IRAS HII region  & \\
& searched for pulsations & PSC, E@H \\  
J1656.5$-$2733  & searched for pulsations & PSC \\   
J1703.6$-$2850  & searched for pulsations & PSC \\   
J1711.0$-$3002 & pulsar-like in the radio, searched for pulsations & Bru23, PSC \\ 
J1714.9$-$3324 & searched for radio pulsations, extended (44.9) & PSC\\
J1714.8-3849 & now ass. with the UNK NVSS J171456$-$384814 & \\ 
J1721.3$-$5257 & searched for pulsations & PSC \\
J1721.7$-$3917 & now associated with the UNK  NVSS J172157$-$391740 & \\
J1727.6$-$2304 & searched for pulsations & PSC \\  
J1729.1$-$3503 & coinc. with WISE G352.932$-$00.374  &  \\
J1730.8$-$3806 & low-probability association with the AGN & \\                    & AT20G J173121$-$380212& \\
J1739.3$-$2531 & searched for radio pulsations & PSC\\
J1740.7$-$2640   & searched for pulsations & E@H \\  
J1742.8$-$2246 & spectral break & Abd22 \\
J1743.7$-$4321 & searched for radio pulsations & PSC\\
J1744.0$-$1311   & searched for pulsations & PSC \\   
J1747.0$-$3505   & searched for pulsations & PSC \\   
J1748.3$-$2906 & coinc. with WISE G000.120$-$00.556  &  \\
J1748.8$-$3915 & searched for pulsations & PSC \\
J1752.8$-$4449 & pulsations detected, MSP PSR J1752$-$4450 &  \\
J1753.8$-$2538 & extended (6.3), searched for pulsations & E@H  \\
J1754.6$-$2933 & searched for pulsations & E@H \\  
J1758.7-4109  & searched for pulsations & E@H \\    
J1800.9$-$2407 & region of the middle-aged SNR W28 (TeV), searched for pulsations & E@H \\  
J1801.6$-$2326 & region of the middle-aged SNR W28  (TeV), extended (15.4) & \\
& searched for pulsations & PSC \\  
J1801.8$-$2358 & HESS J1800-240B, SFR G5.89$-$0.3?,    &  Ham16 \\
& now ass. with UNK NVSS J180200-235857 & \\
J1802.4$-$3041  & searched for pulsations & E@H \\  
J1805.1$-$3618   & searched for pulsations & PSC \\  
J1808.4$-$3358 & searched for radio pulsations & PSC, E@H\\
J1808.5$-$3701 & now ass. with the PSR SAX J1808.4$-$3658 & \\
J1812.8$-$3144 & AGN candidate, VVV J181252.80$-$314443.5 & Don24
\\
& searched for pulsations & PSC \\  
J1814.7$-$3420 & low-probability association with an AGN & \\
J1817.9$-$3334 & AGN candidate,  VVV J181751.39$-$333117.3 & Don24 \\
& searched for pulsations & PSC \\  
J1819.9$-$1530 & possible ass. with a spider candidate & Lu25\\
 & now ass. as SPP with SNR G015.5$-$00.1 & \\
J1823.3$-$1340 & extended (6.6), searched for pulsations & E@H    \\
J1830.8$-$3132   & searched for pulsations & PSC \\  
J1834.7$-$0724 & coinc. with WISE G024.507+00.239  &  \\
J1836.8$-$2354 & ass. with the  globular cluster M22 (NGC 6656) & \\
J1840.4$-$1139 & now ass. as SPP with SNR G021.8-03.0 & \\ 
J1847.2$-$0141 & ass. with HESS J1848$-$018, corresponding to W43?, & Yang20 \\
               & coinc. with WISE G030.796+00.183  & \\
J1852.6+0203   & searched for pulsations & E@H \\   
J1855.2+0456 & ass. with PSR J1855+0455g?  & \\
& searched for pulsations & E@H \\  
J1857.1+0056   & searched for pulsations & PSC, E@H \\
J1858.0+0354 & ass. with a MSFR?, search. for radio pulsations, extended (15.1) & Wan22, PSC, E@H \\
J1900.4+0339 & extended (11.9), coinc. with WISE G037.370$-$00.368 in W47 &   \\
& searched for pulsations & PSC, E@H \\  
J1901.1+0730   & searched for pulsations & E@H \\   
J1901.8$-$0718 & now ass. with the binary  4FGL J1901.8$-$0718 & \\
J1902.2+0448 & coinc. with WISE G038.365$-$00.062  &  \\
J1902.5+0654 & coinc. with WISE G040.154+00.648  &  \\
J1904.7$-$0708 & ass. with PSR J1904$-$0708 & Fan25 \\
J1906.9+0712 & spectral break, extended (10.1), coinc. with WISE G041.074$-$00.162  &  Abd22 \\
 & searched for pulsations & PSC \\   
J1908.7+0812 & ass. with PSR J1908+0811g?, coinc. with WISE G042.227$-$00.067 & \\ & searched for pulsations  & E@H  \\
J1910.2+0904c & coinc. with WISE G043.170$-$00.004 in W49A, &  \\
& now ass. as the UNK NVSS J191011+090510 & \\
J1912.7+0957 & HESS J1912+101, SNR? & Zen21 \\
 J1926.4+1602 & searched for pulsations & PSC \\   
J1929.0+1729 & ass. with HESS J1928+181 (a PWN?), extended (53.7) & \\
& searched for pulsations & E@H \\
J1931.1+1656 & spectral break, extended (28.9),  SNR G52.37$-$0.70? & Abd22,  And17 \\
& searched for pulsations & PSC, E@H \\
J1948.9+3414 &  coinc. with the cataclysmic variable V1449 Cyg & \\
& now associated with UNK  1RXS J194917.1+34104 & \\
J1953.5+3841 & pulsations detected, PSR J1954+3852 & \\
J2027.0+2811 & searched for radio pulsations & PSC, E@H \\
J2038.4+4212 & in Cygnus, coinc. with the SFR DR 21, spectral break,  & Yan24, Abd22 \\
             & extended (219.1), searched for pulsations, now ass. & E@H\\
             & with the UNK NVSS J203901+421941 &  \\
J2041.1+4736 & searched for pulsations & E@H\\             
J2051.7+5051 & pulsations detected, PSR J2051+5050 & Cla24 \\      
J2108.0+5155 & spectral break, extended (42.4), coinc. with LHAASO J2108+5157 & Abd22 \\
 & searched for pulsations & PSC \\ 
J2109.6+3954 & now associated with BLL WISEA J210936.14+395513.5  & \\
J2114.3+5023 & pulsar-like in the radio & Bru23 \\
J2116.2+3701 & pulsations detected, now PSR J2116+3701 & Don23 \\
J2117.9+3729 &  extended, FHES 2116.4+370, searched for pulsations & Ack18, PSC    
\enddata
\tablenotetext{^\dagger}{
For extended sources, the LP\_TSCurv value is given in parentheses.
}
\tablenotetext{^\star}{
A blank entry means that the observation reported in the comment column was made in this work.
2FGES:\, \cite{2FGES}, Abd22:\,  \cite{Abd22}, Ack18:\,  \cite{Ack18}, And17:\, \citep{And17},  Bru23:\, \cite{Bru23}, Cla24:\, TRAPUM collaboration (https://www.trapum.org/discoveries), Dev21:\, \cite{Dev21}, Don23:\,\cite{Don23}, Don24:\,\cite{Don24}, E@H:\,\cite{Ein10}, Fan25:\, \cite{Fan25}, Ge24:\, \cite{Ge24}, Guo17:\, \cite{Guo17}, Ker25:\, \cite{Ker25},  Ham16:\, 
\cite{Ham16}, Liu04:\,  
\cite{Liu04}, Liu25:\,  
\cite{Liu25}, Lu25:\, \cite{Lu25},  Pan24:\, \cite{Pan24}, Pan25:\, \cite{Pan25}, Per24:\, \cite{Per24}, PSC:\, Fermi Pulsar Search Consortium, 
Tur24:\, \cite{Tur24}, Tur26:\, \cite{Tur26}, Wan22: \cite{Wan22}, 
Yan17:\, \cite{Yan17}, Yan24:\, \cite{Yan24}, Yang20:\, \cite{Yang20}, Zen21:\, \cite{Zen21}}
\end{deluxetable*}

\end{document}